\documentclass[10pt,journal,cspaper,compsoc]{IEEEtran}

\ifCLASSOPTIONcompsoc
\else
\fi
\ifCLASSINFOpdf
\else
\fi
\newcommand{\subparagraph}{}

\usepackage[square, comma, sort&compress, numbers]{natbib}
\usepackage{graphicx}
\usepackage{array}
\newcommand{\ApproxSign}{\raise.17ex\hbox{$\scriptstyle\sim$}}
\usepackage[table]{xcolor}
\usepackage{multirow}
\usepackage{titlesec}

\let\labelindent\relax
\usepackage{enumitem}
\setlist[enumerate]{leftmargin=*}

\newcommand{\revised}[1]{{#1}}

\usepackage{array}
\newcolumntype{C}[1]{>{\centering\let\newline\\\arraybackslash\hspace{0pt}}m{#1}}

\usepackage{textcomp}
\usepackage[normalem]{ulem}
\usepackage{url}
\DeclareUrlCommand\ULurl{%
  \renewcommand\UrlLeft{\bgroup}%
  \renewcommand\UrlRight{\egroup}}
\usepackage[linesnumbered,lined,ruled,commentsnumbered]{algorithm2e}
\usepackage{amsmath}
\usepackage[table]{xcolor}
\usepackage{color}

\usepackage{amssymb}
\usepackage{pifont}
\newcolumntype{L}{>{\centering\arraybackslash}m{3.5cm}}
\newcolumntype{G}{>{\centering\arraybackslash}m{10.5cm}}
\begin{document}

%
\title{A Survey of Techniques for Dynamic Branch Prediction}

\author{Sparsh Mittal
 
\IEEEcompsocitemizethanks{\IEEEcompsocthanksitem The author is with IIT Hyderabad, India.  Support for this work was provided by Science and Engineering Research Board (SERB),
India, award number ECR/2017/000622. \protect\\
}
\thanks{}}


%
%

\markboth{}%
{author : A Survey }

\onecolumn
\IEEEcompsoctitleabstractindextext{%

\begin{abstract}
Branch predictor (BP) is an essential component in modern processors since high BP accuracy can improve performance and reduce energy by decreasing the number of instructions executed on wrong-path. However, reducing latency and storage overhead of BP while maintaining high accuracy presents significant challenges. In this paper, we present a survey of dynamic branch prediction techniques. We classify the works based on key features to underscore their differences and similarities. We believe this paper will spark further research in this area and will be useful for computer architects, processor designers and researchers. 
 
\end{abstract}

}

\maketitle

\IEEEdisplaynotcompsoctitleabstractindextext

%
\IEEEpeerreviewmaketitle

%

%
%
%

%


\section{Introduction}

Control-changing instructions, such as branches add uncertainty in execution of dependent instructions and thus, lead to large performance loss in pipelined processors. To offset their overhead, accurate prediction of branch outcomes is vital. Since branch misprediction incurs high latency (e.g., 14 to 25 cycles \cite{seznec2002design}) and wastes energy due to execution of instructions on wrong-path, an improvement in BP accuracy can boost performance and energy efficiency significantly. For example, experiments on real processors showed that reducing the branch mispredictions by half improved the processor performance by  13\% \cite{jimenez2011optimized}.
 
Effective design and management of BPs, however, presents several challenges. Design of BP involves a strict tradeoff between area/energy, accuracy and latency. An increase in BP complexity for improving accuracy may make it infeasible for implementation or offset its performance benefit. For example, a 512KB perceptron predictor may provide lower IPC\footnote{Following acronyms are frequently used in this paper:  basic block (BB), branch history register (BHR), branch predictor (BP),   branch target buffer (BTB),     global history (GH) register (GHR), instruction-per-cycle (IPC), (inverse) fast Fourier transform (IFFT/FFT), least/most significant bit (LSB/MSB), pattern history table (PHT),  program counter (PC), return address stack (RAS), TAGE (tagged geometric history length BP) } 
 than its 32KB version \cite{jimenez2003reconsidering} and a 2-cycle fully accurate BP may provide lower IPC than a 1-cycle partially inaccurate BP \cite{jimenez2000impact}. Further, due to its high access frequency, BP becomes a thermal hot-spot and hence, reducing dynamic and leakage energy of BP is important. Clearly, intelligent techniques are required for resolving these issues. Several recently proposed techniques seek to address these challenges.

\begin{figure} [htbp]
\centering
\includegraphics[scale=0.40]{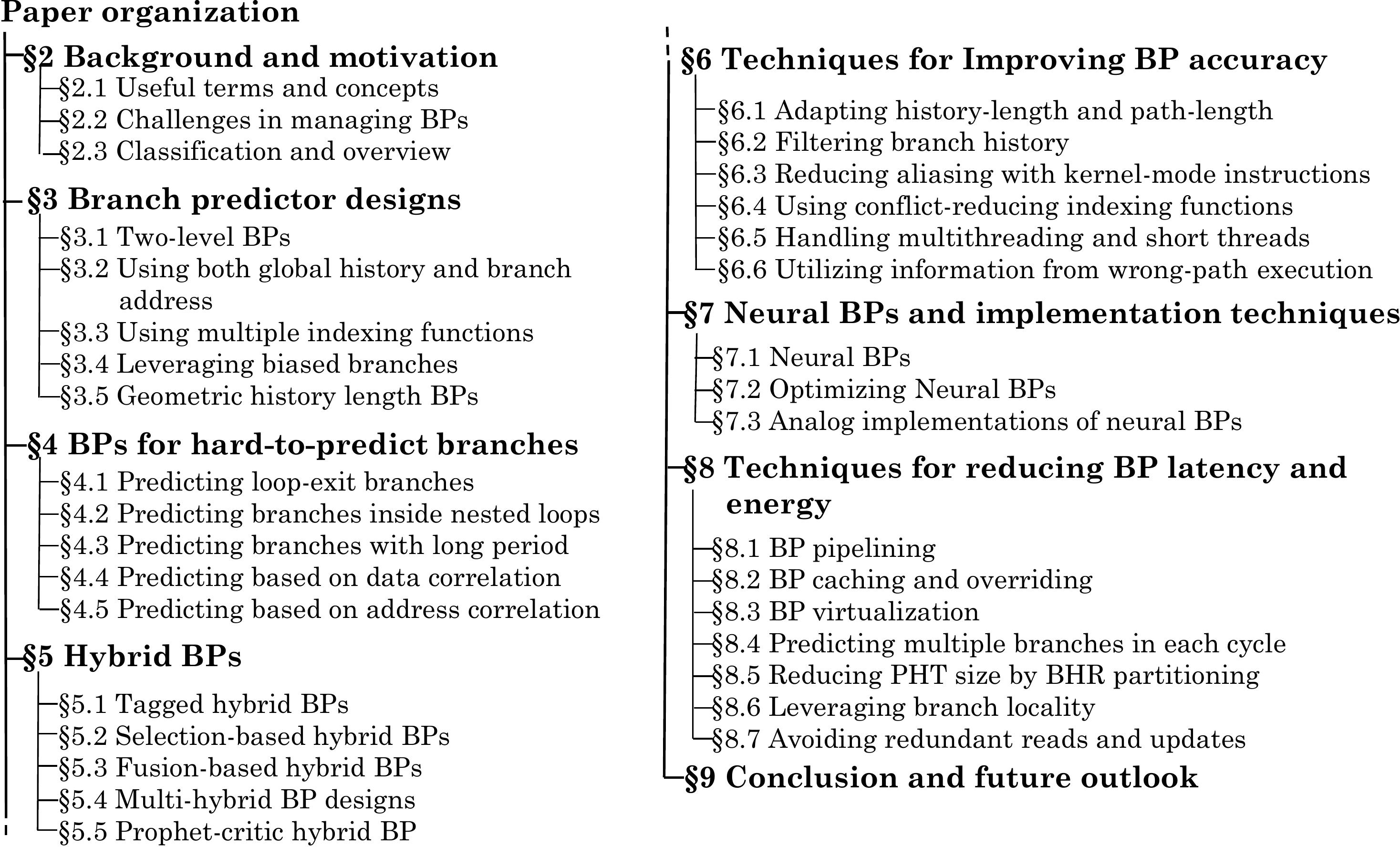}
\caption{ Paper organization  }\label{fig:organization}
\end{figure}

\textbf{Contributions:} In this paper, we present a survey of dynamic branch prediction techniques. Figure \ref{fig:organization} shows the organization of this paper. Section \ref{sec:background} presents a brief background on BPs, discusses the challenges in managing BPs and shows the overview of research works. Sections \ref{sec:bptechniques} and \ref{sec:sideBP} discuss several BP designs for common and specific branches and Section \ref{sec:hybridBPs} discusses hybrid BP designs. Section \ref{sec:improvingaccuracy} discusses techniques for improving BP accuracy. Section \ref{sec:neuralBPs} discusses the neural BPs and techniques to reduce their implementation overheads. Section \ref{sec:latencyenergy} discuses techniques for reducing latency and energy overheads of BPs. We conclude the paper in Section \ref{sec:conclusion} with a mention of future challenges. 
 
\textbf{Scope:} To effectively summarize decades of BP research within the page limits, we had to restrict the scope of this paper in the following way. We focus on branch direction (outcome) prediction which guesses \textit{whether} a conditional branch will be taken or not. We do not include branch \textit{target} prediction or the techniques for indirect or unconditional branches. Static  branch prediction uses only source-code knowledge or compiler analysis to predict a branch \cite{smith1981study} whereas dynamic prediction accounts for time-varying and input-dependent execution pattern of a branch. Since techniques for static prediction and those involving compiler hints to assist dynamic BPs merit a separate survey, we do not include them here and focus on dynamic prediction only.  We  generally focus on key idea of each paper and only include selected quantitative results. 
This paper is expected to be useful for computer architects, chip-designers and researchers interested in performance optimization.

\section{Background and motivation}\label{sec:background}
Here, we present a quick background on BP and refer the reader to prior works for more details on taxonomy \cite{skadron2000taxonomy}, value locality \cite{mittal2017ValueLocality}, static BPs \cite{smith1981study}, comparative evaluation of multiple BPs \cite{jimenez2011optimized,young1995comparative,loh2005simple,evers1996using,jimenez2003reconsidering}  and discussion of commercial BPs \cite{bonanno2013two,jimenez2003reconsidering,fog2011microarchitecture,juan1998dynamic,milenkovic2002demystifying,gwennap1996digital}. We discuss the BP naming scheme \cite{yeh1993comparison,skadron2000taxonomy} in Section \ref{sec:twolevelpredictors}.

\subsection{Useful terms and concepts}

\textbf{Types of branches:} Branches can be classified into conditional or unconditional
and direct or indirect. A direct branch specifies the target address in the instruction itself, whereas indirect branch mentions where the target address is to be found (e.g., a memory or register location).

\textbf{Biased branches:} are either taken or not-taken most of the time. An example of this is checking for  a rare error condition. Biased branches form a large fraction of total branches seen in real programs \cite{gope2014bias}. Biased branches are detected as those showing no change in the direction \cite{gope2014bias} or as those whose perceptron weights have reached to a maximum/minimum value \cite{loh2005reducing}.

\begin{figure} [htbp]
\centering
\includegraphics[scale=0.40]{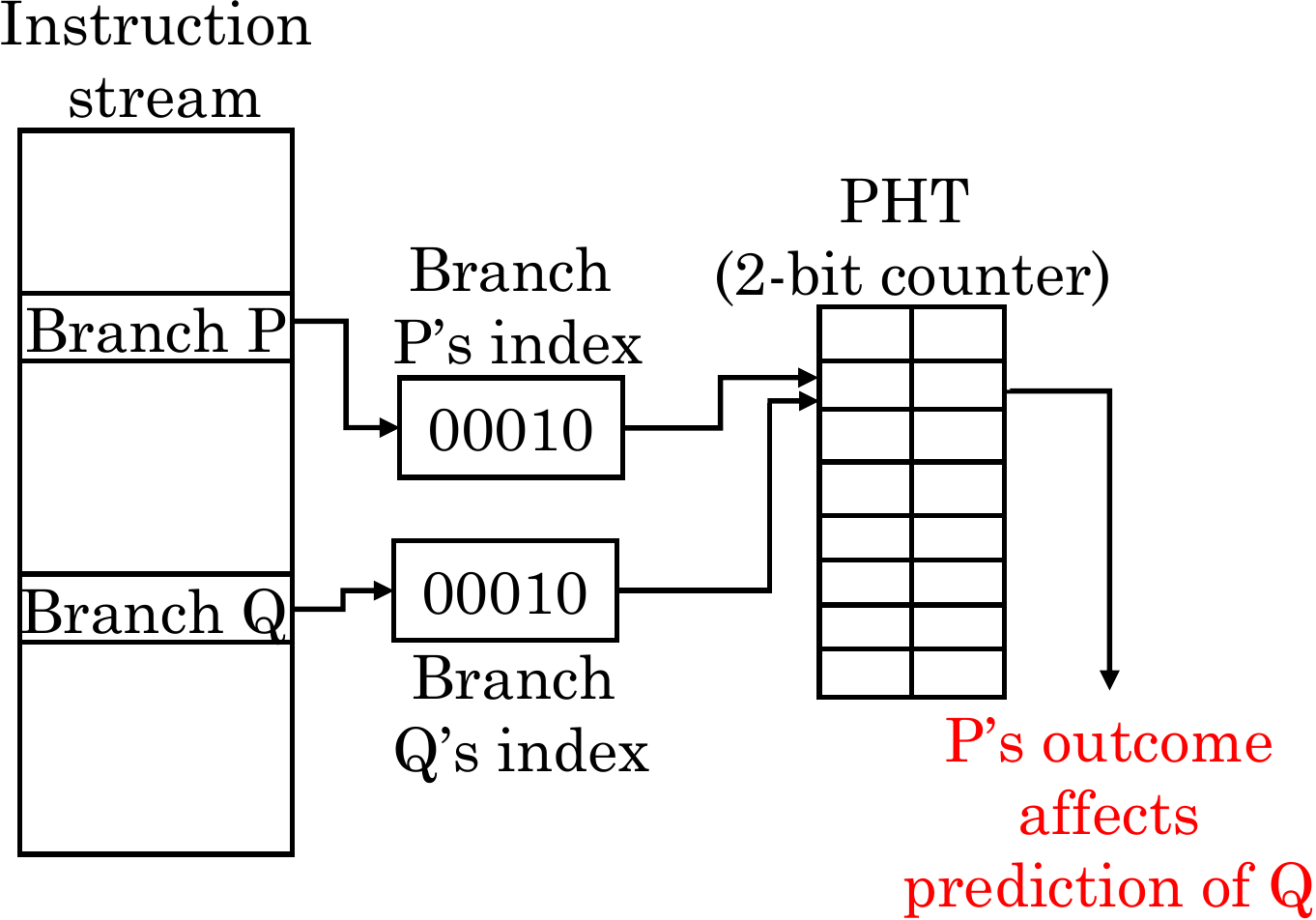}
\caption{\revised{An example of interference between branches \cite{sprangle1997agree}} }\label{fig:interference}
\end{figure}

\textbf{Aliasing or interference:} When multiple unrelated branches use the same predictor entry, they create destructive interference which is termed as aliasing. \revised{Figure \ref{fig:interference} presents an example of interference between branches.} The interference is neutral (harmless) if it does not change the prediction, positive (beneficial) if it corrects the prediction and negative (harmful) if it causes a misprediction. Generally, both the frequency and the magnitude of negative aliasing is higher than that of positive aliasing \cite{young1995comparative}, especially for workloads with large working set.

\textbf{Side predictor: } Commonly used BPs (e.g., TAGE \cite{seznec2006case}) can accurately predict most of the branches, however, they fail to predict other relatively-infrequent classes of branches. To predict such branches, the main BP can be augmented with side (also called auxiliary or complementary or corrector) BPs for improving the overall accuracy.

\textbf{Local and global history:} Some   branches can be predicted by their own execution pattern only and thus, are said to work on `local' (or self or per-branch) history. For example, consider the loop-controlling branch in {\tt for(i=1; i<=4; i++)}. If the loop test is performed at the end of the loop body, it shows the pattern $(1110)^n$ and thus, by knowing the outcome of the branch for last 3 instances, the next outcome can be predicted. 

\begin{figure} [htbp]
\centering
\includegraphics[scale=0.40]{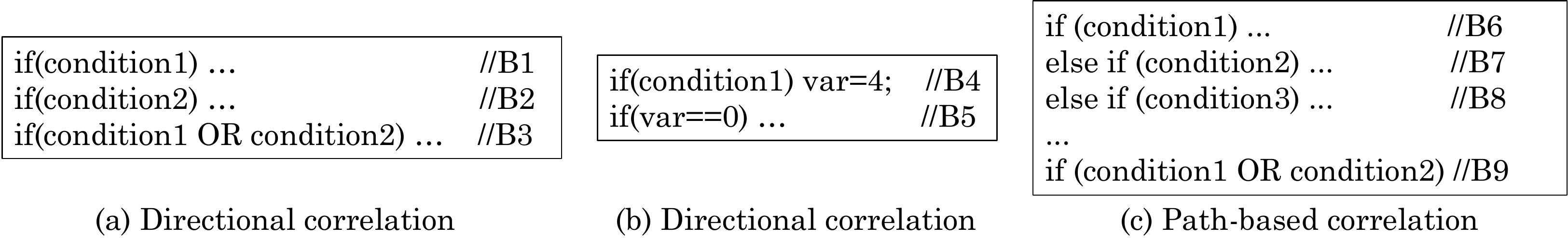}
\caption{ Examples of origin of branch-correlation \cite{evers1998analysis} }\label{fig:OriginOfCorrelation}
\end{figure}

By comparison, if execution of previous branches can provide a hint about the outcome of a candidate branch, such branches are said to be correlated and the BP exploiting this information is said to work on `global history'. Branch correlations may arise due to multiple reasons \cite{evers1998analysis,young1995comparative}:

\begin{enumerate}
\item Two branches may be decided by related information, e.g.,  in Figure \ref{fig:OriginOfCorrelation}(a), if branch B1 is taken, branch B3 will also be taken. 
\item The outcome of one branch changes the condition affecting the outcome of another branch, e.g., in Figure \ref{fig:OriginOfCorrelation}(b), if branch B4 is taken, then B5 will not be taken. This case and the previous case are  examples of ``direction correlation''.  
\item The information about the path through which we arrived at current branch gives us hint about the branches before the correlated branch. For instance, in Figure \ref{fig:OriginOfCorrelation}(c), although the direction of branch B8 is not correlated with that of branch B9, if branch B8 was observed on the path (i.e., was one of the previous $K$ branches), then, outcome of branch B9 can be predicted. Such correlation is referred to as ``in-path correlation''. \emph{Path-history} consists of branches seen on the path to the current branch whereas \emph{pattern history} shows the outcome (direction) patterns  of branches that led to the current branch. Compared to pattern history, use of path history allows better exploitation of correlation \cite{young1995comparative}.
\end{enumerate}

In general, considering higher number of branches provides stronger correlation, e.g., in Figure \ref{fig:OriginOfCorrelation}(a), knowing of the outcome of B1 alone is not sufficient to predict B3, but knowing the outcome of both B1 and B2 is sufficient to predict B3. \revised{Note that  several works on neural BP use perceptron weights as a measure of strength of correlation between branches  \cite{gao2005adaptive,tarjan2005merging,akkary2004perceptron,jimenez2011optimized}.}

\textbf{Basic block:}  \revised{A basic block is a maximal length sequence of branch-free code \cite{cooper2011engineering}. Unless an exception happens, the instructions in a BB always execute together. The control enters a BB at the first instruction and exits at the last instruction.}

\textbf{Metrics: } As for metrics for evaluating BPs, research works have used performance (IPC), number of mispredicted branches (per kilo-instruction), micro-ops fetched and number of flushes per micro-ops, misprediction penalty \cite{nair1995dynamic}, etc.

\subsection{Challenges in managing BPs}\label{sec:challenges}
Effective design and operation of BPs presents several challenges, as we show below.  

\textbf{Latency constraints:}   Since BP lies on critical path, it needs to provide prediction within one cycle from the time the branch address is known. However, due to the complex designs of BPs, use of slow wires and high clock frequency, BP access latency can exceed one cycle \cite{seznec2005analysis,jimenez2002neural}. Apart from BP access latency, BP update latency also has large impact on the performance \cite{loh2006revisiting}.

\textbf{Area and implementation overheads:} Area and power considerations limit storage budget of BPs to tens of kilobytes  (e.g., 32-64KB), which prohibits storing and benefiting from correlations with branches in distant past \cite{seznec201164}. Even worse, blindly increasing BP size may provide marginal benefits, e.g., increasing the size of 2bc+gskew BP \cite{seznec1999dealiased} from 64KB to 1MB provides little improvement in accuracy \cite{seznec2002design}. This happens because a linear rise in global history length increases the BP size exponentially, and yet, extra branches included may not be correlated with branches being predicted which increases aliasing and BP training time. Furthermore, some BP designs may be strongly coupled with the architecture, e.g., data value based BPs may necessitate additional datapaths. Such dependencies   reduce portability and ease of implementation and upgradation.

\textbf{Energy overheads: } Since the BP needs to be accessed in nearly every cycle and its circuitry is aggressively optimized for latency, BP has high energy consumption and becomes a thermal hot-spot \cite{loh2005simple}. As power budget becomes the primary design constraint in all ranges of computing systems \cite{mittal2014SurveyDataCenter}, improvement in energy efficiency of BP has become vital to justify its use in modern processors.

\textbf{BP warmup requirements: } Large BPs may take long time to warm-up and this problem may be especially severe in presence of frequent context-switching. The slow warm-up may even bring their accuracy lower compared to the otherwise less-accurate simple BPs \cite{jimenez2002neural}. 

\textbf{Challenges in use of tags:} Unlike for cache, aliasing is acceptable in BP since a misprediction only impacts performance and not correctness. Hence, a direct-mapped BP can be tag-less. However, tags are required for implementing associativity or removing aliasing. Since each BP table entry is much smaller compared to a cache block (e.g., 2 bits vs 64 bytes), for the same total capacity, the number of entries in BP becomes much larger than the blocks in cache. Hence, the tag-size becomes disproportionately larger than the entry-size and the decoding complexity also becomes higher than that in the cache \cite{jimenez2003reconsidering}. \revised{Further, increasing the tag size beyond a threshold does not improve accuracy \cite{eden1998yags}. }

\textbf{Challenges of hybrid BPs:} Since no solo BP can predict all branch types, hybrid BPs have been proposed which use multiple component predictors along with a meta-predictor (also called selector or chooser) to choose the most-suitable component for each branch.  The limitation of hybrid BPs is that they need to use large meta-predictors for achieving high accuracy. This, however, further reduces the hardware budget and effectiveness of component predictors compared to an equal-budget solo BP.  
 
\textbf{Processor design factors:} Compared to single-issue in-order processors, wide-issue out-of-order processors show lower BP accuracy \cite{jimenez2000impact} since some predictions may be required even before recent branch results can update the GHR.  Also, increasing pipeline depth aggravates branch misprediction cost due to increased branch resolution time. This delays entry of correct path instructions in the pipeline and leads to filling of pipeline with instructions from wrong path \cite{aragon2001selective}, which aggravates the performance loss due to BP.

\subsection{Classification and overview}\label{sec:classification}
 
Table \ref{tab:classification} classifies the research works on several key parameters, e.g., objective of the technique, BP design features and techniques for improving accuracy.

\begin{table}[htbp]
\centering
\caption{A classification based objective, design and optimization features}
\label{tab:classification}
\begin{tabular}{|p{5.0cm}|p{8cm}|}\hline

 \multicolumn{1}{|c}{Category}   & \multicolumn{1}{|c|}{References} \\\hline\hline
\multicolumn{2}{|c|}{Overall goal}   \\\hline
Performance & Nearly-all\\\hline 
Energy &  \cite{bhattacharjee2017using,sethuram2007neural,ayoub2009filtering,xie2013energy,huang2015energy,schlais2016badgr,gao2008address,wang2013practical,amant2008low,baniasadi2004sepas,yang2006power,sendag2008low,loh2005reducing}  \\\hline

\multicolumn{2}{|c|}{BP designs and related features}\\\hline
Neural-network BPs &  \cite{monchiero2005combined,sethuram2007neural,jimenez2001dynamic,jimenez2003fast,saadeldeen2013memristors,wang2013practical,amant2008low,akkary2004perceptron,gao2005adaptive,jimenez2005piecewise,loh2005reducing,egan2003two,jimenez2002neural,jimenez2001perceptron,tarjan2005merging} \\\hline

Hybrid BPs & \cite{gwennap1996digital,seznec2002design,jimenez2001dynamic,baniasadi2004sepas,chaver2003branch,falcon2004prophet,kampe2002fab,baniasadi2002branch,evers1996using,evers2000improving}  \\\hline

Side-predictor &  \cite{seznec2011new,seznec2015inner,albericio2014wormhole,jimenez2011optimized,heil1999improving,gao2008address,sendag2008low,kampe2002fab} \\\hline
Use of branch cache &  \cite{jimenez2011optimized,yeh1993increasing,chang1994branch} \\\hline
Use of pipeline gating &  \cite{akkary2004perceptron,parikh2004power}   \\\hline
Use of static BP &  \cite{chang1994branch,kampe2002fab,evers1996using} \\\hline
Use of profiling for  & 
\revised{identifying biased branches \cite{parikh2002power}, choosing suitable hash function \cite{stark1998variable} or input vector  \cite{gao2005adaptive}} \\\hline
Use of genetic algorithm &  \cite{loh2002predicting} \\\hline

Modeling context switches &  \cite{eden1998yags,evers1996using,seznec1999dealiased,juan1998dynamic} \\\hline
Modeling impact of OS execution &  \cite{sechrest1996correlation,chen1996analysis,li2007aware} \\\hline

\multicolumn{2}{|c|}{Reducing aliasing and improving accuracy}\\\hline

Optimizing biased branches &  \revised{using static BP \cite{chang1996improving,young1995comparative} or side predictor \cite{seznec2011new,seznec1999dealiased} for them, not storing them in BH \cite{gope2014bias,jimenez2011optimized}, not updating all tables \cite{seznec2002design,chang1996improving} and skipping dot-product \cite{loh2005reducing} or BP access \cite{parikh2004power} for them} \\\hline

Predicting loop branches &  \cite{sherwood2000loop,sendag2008low} \\\hline
Correlating on data values  & \cite{alotoom2010exact,chen2003dynamic,gao2008address,heil1999improving,aragon2001selective} \\\hline
Modifying/filtering global history &  \cite{porter2009creating,ayoub2009filtering,kampe2002fab,xie2013energy,chang1996improving} \\\hline
Adapting history length &  \cite{seznec2005analysis,juan1998dynamic} \\\hline 
\end{tabular}
\end{table}

\section{Branch predictor designs}\label{sec:bptechniques}

The simplest 2bc (2-bit counter) BP uses some of the branch address bits to select a 2-bit counter from a 1D branch history table. The value of this counter decides the prediction. More complex BPs use both branch history and current branch address in a two-level organization (Section \ref{sec:twolevelpredictors}) or combine them in different ways to obtain index of a single-level predictor table (Section \ref{sec:gsharegselect}). Other predictors seek to improve accuracy by using multiple predictor tables accessed with different indexing functions (Section \ref{sec:multipleindexing}), exploiting biased branches (Section \ref{sec:biasedbranches}) and using geometric history length BPs to track very large history lengths (Section \ref{sec:geometriclength}).   We now discuss several BP designs.

\subsection{Two-level predictors}\label{sec:twolevelpredictors}

Yeh et al. \cite{yeh1991two,yeh1992alternative} propose a two-level BP which uses a  branch history register (BHR) table and a ``branch pattern history  table'' (PHT). Most recent branch results are shifted into BHR. History pattern bits show the most recent branch outcomes for a specific content in the BHR. For a branch, its instruction address is used to index BHR table and the corresponding BHR contents are used to index PHT for making predictions. All history registers reference the same PHT and hence, it is termed a global PHT. History pattern of last $P$ results of a branch decide its prediction, hence, BHR has $P$ bits, which can store $2^P$ different patterns. PHT has $2^P$ entries each of which can be indexed by a different history pattern.

In PHT, branch behavior for the recent $S$ occurrences of a given pattern of these $P$ branches is stored. The branch is predicted by seeing the branch response for recent $S$ instances of the pattern. For instance, let $P=7$ and the outcome of last $P$ branches was 1010011 (1=taken, 0=not-taken). If $S = 5$ and in each of the last 5 times, the past seven branches showed the pattern 1010011, the branch outcome switched between taken and not-taken, then the level-2 history will be  10101. Then, the BP predicts current branch as `not-taken'.

They propose three implementations of their BP, viz., GAg, PAg and PAp. GAg scheme uses only one global BHR and one global PHT for all branches, however, this leads to aliasing in both levels. PAg scheme uses one BHR for each static branch to record their branch history individually which reduces aliasing in the first-level table, hence, this scheme is termed as `per-address' branch prediction with global PHT. To remove the aliasing in second-level table also, their PAp scheme uses a PHT for each branch along with one BHR for each static branch. In PAg and PAp, the BHR table can be implemented as a set-associative or a direct-mapped structure. \textit{Results:} Due to interference effect, the decreasing order of performance is PAp, PAg and GAg. For achieving same level of accuracy, PAg and PAp incur least and highest hardware cost, respectively. This is because, GAg requires long history register whereas PAp requires multiple PHTs.

\begin{figure} [htbp]
\centering
\includegraphics[scale=0.40]{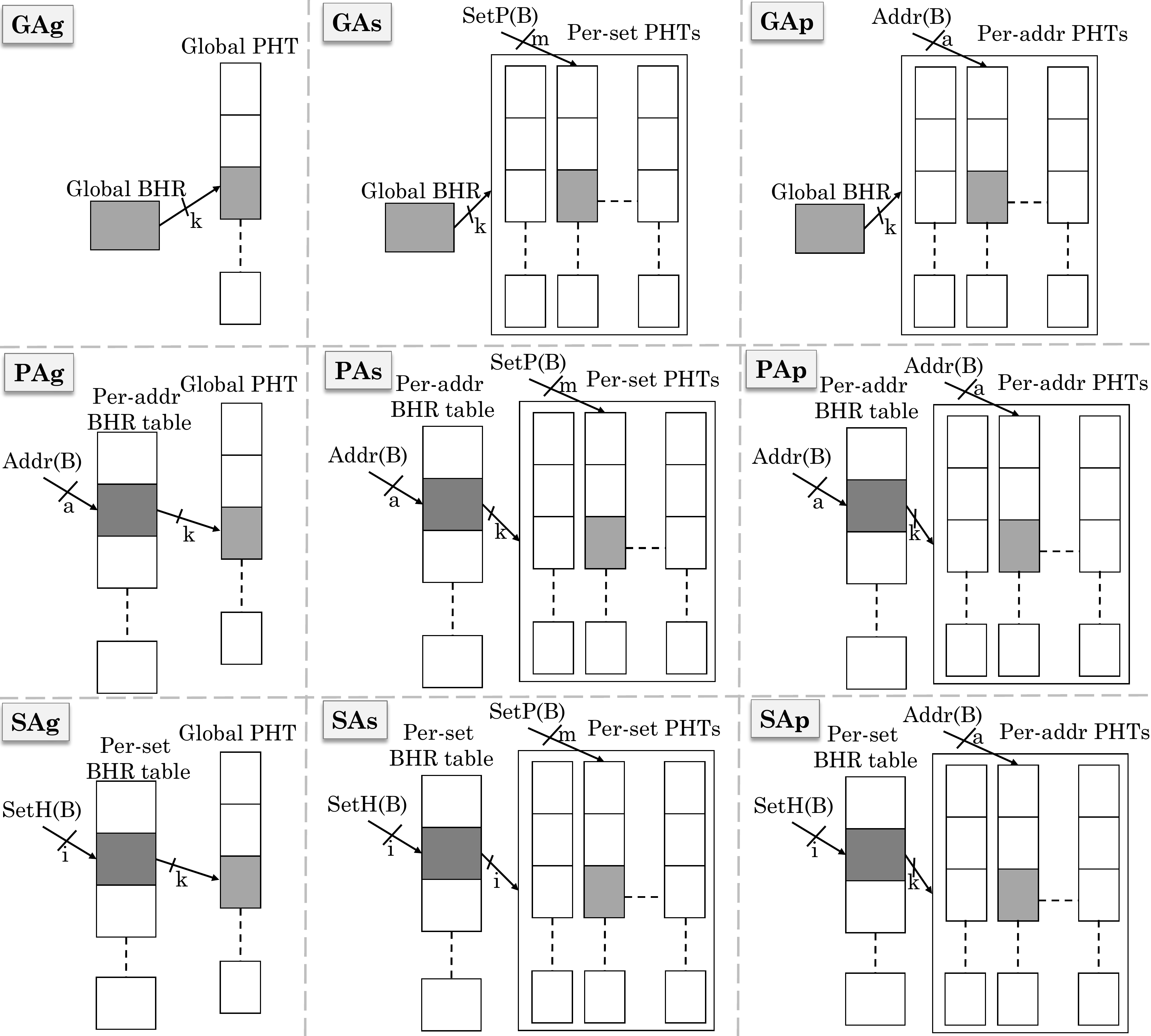}
\caption{Nine variants of two-level BPs \cite{yeh1993comparison}}\label{fig:TwoLevel9BPs}
\end{figure}

Yeh et al. \cite{yeh1993comparison} study and compare nine variants of two-level BP designs which are shown as [G/P/S]A[g/p/s], depending on how branch history (level-1) and pattern history (level-2) are maintained and mutually associated. These designs are shown in Figure \ref{fig:TwoLevel9BPs} and the meanings of the symbols G/P/S and g/p/s are shown in Table \ref{tab:nineBPsmeaning}. The second letter in the name indicates whether the PHT is \underline{a}daptive (i.e., dynamic), or \underline{s}tatic.

\begin{table}[htbp]
  \centering
  \caption{Distinguishing characteristics of each BP in level-1 and 2 tables \cite{yeh1993comparison}}
    \begin{tabular}{|c|C{3cm}|C{3cm}|C{3cm}|}
    \hline
    Level-1 & G     & P     & S  \\
    \hline
    $k$ branches stored are:  & last $k$ branches actually seen & last $k$ instances of the same branch & last $k$ instances of branches from the same set \\
    \hline\hline
    Level-2 & g     & p     & s \\
    \hline
    One PHT for:  & all branches & each branch & a set of branches \\
    \hline
    \end{tabular}%
  \label{tab:nineBPsmeaning}%
\end{table}%

Global history BP (also referred to as ``correlation BP'' \cite{pan1992improving}) needs only one BHR. Per-address history BP uses one BHR for each static conditional branch and thus, prediction of a branch is made based on its own execution history only. `S' refers to partitioning of branch addresses into sets, e.g., branches in a 1KB block (256 instructions) may be members of the same set. The `SetP(B)' indexing function (middle 3 figures in Figure \ref{fig:TwoLevel9BPs}) is formed by combining set-index and low-order bits of branch address.  In per-set history BP, all branches in a set update the per-set BHR and thus, they influence each others' prediction outcome.

\textit{Results:} They find that in applications with several {\tt if/else} constructs, branches correlate strongly with earlier branches and hence, global history BPs are  effective for such applications. However, global history BPs incur large hardware cost since they need longer branch history or multiple PHTs to mitigate aliasing. Per-address history BPs provide high accuracy for applications containing recurring loop-control branches since their periodic branch characteristics are accurately captured by per-address BHR tables. Further, due to reduced aliasing, they need smaller tables which reduces their storage cost. Per-set history BPs provide high accuracy for both the above types of applications, however, their hardware cost is higher than even global history BPs since they need separate PHTs for every set.  They also find that, in general, among low- and high-cost BPs, PAs and GAs are most cost effective.

\begin{figure} [htbp]
\centering
\includegraphics[scale=0.40]{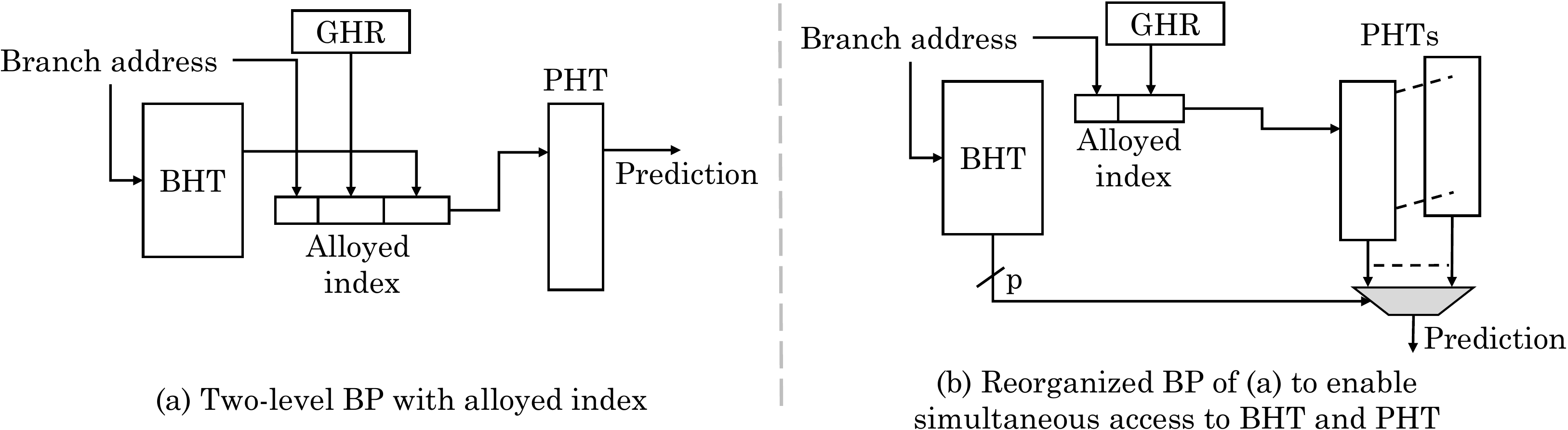}
\caption{Alloyed-index BP \cite{skadron2000taxonomy} }\label{fig:alloyedindex}
\end{figure}

Skadron et al. \cite{skadron2000taxonomy} note that other than aliasing, several mispredictions are caused due to tracking incorrect-type of history for a branch (e.g., global history in place of local history and vice versa). They propose a BP which  is especially effective in reducing such mispredictions. 
Their BP adds a GHR to the two-level BP, as shown in Figure \ref{fig:alloyedindex}(a). Then, both global and local history bits are ``alloyed'' in one PHT index. This allows using both local and global history without requiring a meta-predictor or division of a BP in multiple predictor components as in a hybrid BP. Extending the terminology of Yeh et al. \cite{yeh1993comparison}, their BP is termed as a MAs BP, where `M' refers to merging of local and global histories. Compared to PAs and GAs BPs, their MAs BP requires fewer bits to achieve the same level of anti-aliasing since the merging of histories already reduces aliasing. Bimodal BP \cite{lee1997bi} is a special case of MAs BP which uses just one bit of local history.
A limitation of their BP is the requirement of two serial accesses which increases the overall latency. To resolve this, they split PHT in multiple tables which can be accessed concurrently, as shown in Figure \ref{fig:alloyedindex}(b). This is feasible in case of small number of local-history bits (e.g., $p\le 4$). This design allows accessing PHT and BHT (branch history table) in parallel. The limitation of this design, however, is the overhead of   accessing a large number of tables and a multiplexer. Their alloyed BP provides higher accuracy than hybrid BPs at small area budget and comparable accuracy  at large area budget. Further, it provides better accuracy than two-level designs.

Sechrest et al. \cite{sechrest1996correlation} note that for small-budget BPs, aliasing can degrade accuracy significantly and even nullify the advantage of modeling inter-branch correlation in global history based BPs, e.g., gshare \cite{mcfarling1993combining} and GAs \cite{yeh1993comparison}. Thus, the need to avoid aliasing necessitates increasing the size of global history based BP. For local-history based BPs such as PAs, interference is more prononunced in the buffer that stores branch history (first-level) than in the predictor table (second-level). Overall, achieving high BP accuracy with large programs requires sufficient storage resources either in the first or the second-level table.

\begin{figure} [htbp]
\centering
\includegraphics[scale=0.40]{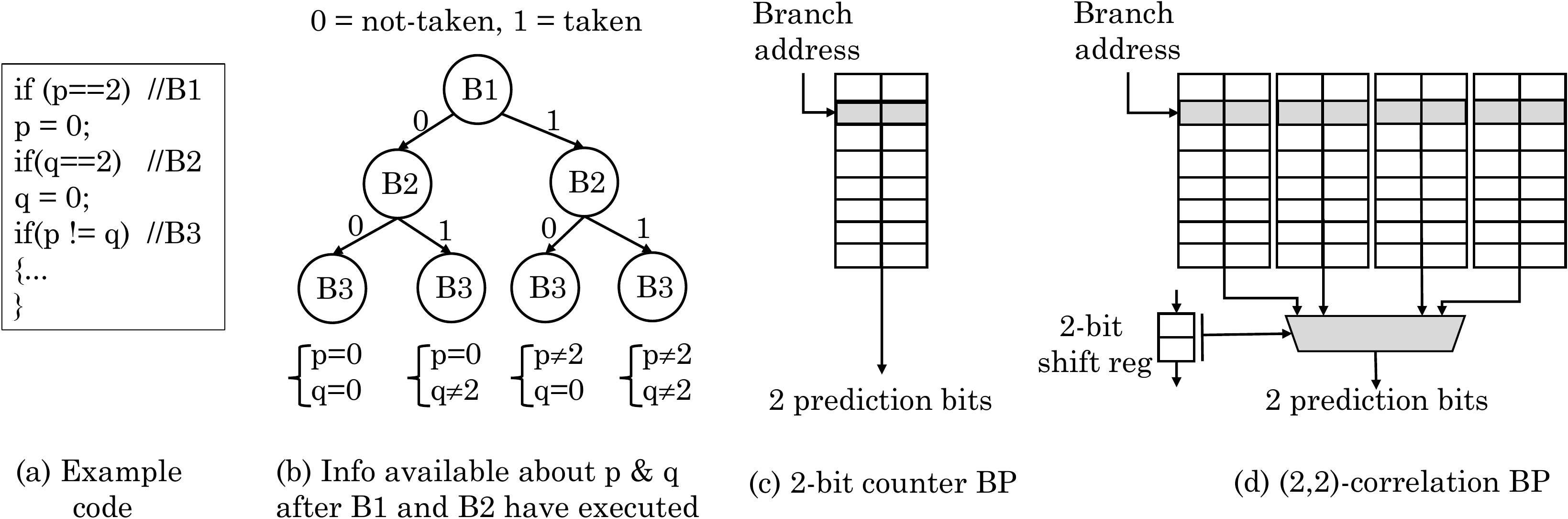}
\caption{(a) A code segment with correlated branches (b) information provided by branch correlation (c) a  conventional 2-bit counter BP (d) The correlation BP proposed by Pan et al. \cite{pan1992improving}}\label{fig:branchcorrelation}
\end{figure}

Pan et al. \cite{pan1992improving} note that due to correlation between branches, the outcome of a branch can be predicted by not only its own history but that of other branches also. For example, in Figure \ref{fig:branchcorrelation}(a), branch B3 is correlated with B1 and B2. After B1 and B2 have executed, some information for resolving B3 is already known. As shown in Figure \ref{fig:branchcorrelation}(b), if conditions at B1 and B2 were true, then, condition at B3 can be accurately predicted as false.  However, BPs based only on self-history, shown in Figure \ref{fig:branchcorrelation}(c), are unable to use this information. By dividing history of B3 into four subhistories and selectively using proper subhistory, randomness of B3 can be reduced. In general, their technique examines past $K$-branches to split the history of a branch in $2^K$ subhistories and then predicts independently within each subhistory using any history-based BP, e.g., $N$-bit counter-based BP. They use a $K$-bit shift register for storing the result of past $K$ branches. The register identifies $2^K$ subhistories of a branch and for any subhistory, its $N$-bit counter is used for predicting the outcome which provides $N$ prediction bits. BP is updated as in the $N$-bit counter-based BP. Figure \ref{fig:branchcorrelation}(d) shows their BP design assuming $K=2$ and $N=2$. Their technique achieves high accuracy with only small hardware overhead.

Chen et al. \cite{chen1996analysisdata} present a conceptual model of branch prediction based on data compression to allow BP research to benefit from the research on data compression. They show that the ``two-level'' or correlation based BPs are  simple versions of ``prediction by partial matching'' (PPM) \cite{cleary1984data}, an optimal predictor in data compression. They show that PPM predictor provides small improvement over two-level predictor  for small sized BTBs. As PPM is optimal, substantial asymptotic improvements in two-level BPs are unlikely as long as information presented to the BPs is not modified. Still, by using the predictors proposed in compression field, small improvements can be obtained in low-budget predictor designs. Some researchers have proposed  PPM-like BPs \cite{gao2007pmpm,michaud2005ppm}.

\subsection{Using both global history and branch address}\label{sec:gsharegselect}

\revised{Mcfarling \cite{mcfarling1993combining} notes that for a given BP size, global history BP has lower accuracy than the local history BP. Also, it provides higher accuracy than bimodal BP only when BP size exceeds 1 KB. This is because for small BP size, the branch address bits employed in bimodal BP can effectively distinguish between different branches. However, with increasing number of counters, every frequent branch is mapped to a different counter, and hence, for very large tables, the information value of each extra address bit approaches zero.
    
For global history BP, the information value of the counters increases with increasing table size. This allows the global BP to distinguish different branches, although less effectively than the branch address. However, the global BP can store more information than merely identifying the current branch and as such, with increasing BP size, it outperforms the bimodal BP.} 
    
To reduce aliasing by bringing the best of global history and branch history information together, Mcfarling \cite{mcfarling1993combining} presents ``gselect''  and ``gshare'' BPs. In gselect, the (low-order) address and global-history bits are concatenated whereas in gshare, these bits are XORed to get the table index.  Table \ref{tab:gsharegselect} shows an example of two branches, each having only two common global histories. Clearly, while gshare is able to separate the 4 cases, gselect is unable to do so. \revised{This is because by virtue of using XOR function, gshare forms a more randomized index.}  For the same hardware cost (8b index in the example of Table \ref{tab:gsharegselect}), the gselect BP cannot use higher-order bits to distinguish different branches and hence, gshare BP provides higher accuracy for large BP sizes. For small sizes, however, gshare provides lower accuracy since addition of global history worsens the already-high contention for counters. They also propose two-component hybrid BPs (e.g., bimodal \cite{lee1997bi}+gshare) along with a meta-predictor which selects the best BP out of the two for each branch. Hybrid BP provides higher accuracy than component BPs.

\begin{table}[htbp]
  \centering
  \caption{Index with gselect and gshare scheme for different addresses/histories. \revised{Only gshare is able to distinguish all four cases.}}
    \begin{tabular}{|c|c|c|c|}
    \hline
    
Branch address & Global history & gselect index (4b+4b) & gshare index (8b XOR 8b) \\\hline
   00000000 &  00000001 &  00000001 &  00000001  \\
    00000000 & 00000000  & 00000000  & 00000000 \\
    11111111& 00000000 &11110000& 11111111 \\
    11111111& 10000000& 11110000 &01111111 \\\hline
    \end{tabular}%
  \label{tab:gsharegselect}%
\end{table}%

\subsection{Using multiple indexing functions}\label{sec:multipleindexing}
Michaud et al. \cite{michaud1997trading} note that aliasing in BP tables is similar to cache misses and thus, aliasing can be classified as conflict, capacity and compulsory similar to that in caches. Since use of tags for removing aliasing introduces large overhead, they propose a ``skewed'' BP design, similar to the skewed-associative cache design \cite{seznec1993case}. Since exact occurrence of conflicts depends on the mapping function, their BP uses multiple odd (e.g., 3) number of BP banks but uses different hash functions obtained from same information (e.g., global history and branch address). Each bank acts as a tagless predictor and is accessed in parallel. Every bank provides a prediction and the final prediction is chosen using majority voting based on the idea that branches that interfere in one bank are unlikely to do so in other banks. \revised{Notice that for BPs divided in multiple banks/tables, different banks can differ in the indexing functions used \cite{ma2006using,michaud1997trading} or the length of history used by them \cite{seznec2005analysis,seznec2006case}.} 

For updating their BP, they consider a `total update' policy where all banks are updated and a `partial update' policy where a bank providing wrong prediction is not updated when the final prediction is correct but when the final prediction is incorrect, all the banks are updated.  Their BP with 3 banks achieves same accuracy as a large 1-bank BP while requiring nearly half storage resources. This is because their skewed BP  increases redundancy by using multiple tables which increases capacity aliasing while reducing conflict aliasing. Further, partial update policy performs better than total update policy since not updating the bank providing incorrect prediction allows it to contribute to the correct prediction for some other stream which increases overall effectiveness of the BP.

\subsection{Leveraging biased branches}\label{sec:biasedbranches}

Chang et al. \cite{chang1994branch}  propose classifying branches to different streams and predicting a branch using a BP which is most suited for it. They classify the branches in different classes based on their taken-rate.  They experiment with multiple BPs, viz., profile-driven, 2-bit counter and three two-level BPs, viz.,  PAs, GAs and a modified GAg (gshare). They note that biased branches are better predicted by BPs with short BHRs due to their quick warming-up and the fact that for a fixed hardware cost, smaller history register implies larger number of PHTs which reduces aliasing. In contrast, mixed-direction (i.e., unbiased) branches are better predicted by BPs with long BHRs due to their ability to distinguish between larger number of execution states and to effectively hold histories of correlated branches. 

\begin{figure} [htbp]
\centering
\includegraphics[scale=0.40]{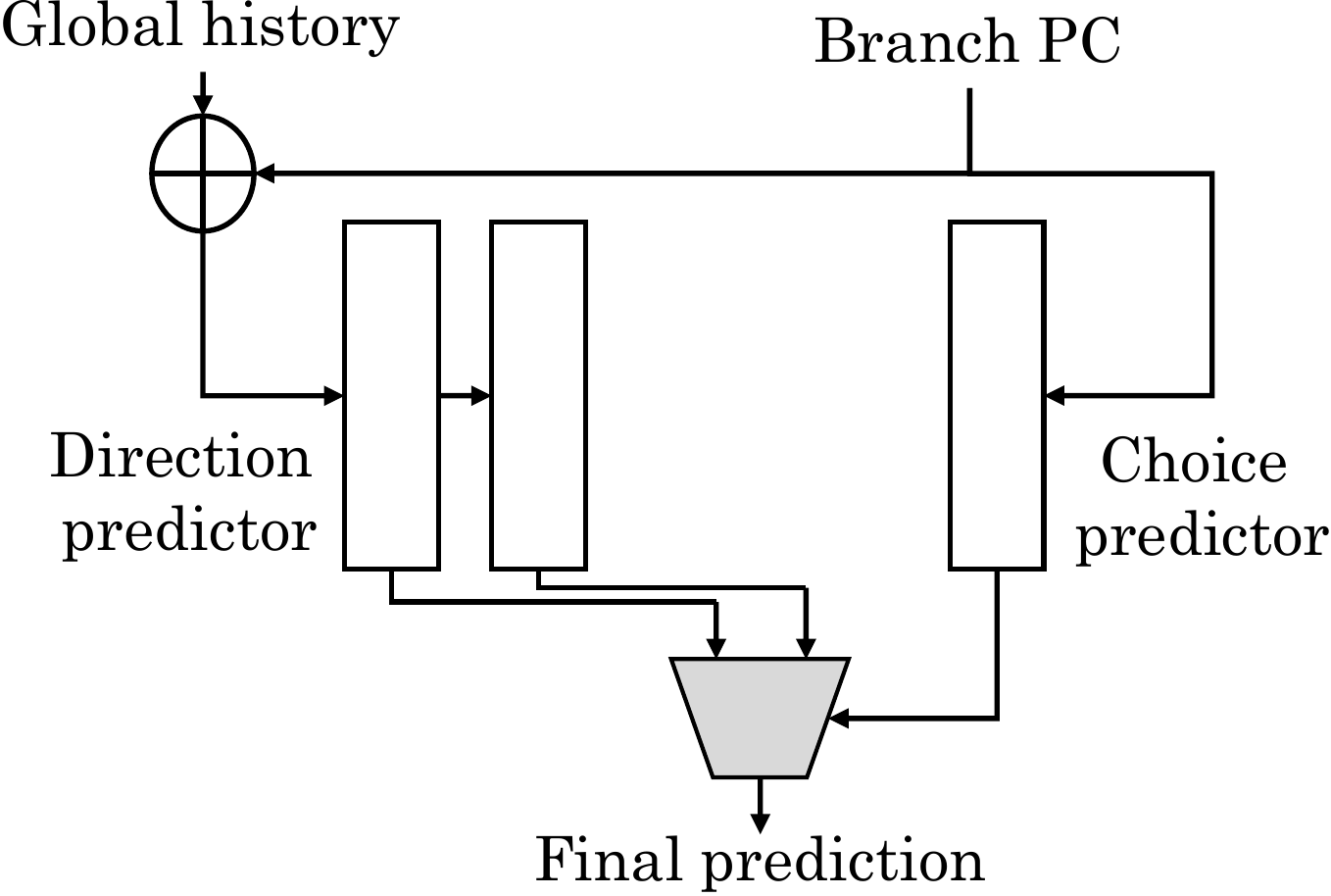}
\caption{Bimodal BP  \cite{lee1997bi}}\label{fig:bimodalbp}
\end{figure}

Lee et al. \cite{lee1997bi} present the ``bi-modal'' BP which is shown in Figure \ref{fig:bimodalbp}. Their BP  divides the level-2 counter table in two parts, viz. taken and not-taken `direction predictors'. For any history pattern, one counter is chosen from each part. From these, one counter is finally chosen to make the prediction, based on the input of another `choice predictor'. Choice predictor is indexed by branch address only. The classification of GH patterns in two groups is done based on per-address bias of the choice predictor.    They use partial-update policy, where only the selected counter is updated based on the branch result. They always update the choice predictor, except when the choice is opposite of the branch result but the final prediction of the selected counter is right. \revised{Overall, bi-modal BP separates branches showing negative aliasing and keeps-together those showing neutral/beneficial aliasing.} 
By virtue of removing harmful aliasing, their technique achieves high prediction accuracy.

\begin{figure} [htbp]
\centering
\includegraphics[scale=0.40]{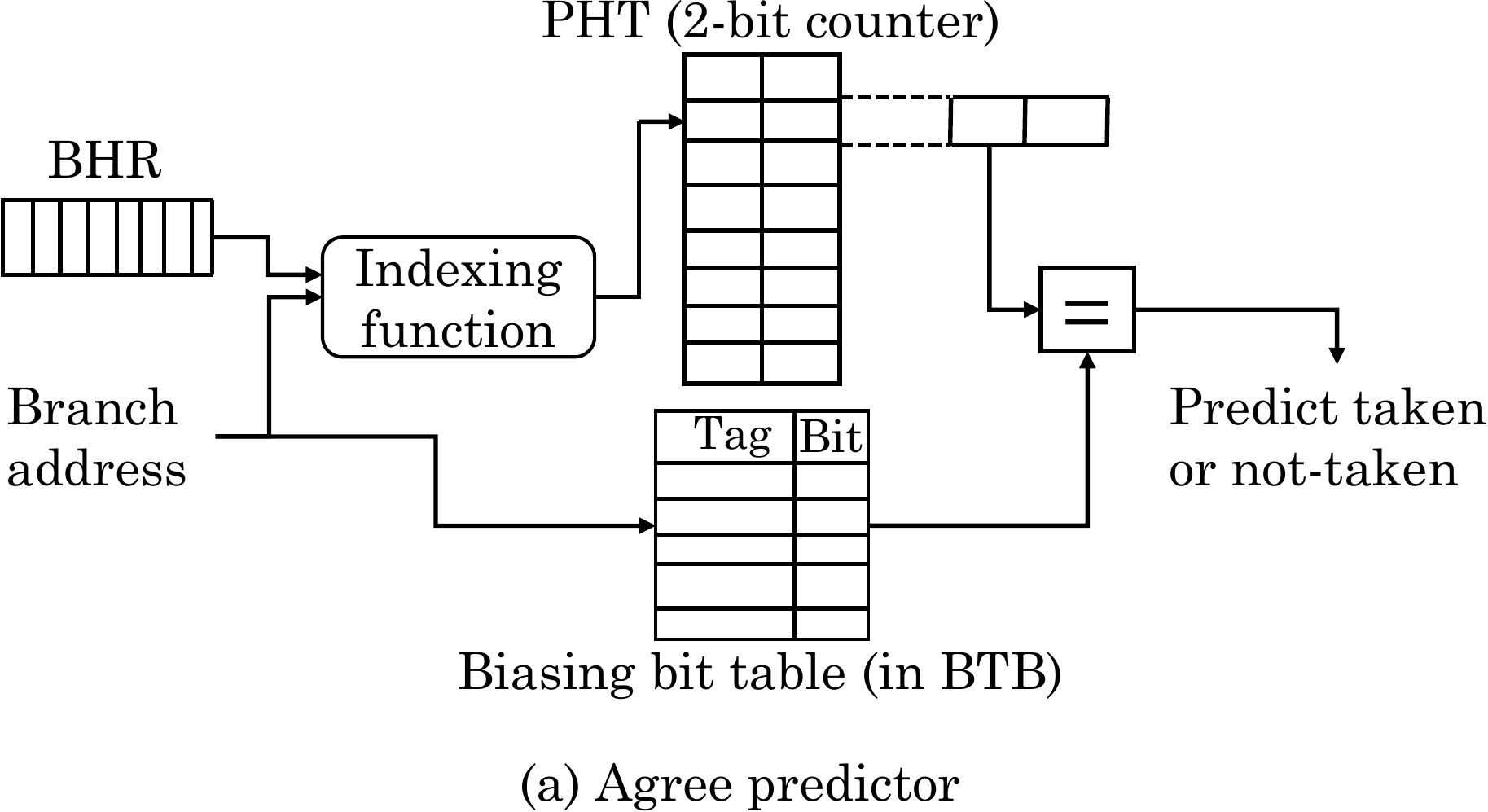}
\caption{Agree predictor \cite{sprangle1997agree} }\label{fig:agreepredictor}
\end{figure}

Sprangle et al. \cite{sprangle1997agree} propose a technique which converts harmful interference to beneficial or harmless cases by changing the interpretation of PHT entry.  Figure \ref{fig:agreepredictor} shows their BP, which is termed as agree predictor. For each branch, they add a biasing bit  which predicts the most likely outcome of the branch. Instead of predicting the direction of a branch (as in traditional scheme), the PHT counters in their technique predict whether the branch outcome will agree with the biasing bit. When the branch is resolved, PHT counters are increased if the branch direction matched the biasing bit and vice versa. They set biasing bit to the direction of the branch in first execution. With proper choice of biasing bit, the counters of two branches mapping to same PHT entry are more likely to be updated in the same direction viz. agree state.

  Mathematically, let $T$, $NT$, $A$ and $D$ show probability of taken, not-taken, agree and disagree state for any branch, respectively. Then, the probability of harmful interference in conventional predictor is $(T_{B1}*NT_{B2}+NT_{B1}*T_{B2})$, whereas that in their predictor is $(A_{B1}*D_{B2}+ A_{B2}*D_{B1})$. 
 For example, if  $T_{B1}$ = 85\% and $T_{B2}$ = 15\%, then $A_{B1}$ = 85\% and $A_{B2}$ = 85\%, and thus, harmful interference is reduced from 74\% to 25\%. Another benefit of their technique is that PHT entries of a new branch are expected to be in warmed-up state (viz. agree state) already which improves accuracy.

\subsection{Geometric history length BPs}\label{sec:geometriclength}

Seznec \cite{seznec2005analysis} presents GEHL (geometric history length) BP which allows capturing both recent and old correlations. GEHL BP uses $M$ (= 4 to 12) predictor tables T$i$ ($0 \le i < M$) whose indices are obtained by hashing global path/branch history with the branch address. Each table stores predictions as signed counters. Prediction is performed as shown in Figure \ref{tab:GEHLtable}. 
Different tables use different history lengths, as shown in Table \ref{tab:GEHLtable}. Using history lengths in geometric series permits using long history lengths for some tables, while still using most of the storage for tables with short history lengths. \revised{For example, assuming an 8-component BP, if $\alpha=2$ and $L(1) =2$, then $L(i)$ form the series $\{0,2,4,8,16,32,64,128\}$ for $0\le i<M$. Thus, 5 tables are indexed using at most 16 history bits, while still capturing correlation  with  128-bit history.}

\begin{table}[htbp]
  \centering
  \caption{\revised{Prediction and indexing process in GEHL BP \cite{seznec2005analysis}}}
    \begin{tabular}{|p{2.0cm}|p{11.0cm}|}\hline    
     Prediction & One counter C($i$) is read from each table T$i$ and they are added as follows: $Sum = M/2 + \sum_{i=0}^{M-1} C(i)$. If $Sum \ge 0$, prediction is taken and vice-versa \\\hline
   
     Table indexing & Table T$0$ is indexed using branch address. Tables $1\le i<M$ are indexed using history length of: L($i$) = $\lfloor (\alpha^{i-1}\times L(1)+0.5)\rfloor$ \\ \hline
    \end{tabular}%
  \label{tab:GEHLtable}%
\end{table}%

The predictor is updated only on a misprediction or when absolute value of $Sum$ is below a threshold. Their BP allows adapting history lengths such that in case of high aliasing on updates, short history length is used and vice versa. Also, the update threshold is adapted with a view to keep the number of updates on mispredictions and those on correct predictions in the same range. Using ahead pipelining, the latency of their BP can be kept low. The limitation of their BP is its complexity and the need of checkpointing a large amount of state that must be restored on a misprediction. Overall, GEHL BP provides high accuracy. 

\begin{figure} [htbp]\centering
\includegraphics[scale=0.40]{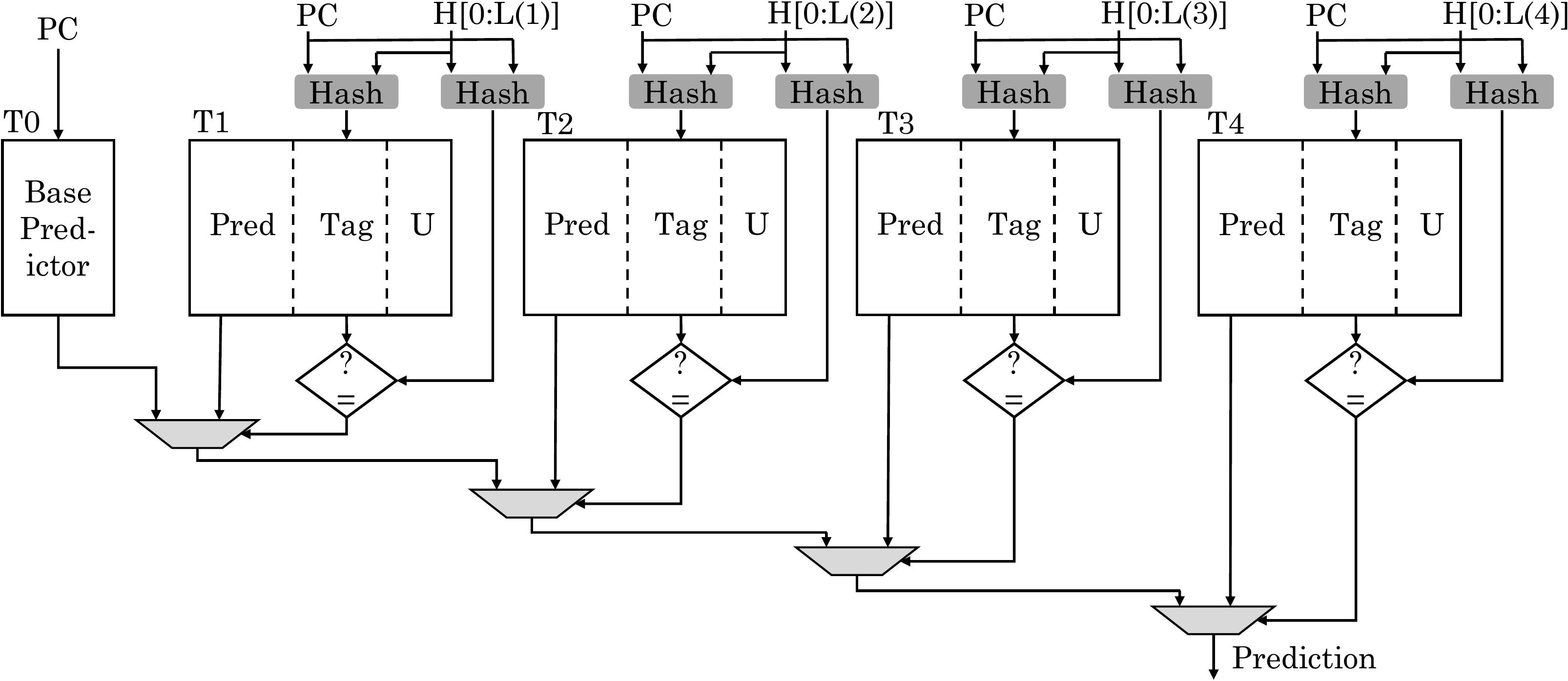}
\caption{TAGE BP \cite{seznec2006case} }\label{fig:TAGE}
\end{figure}

Seznec et al. \cite{seznec2006case} present TAGE (tagged geometric history length) BP which is an improved version of the tagged PPM-like BP \cite{michaud2005ppm}. The TAGE BP uses a base predictor (T0) which provides basic prediction and multiple partially-tagged components T$i$ ($1\le i \le M$). T0 can be a 2-bit counter bimodal table.  T$i$ predictor is indexed with a global history length L($i$) = $\lfloor (\alpha^{i-1}\times L(1)+0.5)\rfloor$. The T$i$ predictors have a signed `prediction' counter whose sign gives the prediction, a tag and a ``useful'' counter (U). Figure \ref{fig:TAGE} shows the TAGE BP with 5 components. For making a prediction, both base and tagged BPs are simultaneously accessed. If multiple tagged BPs hit, the one with longest history length (say Tc) provides the prediction, otherwise, the prediction of base predictor is used. For example, if T2 and T4 hit but T1 and T3 miss, then, T4's prediction is used ($c =4$) and prediction of T2 is called ``alternate prediction''.  If no component hits, the default prediction acts as the alternate prediction. 

As for updating, when the alternate prediction is different from the final prediction, the `U' counter is increased/decreased by one if the final prediction is correct/incorrect, respectively. Periodically, `U' counter is reset so that entries are not marked useful for ever. The `pred' counter of Tc gets updated on correct and incorrect predictions. Additionally, on an incorrect prediction, if $c<M$, an entry is allocated in a predictor Tk ($c<k<M$) \cite{seznec2006case}. If the prediction of a Tc component is provided by such ``newly allocated'' entries, alternate prediction is used as final prediction since newly allocated entries tend to provide wrong prediction for some time due to lack of training. Their BP outperforms previous predictors \cite{seznec2005analysis,michaud2005ppm} and thus,  for BPs using geometric history lengths, partial-tagging \cite{seznec2006case} is more cost-effective in selecting the final prediction compared to the adder tree \cite{seznec2005analysis}. Also, using more than 8 tagged predictors does not provide any benefit. They also show that observing the Tc component and prediction counter value allows obtaining an estimate of misprediction probability \cite{seznec2011storage}. 

\textbf{Comparison:} For comparable hardware budget, gshare \cite{mcfarling1993combining} and perceptron BPs \cite{jimenez2001dynamic} track correlations with nearly 20 and 60 branches, respectively \cite{huang2015energy}, whereas GEHL and TAGE track correlations with nearly 200 and 2000 branches, respectively. \revised{Among the solo-BPs, the TAGE predictor is considered the most accurate BP \cite{seznec2011new}, and by combining it with side predictors, even more accurate hybrid BPs have been obtained \cite{seznec2011new,seznec2015inner,seznec2007256,seznec2011isltage}.} Also note that while TAGE BP \cite{seznec2005analysis} uses multiple tagged predictors, other BPs \cite{eden1998yags} use only one tagged predictor.

\section{BPs for hard-to-predict branches}\label{sec:sideBP}

Several branches cannot be predicted by regular BPs such as TAGE BP. These branches may be   loop-exit branches (Section \ref{sec:loopexitbranches}), those inside nested loops (Section \ref{sec:nestedloopbranches}), those having large period (Section \ref{sec:longperiodbranches}), etc. Similarly, some branches can be better predicted based on data correlation or address correlation (Sections \ref{sec:datacorrelation}-\ref{sec:addresscorrelation}). We now discuss BPs which especially target such hard-to-predict branches and hence, they are  suitable to be used as side predictors.

\subsection{Predicting loop-exit branches}\label{sec:loopexitbranches}
 
\begin{figure} [htbp]
\centering
\includegraphics[scale=0.40]{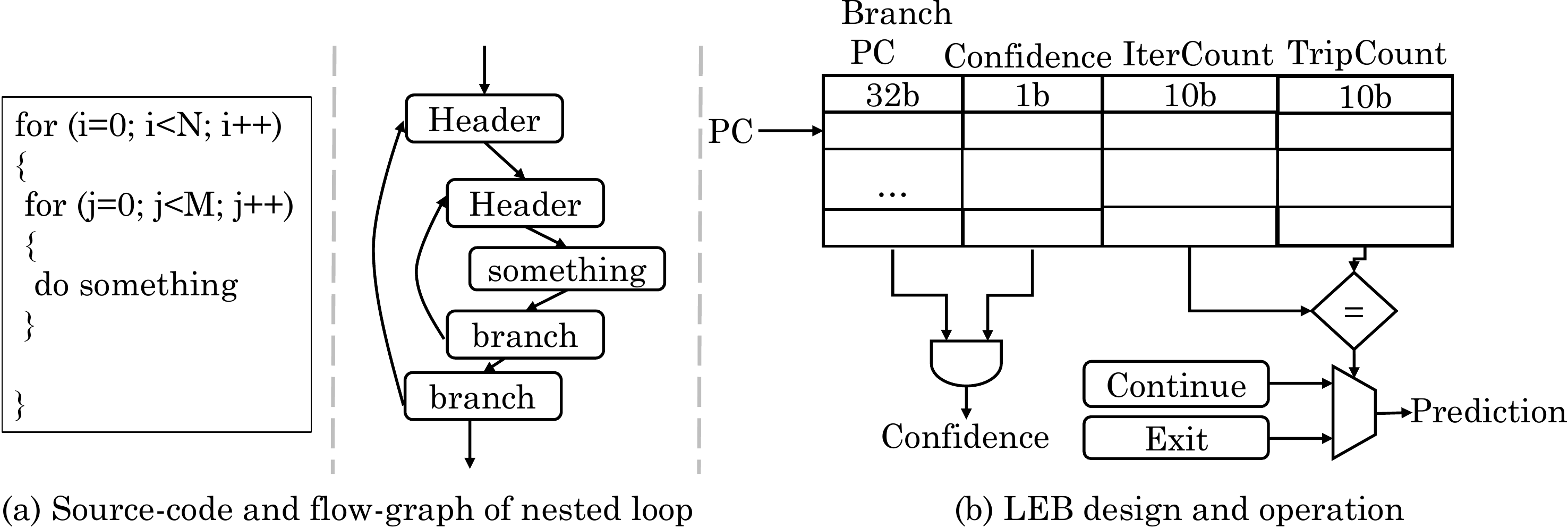}
\caption{(a) Illustration of loop branches and (b) working and operation of LEB \cite{sherwood2000loop}}\label{fig:loopExitBuffer}
\end{figure}

Sherwood et al. \cite{sherwood2000loop} note that loop exits cannot be  predicted by local history BP if the loop-count ($N$) is higher than the local history size. They can be predicted using global history only if the global history size is higher than $N$  or there is a distinct branching sequence before the loop exit. They present a loop exit buffer (LEB) based BP for predicting loop branches. Figure \ref{fig:loopExitBuffer}(a) shows an example of a loop branch. LEB detects a backward branch which is mispredicted by the primary BP, as a loop branch and inserts it in the LEB which is shown in Figure \ref{fig:loopExitBuffer}(b).  As for the overall BP design, the primary BP provides a prediction, which is overridden by their LEB-based BP in case its prediction has high confidence.

In LEB, the TripCount field stores the number of successive times the loop-branch was taken before the last not-taken; and the confidence bit shows that the same loop TripCount has been observed at least twice consecutively. For any branch hitting in LEB, if IterCount is not equal to TripCount, IterCount is incremented by one and thus, IterCount tracks the number of times the branch has been consecutively taken. When IterCount equals TripCount and the confidence bit equals one, a loop-exit is predicted. When a branch is resolved, TripCount and confidence bit of a not-taken loop-branch are updated. They also use extra counters to recover from branch mispredictions (not shown in Figure \ref{fig:loopExitBuffer}(b)). Their loop predictor can predict with up to 100\% accuracy after a brief warmup period. \revised{However, in case of frequently changing loop-count, their technique will have poor coverage and/or accuracy.}

Sendag et al. \cite{sendag2008low} propose a design which uses a complementary BP (CBP) along with the conventional BP. The CBP only focuses on commonly mispredicted branches, whereas the conventional BP speculates on predictable branches.   They use a ``branch misprediction predictor'' (BMP) which uses the number of committed branches between consecutive branch mispredictions for any index, to predict the time of next subsequent branch misprediction. Based on this, the direction of this expected misprediction is changed, allowing progress on the correct-path. Most mispredictions removed by BMP are due to (1) loop branches with changing loop-counts that are too long for a BP and (2) early loop-exit, e.g., due to {\tt break} instruction. BMP can work with any BP. Their technique improves overall prediction accuracy and saves energy.

\begin{figure} [htbp]
\centering
\includegraphics[scale=0.40]{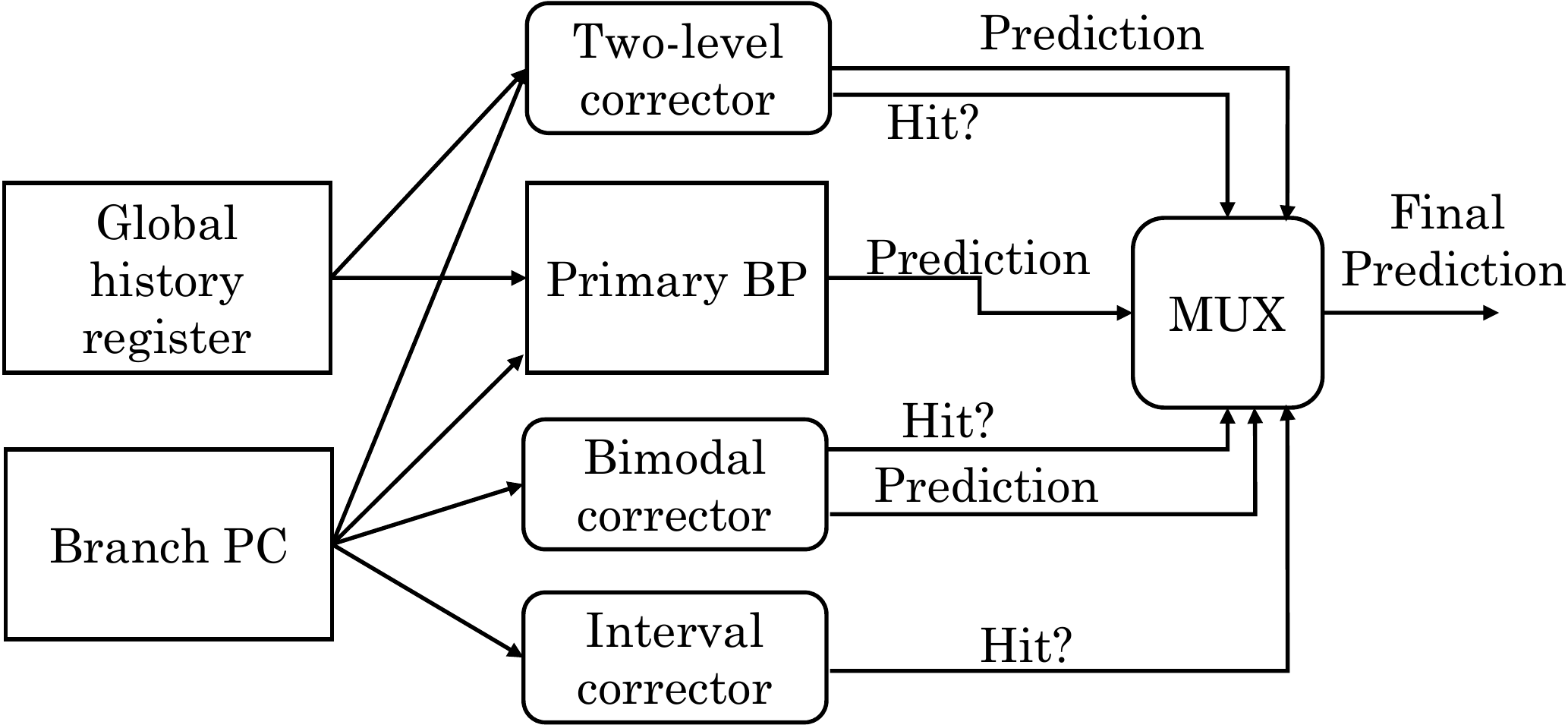}
\caption{The overall BP design proposed by Lai et al. \cite{lai2005improving} }\label{fig:laicorrectors}
\end{figure}

Lai et al. \cite{lai2005improving}  propose a design where multiple correctors are used to correct the prediction of branches which are mispredicted by the primary BP. Each corrector focuses on a distinct pattern and thus, can correct only the branches it remembers using the tags. In each corrector, when required, the entry with least confidence is replaced. They use gskew BP \cite{michaud1997trading}  with partial update policy as the primary BP. All correctors keep a confidence value and the correction is made only if the confidence of their prediction is higher than a threshold. As shown in Figure \ref{fig:laicorrectors}, they use three correctors (1) an interval corrector which corrects loop-exit branches. This corrector assumes that after a fixed interval, the primary BP will mispredict and hence, it corrects primary BP after a fixed interval. They use two interval values to handle early loop-exits. Depending on whether the primary prediction is correct or incorrect, the confidence is decreased or increased. If primary BP makes a misprediction and the current counter does not match any of the two interval values, the current counter replaces the interval value with lower confidence. Also, the confidence of interval value with higher confidence is decreased by one. \revised{Notice that both their interval corrector and the BMP of Sendag et al. \cite{sendag2008low} work by predicting when a misprediction will happen and then avoiding that misprediction.}

(2) The bimodal corrector tracks local history or bimodal patterns. 
(3) Two-level corrector tracks two-level patterns \cite{yeh1992alternative} and its reversal/update scheme is similar to other correctors. Due to using global histories, it achieves high accuracy and hence, its confidence threshold is set to be lower than that of other correctors. In terms of decreasing accuracy, the correctors are: two-level, interval and bimodal. Their overall BP design achieves high accuracy. The limitation of their design is that primary BP and correctors operate in parallel and they are found to have same predictions  for more than half the times, which leads to redundancy and energy wastage.

\subsection{Predicting branches inside nested loops}\label{sec:nestedloopbranches}
 
\begin{figure} [htbp]
\centering
\includegraphics[scale=0.40]{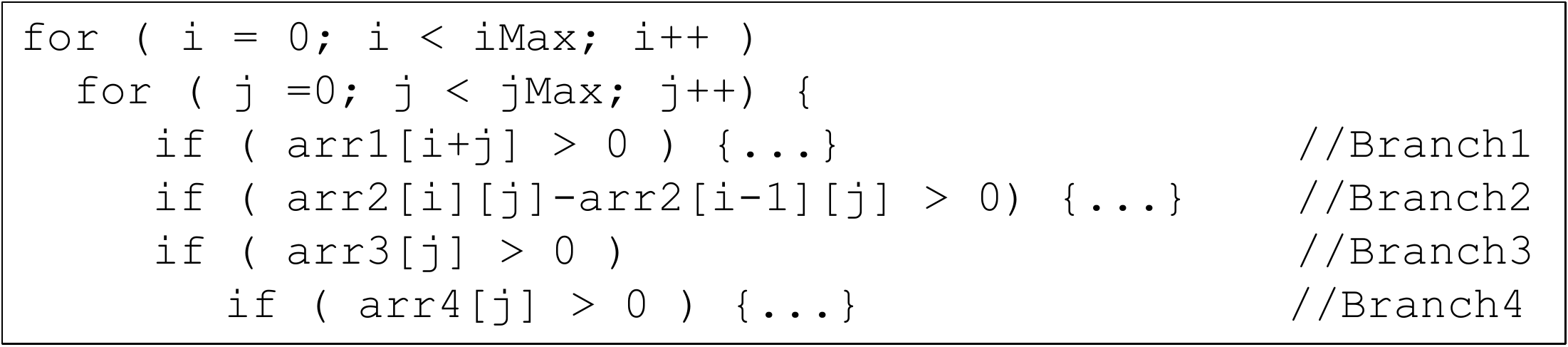}
\caption{An illustration of branches whose results correlate with previous
iterations of the outer loop \cite{seznec2015inner,albericio2014wormhole} }\label{fig:imliseznecinner}
\end{figure}

Albericio et al. \cite{albericio2014wormhole} note that results of many branches of nested loops are correlated with those of prior iterations of the outer loop and not recent results of the inner loop. \revised{Mathematically, for a branch in the inner loop, its {\tt Outcome[i][j]} is correlated with {\tt Outcome[i-1][j+D]}, where {\tt D} is a small number such as 0, -1 or +1.  Consider Figure \ref{fig:imliseznecinner} where we assume that the values of arrays {\tt arr1}, {\tt arr2}, {\tt arr3} and {\tt arr4} are not changed inside the loops. Notice that for Branch1, {\tt Outcome[i][j]} equals {\tt Outcome[i-1][j+1]} and for Branch2, {\tt Outcome[i][j]} is weakly correlated with {\tt Outcome[i-1][j]}. For both Branch3 and Branch4, {\tt Outcome[i][j]} equals {\tt Outcome[i-1][j]}. }

\begin{figure} [htbp]
\centering
\includegraphics[scale=0.40]{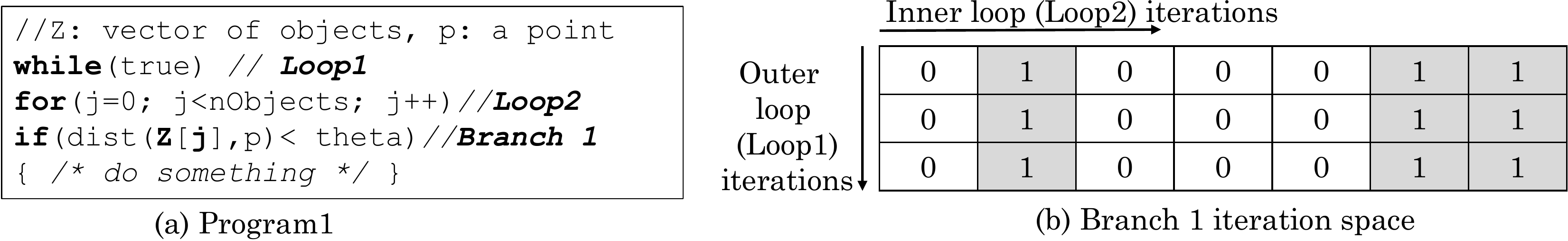}
\caption{Program 1 and the outcome of its branch 1 \cite{albericio2014wormhole}. The outcome of branch remains same in different iterations of outer loop.}\label{fig:wormhole1}
\end{figure}

On viewing as a linear history, the pattern present in the outcome stream of such branches is not clear and hence, these branches appear hard-to-predict. However, on storing the history as multidimensional matrix  instead of one-dimensional array, strong correlation can be observed. \revised{Figure \ref{fig:wormhole1}(b) shows an example of storing the correlation in 2D matrix.} They propose a BP which can identify patterns in branches showing multidimensional history correlations. They use their BP as a side predictor to ISL-TAGE BP  \cite{seznec2011isltage}. Using ISL-TAGE BP, they identify candidate branches in the inner loop along with their current and total iteration counts. For branches which behave almost same in each iteration, their behavior in first iteration is noted and using them, prediction is made for subsequent iterations. Figure \ref{fig:wormhole1}(a) shows a sample program code, where branch 1 depends only on {\tt j} value and hence, for the same value of {\tt j} in different iterations of outer loop, the outcome of the branch remains same. This is evident from Figure \ref{fig:wormhole1}(b) which shows the outcome of the branch for different iterations of outer and inner loops.

\begin{figure} [htbp]
\centering
\includegraphics[scale=0.40]{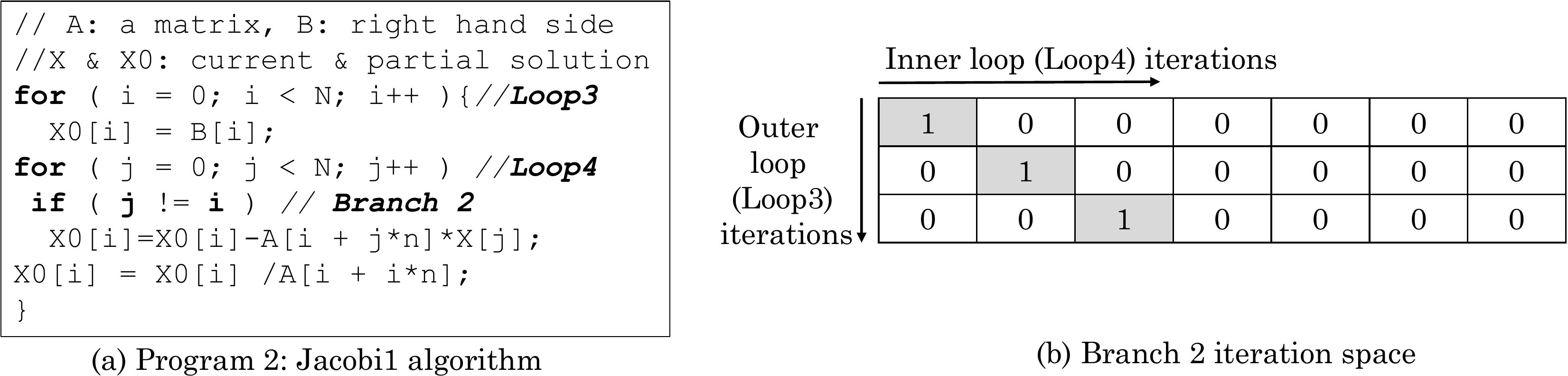}
\caption{Program 2 and the outcome of its branch 2 \cite{albericio2014wormhole}. The branch outcome shows diagonal pattern.}\label{fig:wormhole2}
\end{figure}

For other branches, e.g., those showing diagonal pattern, their BP identifies correlations based on a part of history consisting of bits from the previous iteration and the present iteration streams. As an example, for branch 2 in Figure \ref{fig:wormhole2}(a), the branch outcome depends on both {\tt i} and {\tt j}. Hence, its outcome shows diagonal pattern, which is shown in Figure \ref{fig:wormhole2}(b). For such a branch, the outcome in first iteration is noted. Then, for the second iteration, ``100'' is taken from the first iteration along with ``0'' in the current iteration to form the history ``0100''. Based on this, next outcome of the branch is predicted as ``1'', as shown in Figure \ref{fig:wormhole2}(b).

The outcome of their ``wormhole'' BP overrides that of ISL-TAGE if the confidence in the prediction is high. Many applications spend large fraction of time in nested loops which makes their BP useful. As a side predictor, their BP reduces the mispredictions significantly.

Seznec et al. \cite{seznec2015inner} present a technique to track branch correlations present in nested loops  using ``inner-most loop iteration'' (IMLI) counter based BPs.  IMLI counter is the iteration-number of the loop that encapsulates the branch. 
IMLI-count is the number of times a backward conditional branch (i.e., a loop-exit branch) has been successively taken. Using this, IMLI-count can be found at fetch time.  They propose two components that use this information.  For some hard-to-predict branches,  the inner-most iteration count is tested in the inner loop and hence, their behavior depends only on the inner-most loop counter, i.e., {\tt Outcome[i][j] = Outcome[i-1][j]}. To capture such ``same iteration correlation'', their first component use a table which is indexed with IMLI counter and PC. While wormhole predictor \cite{albericio2014wormhole} tracks only those correlations existing in regular loops with fixed iteration counts and executed on every iteration of inner loop, their BP does not have this limitation. 

However, wormhole BP can predict correlation between {\tt Outcome[i][j]} and {\tt Outcome[i-1][j-1]} \revised{(i.e., diagonal pattern)} which cannot be captured by this first component. To address this, their second component uses a table to store branch outcomes such that while predicting for {\tt Outcome[i][j]}, both {\tt Outcome[i-1][j]} and {\tt Outcome[i-1][j-1]} can be retrieved. Their IMLI-based BP components can be used with a GH-based BP and on using them, further benefit from using local history becomes small. Further, the speculative state of their proposed BP is easily manageable and using their BP  as a side predictor with any TAGE or neural based BP provides high accuracy.

\subsection{Predicting branches with long period}\label{sec:longperiodbranches}
Kampe et al. \cite{kampe2002fab} note that several branches occur with periods much higher than the number of history bits generally used in BPs (e.g., 32 bits), and to record the entire period of all branch execution patterns for making a prediction, $2^{13}$ bits per branch are required. By translating the history from time-domain to frequency domain, the size of history register can be significantly reduced, e.g., only 52 bits are required for representing a history pattern of $2^{13}$ bits. Based on this, they propose a discrete Fourier transform (DFT) based side-BP for predicting branches with long periods that do not correlate with other branches. Their BP is used in a hybrid BP with 4 other BPs, viz., a static BP, the 2-bit dynamic BP \cite{smith1981study}, and the PAp and GAp dynamic BPs \cite{yeh1993comparison}. Due to latency-overhead of DFT, they apply it to a branch only if its accuracy with any of the four above-mentioned BPs is less than 99.99\%.

\begin{figure} [htbp]
\centering
\includegraphics[scale=0.40]{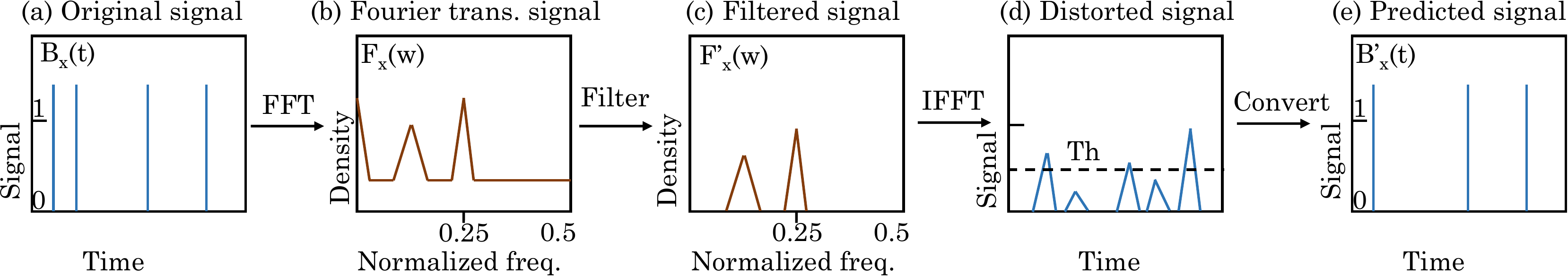}
\caption{Fourier transform based BP (trans. = transformed, Th = threshold, freq. = frequency) \cite{kampe2002fab} }\label{fig:fourier}
\end{figure}

Using 1/0 to represent taken/not-taken branches (respectively), they first transform the branch history pattern into frequency domain using DFT (Figure \ref{fig:fourier}(a) $\rightarrow$ \ref{fig:fourier}(b)). Then, only frequency components with largest peaks are kept due to their high contribution to the total probability. Remaining components are removed as shown in Figure \ref{fig:fourier}(c). This filters noise (non-periodic events) from the history, which is another advantage of their technique. Then, the time domain equivalent of unfiltered high peaks is utilized for predicting the original branch pattern. Then, all the frequency components ($A_n sin(\omega_n t)$) are added and if the sum is higher than the threshold (Figure \ref{fig:fourier}(d)), the branch at time $t$ is predicted to be taken (Figure \ref{fig:fourier}(e)). 
IFFT refers to adding the frequency components. The limitation of their BP is that it does not show high accuracy for branches showing large difference from a 50\%-50\% not-taken/taken rate since such patterns require many frequency signals for accurate reconstruction. On using only one frequency signal, the taken rate becomes much higher than that of the actual branch due to the width of sine wave for the threshold used. In practice, this limitation does not have large impact since such branches are already well-predicted by other BPs. When used as a component of hybrid BP, their BP reduces misprediction rate significantly.

\revised{Note that both: storing the history in frequency-domain \cite{kampe2002fab} or using \textit{geometric} history length tables \cite{seznec2005analysis,seznec2006case} are solutions for branches with large period. The former method focuses on branches requiring very long history whereas the latter method focuses on branches requiring short- and medium-long histories. 
}

\subsection{Predicting based on data correlation}\label{sec:datacorrelation}

\begin{figure} [htbp]
\centering
\includegraphics[scale=0.40]{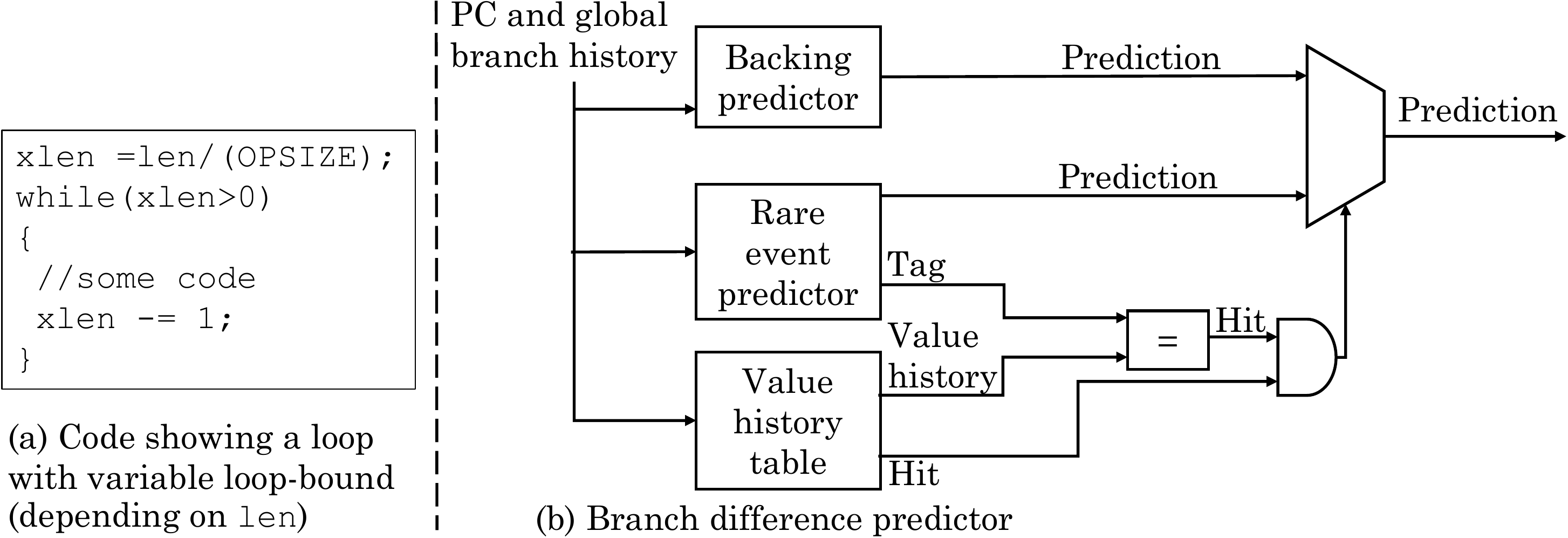}
\caption{\revised{(a) A loop which runs variable number of times} (b) Branch difference predictor \cite{heil1999improving} }\label{fig:branchdifference}
\end{figure}

Heil et al. \cite{heil1999improving} present a BP which predicts based on the correlation of data values just as traditional BPs correlate on GBH. \revised{For example, consider the loop in Figure \ref{fig:branchdifference}(a)  which executes for 21 iterations on average for a workload. This is much higher than the length of local or global BHR. Further, the loop-count varies depending on the value of {\tt len} and hence, the loop predictors based on the assumption of a fixed loop-bound (e.g., \cite{sherwood2000loop}) generally mispredict the loop-exit branch.   However, since loop-counter ({\tt xlen}) reaches a specific value (0),  using  loop counter as an input to BP allows accurate prediction of this branch.}  Since several branch instructions compare two register values and check the sign or detect equal value, delta values correlate strongly with the branch result \cite{mittal2016SurveyCPURF}. Hence, they store the delta between the source register values instead of the values themselves which also saves space. The deltas for each static branch are stored in value history table and for this reason, their BP is termed as  ``branch difference predictor''.    Figure \ref{fig:branchdifference}(b) shows the design of their BP.

However, even with the deltas, the number of patterns becomes large. They note that one static branch instruction may generate several values, of which only few branch deltas are prominent. Hence, they use a backing predictor  which speculates most cases without using delta values and for predicting the remaining values, they use ``rare event predictor'' (REP), a tagged cache-like design for hard-to-predict rare branches. REP preferentially replaces patterns leading to correct predictions that differ from the outcome of backing predictor. REP is updated only on a misprediction in backing predictor. Backing predictor is updated only when it makes a prediction and this reduces aliasing and improves correlations.

Chen et al. \cite{chen2003dynamic} present a value-based BP design that works by  tracking dependent-chain of instructions. They note that if every register value associated with resolution of a branch in the data dependence chain has same value as in the previous instance, the branch-result will be the same.  However, since all the branch register values required for making such value-based prediction are generally not available in-time, use of  register values in the dependence chain is highly useful. The branch behavior on a specific path usually remains consistent which can be easily learnt by a 2-bit saturating counter. They use both PC and register ID to get the index into table. Further, to distinguish instances of same path having different register values, they use hashed value of registers as the tag. Thus, their technique uses both value and path-based information for classifying branches. In deeply-pipelined superscalar processors, dependence-chains may span multiple loop iterations and if dependent-registers remain same in every iteration, the path information becomes ambiguous. To avoid ambiguity between iterations, their BP uses maximum number of instructions spanned by dependence chain in the tag. Their BP provides higher accuracy than an iso-sized hybrid BP.

\begin{figure} [htbp]
\centering
\includegraphics[scale=0.40]{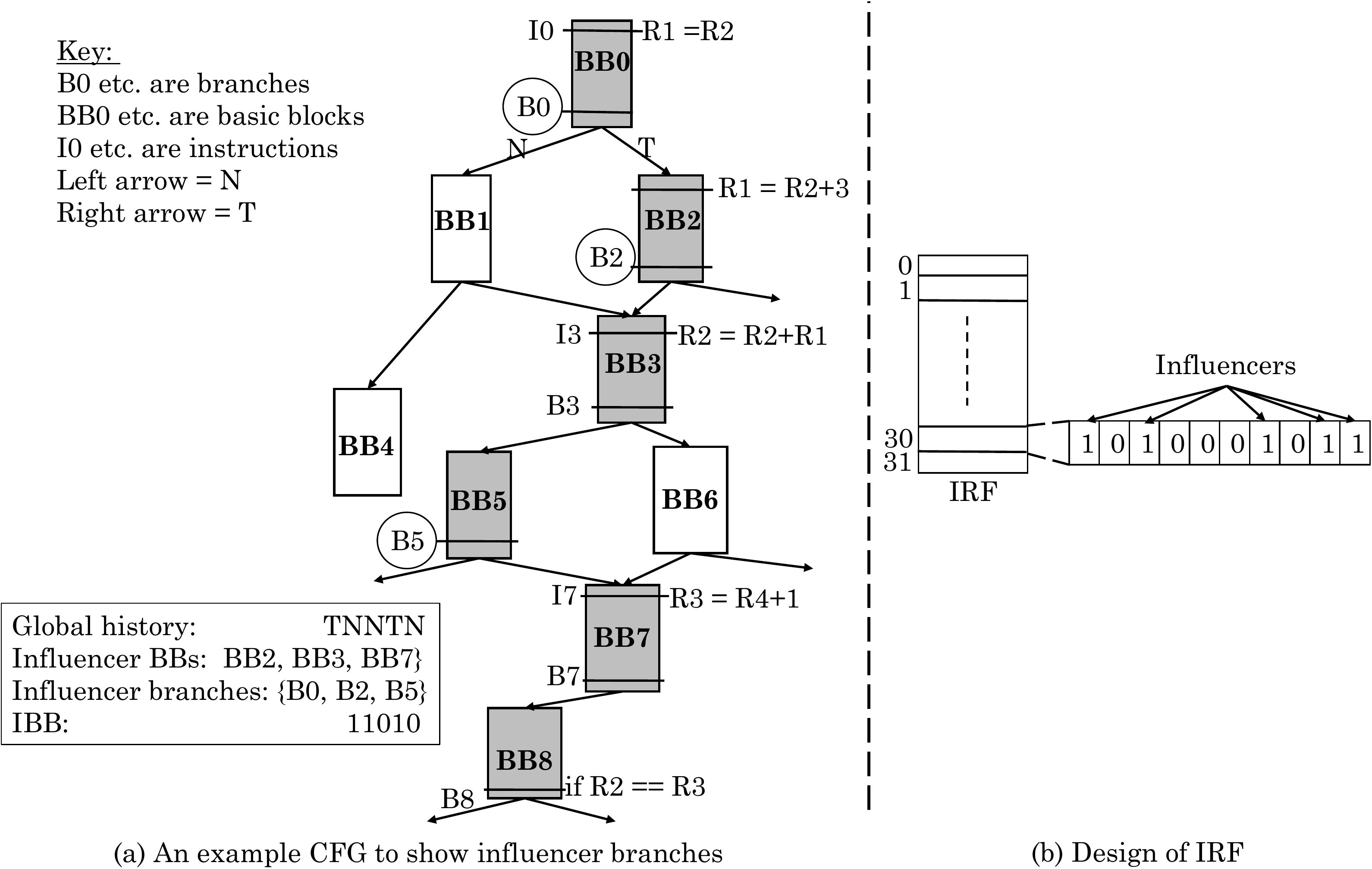}
\caption{\revised{(a) An example CFG (b) design of IRF \cite{thomas2003improving}} }\label{fig:influencer BB}
\end{figure}

Thomas et al. \cite{thomas2003improving} present a technique which uses  runtime dataflow information to ascertain the ``influencer branches'' for improving BP accuracy. If a branch $B_i$ decides whether certain operations directly affecting the source operands of an upcoming dynamic branch $B_0$ get executed, then $B_i$ becomes influencer branch for $B_0$. \revised{For instance, in Figure \ref{fig:influencer BB}(a), assume the control is flowing through the shaded path and we want to predict branch B8. B8 depends on registers R2 and R3. R2 value is generated in BB3 which takes R1 value generated in BB2. The R3 value consumed by B8 is generated in BB7. Thus, BB2, BB3 and BB7 are influencer BBs for B8. Also, the branches which determined the flow through these BBs are B0, B2, B5 and hence, these branches are called influencer branches for B8.}

Influencer branches show stronger correlation with their influenced branch than rest of the branches in the history and this is confirmed by comparison with the weights of a perceptron BP \cite{jimenez2001dynamic}. They track influencer branches for the latest instructions writing into each architectural register and store this information in an ``influencer register file'' (IRF). \revised{Figure \ref{fig:influencer BB}(b) shows the design of IRF which has one entry for every architectural register.} 
For a conditional branch, the influencer branch details are inherited from producers of its source operand. For a conditional branch, IRF entries of its source registers are read and OR'ed to obtain ``influencer branch bitmap'' (IBB). 

\begin{figure} [htbp]
\centering
\includegraphics[scale=0.40]{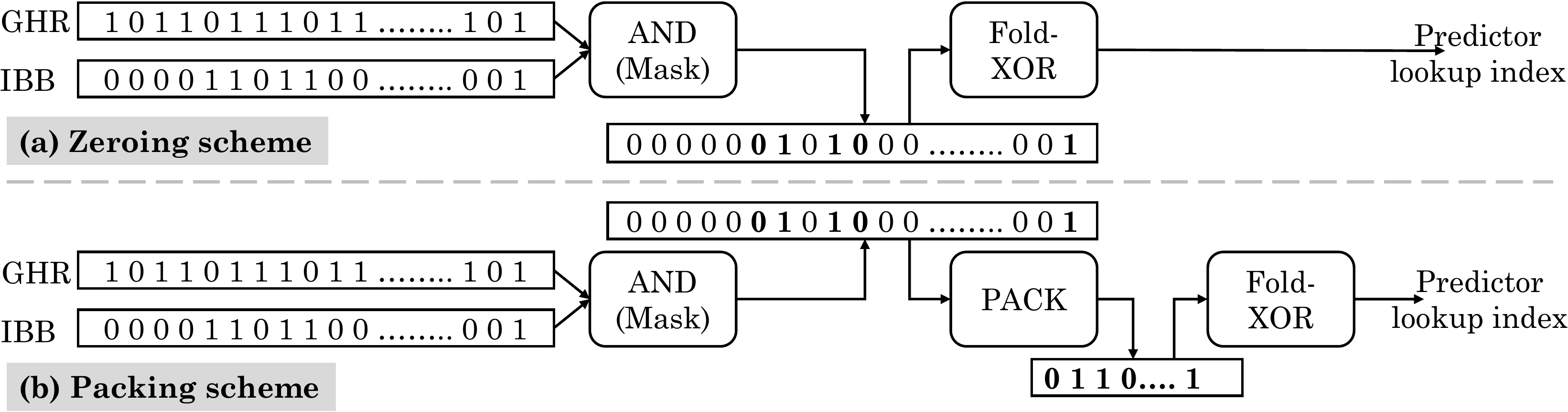}
\caption{ Use of influencing branch information for prediction in  (a) Zeroing and (b) Packing schemes \cite{thomas2003improving} }\label{fig:ZeroingPacking}
\end{figure}

They propose two techniques for using this information for making prediction.  In the ``Zeroing scheme'', shown in Figure \ref{fig:ZeroingPacking}(a) all non-influencer bits in the GHR are masked by ANDing them with IBB. Thus, influencer branches retain their positions in the global history, whereas non-influencer branches are shown as zero. Then, by using fold and XOR operations, masked GHR value is hashed down to the number of bits required for the predictor index. In the ``Packing scheme'', the only difference with ``Zeroing scheme'' is that  after masking the non-influencer bits, they are removed completely which compacts the subsequent influencers. This scheme in shown in Figure \ref{fig:ZeroingPacking}(b).

In the overall design, the line BP provides a 1-cycle prediction for fetching next instruction. The predictions of primary BP and corrector BP are obtained later one-after-another and they override respective previous predictors in case of disagreement; since by virtue of using long global history, corrector BP can make more accurate predictions. Corrector BP is modified form of REP \cite{heil1999improving} and it predicts only for influencer histories that can consistently correct the mispredictions of primary BP.  Using their technique with a perceptron BP improves accuracy with much less hardware cost than increasing the size of perceptron BP.

\subsection{Predicting based on address correlation} \label{sec:addresscorrelation}  
    
\begin{figure} [htbp]
\centering
\includegraphics[scale=0.40]{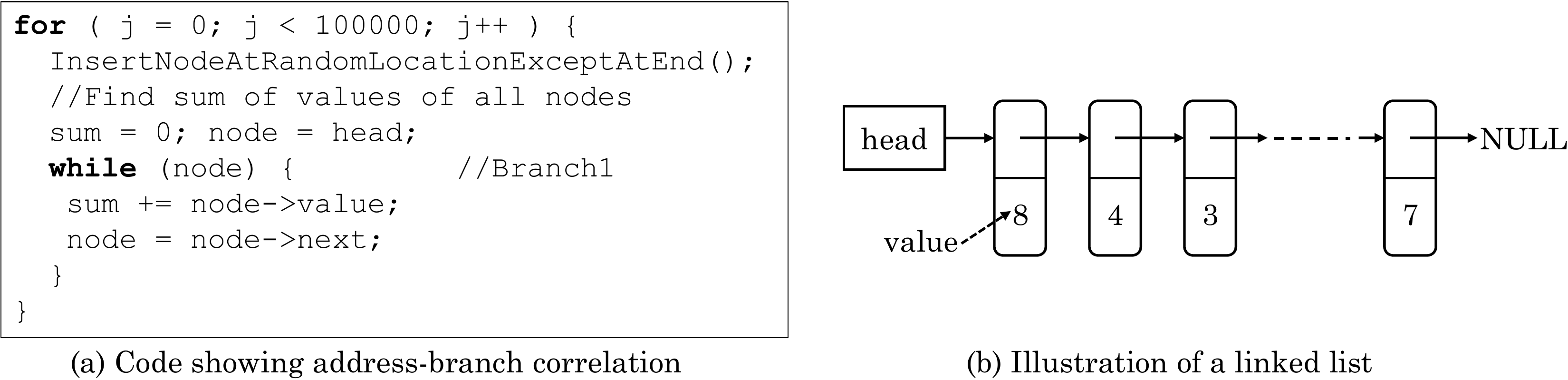}
\caption{Address branch correlation \cite{gao2008address}}\label{fig:addressbranchcorr}
\end{figure}

Gao et al. \cite{gao2008address} note that for branches depending on long-latency cache misses, if loaded values show irregular pattern, BPs based on leveraging branch history correlation show poor accuracy. They propose leveraging address-branch correlation to improve accuracy for such branches. For several programs (especially memory-intensive programs with significant pointer-chasing),  addresses of key data structures do not change for a long duration. For example, for the code shown in Figure \ref{fig:addressbranchcorr}, the address of last node in a linked list is likely to remain same until another node is appended to it. Thus, the outcome of a branch depending on  end of the list, can be determined based on the load address itself without requiring the actual value. This allows much faster resolution of the branch since the load address can be ascertained much before the value. By focusing on only few (e.g., 4) branches showing address-branch correlation, most of the branch misprediction latency can be alleviated. Using their 9KB BP as a side predictor with a 16KB TAGE predictor provides better performance and energy efficiency than a 64KB TAGE predictor. \revised{The limitation of their BP is that it shows poor accuracy for codes which frequently append to the linked list.}

\section{Hybrid BPs} \label{sec:hybridBPs}
In this section, we discuss hybrid BP designs. Table \ref{tab:hybrid} highlights essential ideas of several hybrid BPs.

\begin{table}[htbp]
\centering
\caption{\revised{Key ideas of hybrid BPs}}
\label{tab:hybrid}
\begin{tabular}{|p{6cm}|p{7.5cm}|}\hline
\multicolumn{2}{|c|}{Number of predictors/tables that the hybrid BP allows to choose from}\\\hline
Two &  \cite{mcfarling1993combining,seznec1999dealiased,falcon2004prophet,lee1997bi} \\\hline
Three & \cite{seznec1999dealiased} \\\hline  
Arbitrary number  &  \cite{evers1996using,seznec2005analysis,seznec2006case,seznec2011storage} \\\hline

\multicolumn{2}{|c|}{Accessing choice predictor using }\\\hline

Branch address only & \cite{lee1997bi} \\\hline 
Both branch address and global history &  \cite{seznec1999dealiased} \\\hline
\multicolumn{2}{|c|}{From the predictions of multiple components, the single final prediction can be chosen based on }\\\hline

Bias of each branch & \cite{lee1997bi,chang1994branch,seznec2002design} \\\hline
Using choice predictors &  \cite{seznec1999dealiased,gwennap1996digital}\\\hline
Majority voting & \cite{michaud1997trading,loh2002predicting,chaver2003branch} \\\hline
Giving preference to & tagged predictor with longest history  \cite{seznec2006case}, to predictor having highest confidence \cite{evers1996using}, to the critic BP in prophet-critic hybrid BP design \cite{falcon2004prophet} and to the larger BP in overriding BP designs \\\hline
Combining strategies & Adding  \cite{seznec2005analysis} or fusing \cite{loh2002predicting} the predictions of the components  to obtain final prediction \\\hline
\end{tabular}
\end{table}

\subsection{Tagged hybrid BPs}\label{sec:taggedhybrid}

\begin{figure} [htbp]
\centering
\includegraphics[scale=0.40]{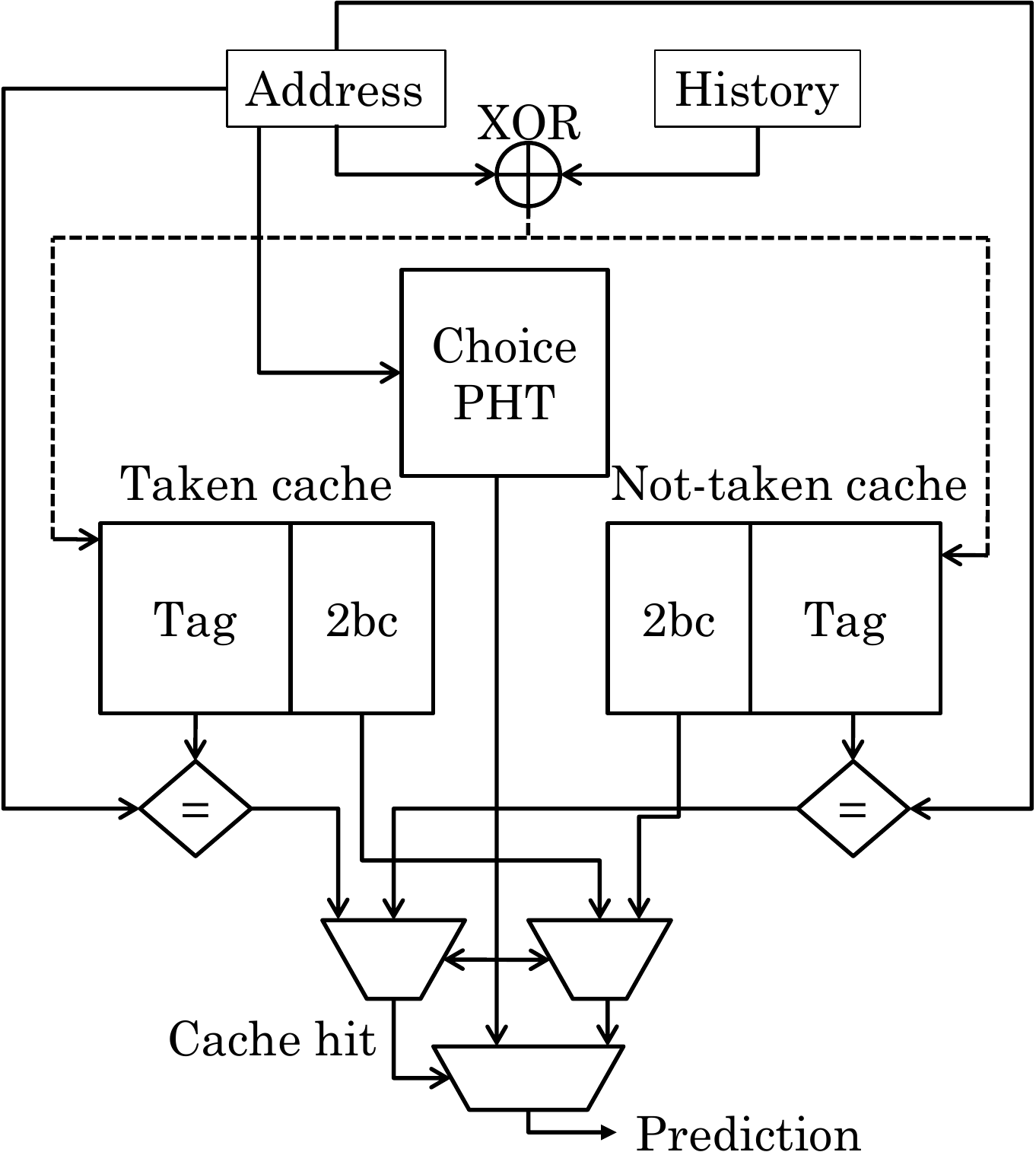}
\caption{The tagged BP proposed by Eden et al. \cite{eden1998yags}}\label{fig:yags}
\end{figure}

Eden et al. \cite{eden1998yags} propose a BP which is hybrid of gshare and bimodal BPs.  
Their technique divides the PHT into two branch streams corresponding to taken and not-taken bias and stores them in choice PHTs, as shown in Figure \ref{fig:yags}. Further, in direction PHTs, it stores the occurrences which do not agree with the bias. This lowers the information content in direction PHTs and allows reducing its size below that of choice PHT. To identify such occurrences in direction PHTs, they  store 6-8 least significant bits of the branch address as the tags in every entry.   These narrow  tags almost completely remove aliasing, especially after the context-switching.

For any branch, first the choice PHT is consulted and if it indicates `not-taken', then the `taken' direction PHT (also called a cache due to use of the tags) is referenced to see whether this is a special case where bias and prediction disagree. On a hit or miss in the taken cache, its own outcome or that of choice PHT (respectively) is used as prediction. Converse action is taken if the PHT access for a branch indicates `taken'. The addressing and updating of choice PHT happens as in the bimodal choice PHT \cite{lee1997bi}.  The `taken' cache is updated if its prediction was used or if the choice PHT indicates `not-taken' but the branch result was taken. Converse is true for updating the `not-taken' cache.
To remove aliasing for branch occurrences that do not agree with the bias of the branch, they design direction caches as two-way set-associative cache. They use LRU replacement policy, except that an entry in `not-taken' cache indicating `taken' is preferentially replaced for removing redundant information since this information is also available in the choice PHT. Their technique achieves high accuracy for the same reason as the bimodal BP and also due to the use of tags.  The limitation of their technique is that it provides only small improvement over bimodal predictor. Also, beyond a certain tag size (6 bits), there is no improvement in accuracy.

\subsection{Selection-based Hybrid BPs}\label{sec:selectionhybridBPs}

Chang et al. \cite{chang1994branch} note that since both biased and non-biased branches appear frequently, a single BP cannot provide high overall accuracy. They propose and evaluate multiple hybrid BP designs (1) a GAs BP where short and long BHR is used for biased and non-biased branches, respectively (2) using profile-guided BP for biased branches and gshare \cite{mcfarling1993combining} for non-biased branches (3) 2bc+gshare (4) PAs+gshare  (5) profile-guided BP for biased branches and PAs+gshare hybrid for non-biased branches. They show that these hybrid BPs provide higher accuracy than solo-BPs. Also, using static BPs for biased branches allows dedicating higher storage budget to dynamic BPs. 

\begin{figure} [htbp]
\centering
\includegraphics[scale=0.40]{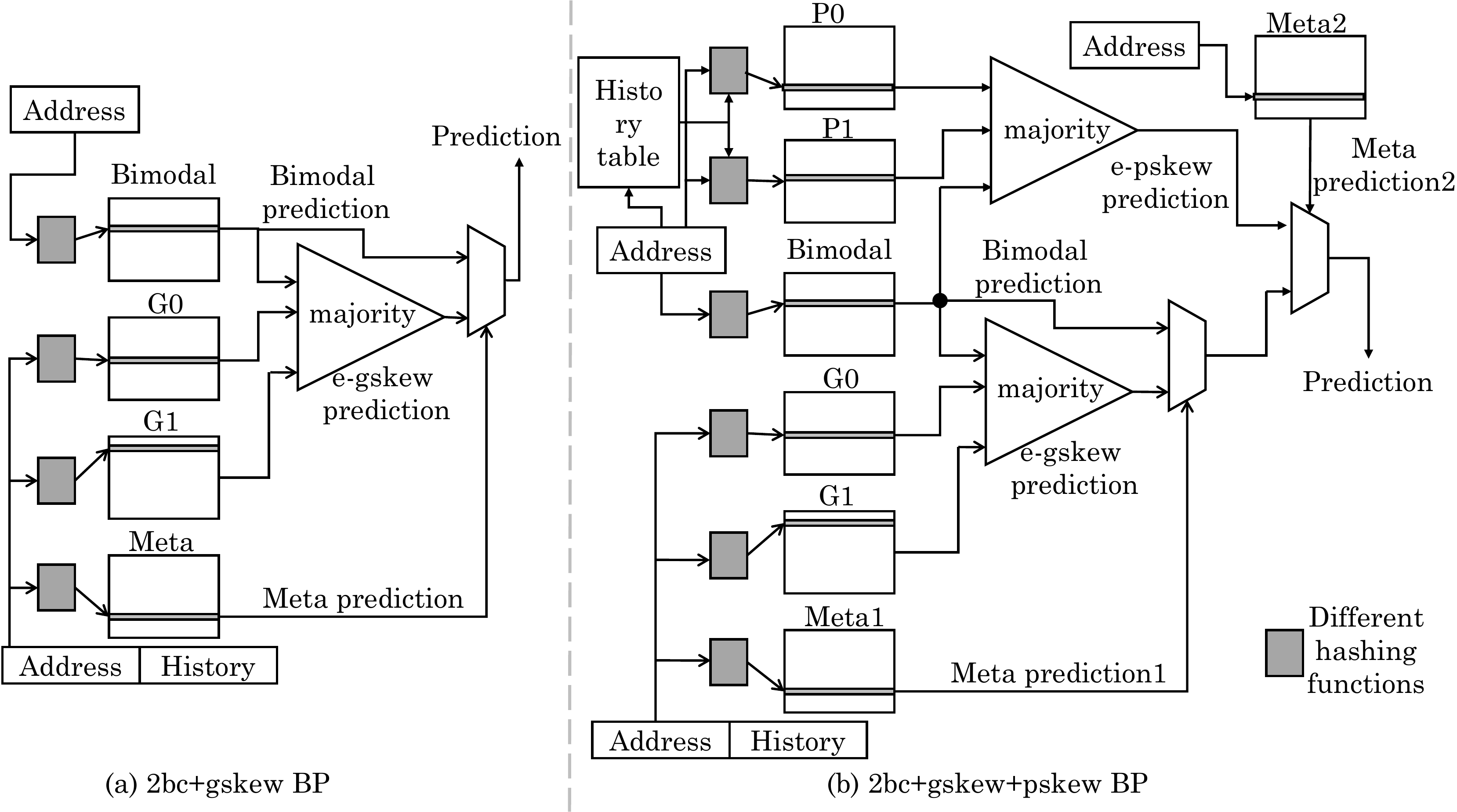}
\caption{(a) 2bc+gskew and (b) 2bc+gskew+pskew BPs \cite{seznec1999dealiased}}\label{fig:2bc-gskew-pskew}
\end{figure}

Seznec et al. \cite{seznec1999dealiased} design two hybrid BPs, namely 2bc+gskew and 2bc+gskew+pskew. 2bc+gskew has 4 predictor banks: 3 banks of e-gskew \cite{michaud1997trading} and a meta predictor (refer Figure \ref{fig:2bc-gskew-pskew}(a)). Bimodal predictor is already part of e-gskew. Meta predictor is accessed using a combination of branch address and global history. They use partial update policy whereby the three banks of e-gskew are updated on an incorrect prediction. Also, for a correct prediction, if the prediction was provided by bimodal BP, only this is updated, but if it was provided by the e-gskew, only the banks that provided the prediction are updated. The meta predictor is updated only in case of disagreement between the two predictors. They show that 2bc+gskew provides higher accuracy than e-gskew since bimodal BP provides large fraction of predictions due to which other two banks are not updated and thus, not polluted by branches not requiring the use of GHR for accurate prediction. Also, in most cases, bimodal and e-gskew BPs agree and hence, meta predictor is neither required nor updated, and hence, aliasing in meta predictor does not harm accuracy. In 2bc+gskew+pskew BP shown in  Figure \ref{fig:2bc-gskew-pskew}(b), e-pskew component is the per-address history BP for accurately predicting branches that benefit from per-address history instead of global history; thus, 2bc+gskew+pskew has three constituents: bimodal, per-address history and global history. Due to this, 2bc+gskew+pskew BP provides higher accuracy than 2bc+gskew BP although it also incurs higher latency.

Seznec et al. \cite{seznec2002design} present design of BP in Alpha EV8 processor which takes 352Kb space. Alpha EV8 fetches up to two, 8-instruction blocks in each cycle, thus, up to 16 branches may need to be predicted in each cycle. Hence, they use a GH BP, since a local history BP would require high resources (e.g., 16-ported predictor table) and have high complexity (e.g., competition for predictor entries from different threads). Their BP is based on 2bc+gskew hybrid BP \cite{seznec1999dealiased}. In this hybrid BP, biased branches are precisely predicted by bimodal predictor and hence, for such branches, other tables are not updated. This partial update scheme avoids aliasing due to biased branches and also simplifies implementation. This also allows separating predictor and hysteresis arrays, and to meet area budget, hysteresis array for meta and G1 predictors (refer Figure \ref{fig:2bc-gskew-pskew}(a)) are reduced to half the size. Use of partial update reduces the writes to hysteresis array which lowers the aliasing effect. Also, they use larger history length for G1 than G0 and smaller predictor table for bimodal than for G0, G1 and meta predictors. 

The BP is implemented as a 4-way bank-interleaved with single-ported memory cells and bank conflicts are completely avoided by computing bank numbers in a way to ensure that two consecutive fetch blocks access different banks. They show that despite implementation constraints, the accuracy of their  BP is comparable to other iso-size GH-based BPs.

\subsection{Fusion-based hybrid BPs}
\begin{figure} [htbp]
\centering
\includegraphics[scale=0.40]{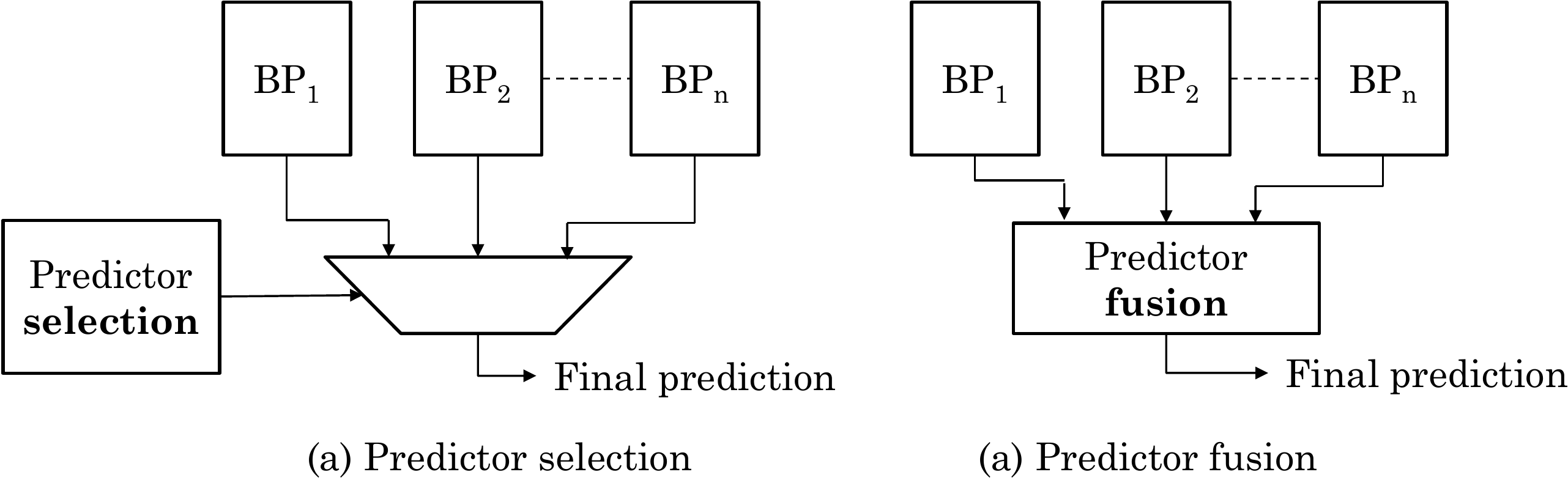}
\caption{(a) Selection of one BP from $N$ BPs (b) Combining information from all $N$ BPs \cite{loh2002predicting}}\label{fig:predictorselectionfusion}
\end{figure}

Loh et al. \cite{loh2002predicting} note that hybrid BPs that choose one constituent BP ignore the information from unselected BPs. They propose a BP fusion approach which makes final prediction based on information from all constituents. These approaches are shown in Figures \ref{fig:predictorselectionfusion}(a) and \ref{fig:predictorselectionfusion}(b) respectively. 

\begin{figure} [htbp]
\centering
\includegraphics[scale=0.40]{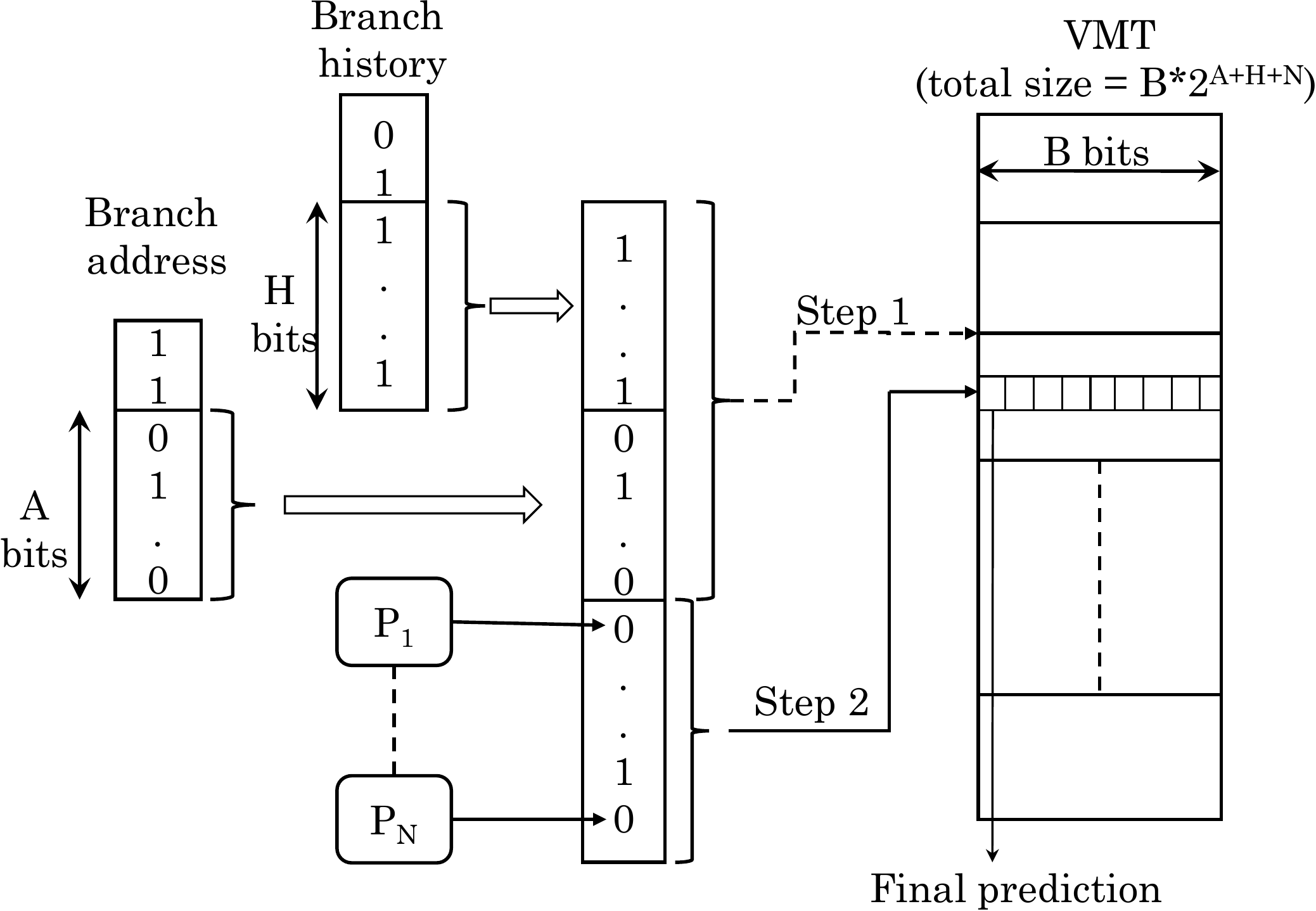}
\caption{Combining predictions of all constituent BPs to arrive at the final prediction \cite{loh2002predicting}. Each entry of mapping table is a $B$-bit counter. }\label{fig:LohFusionVMT}
\end{figure}

Their proposed predictor has $N$ constituent BPs along with a ``vector of mapping tables'' (VMT) for mapping their predictions to final prediction. Figure \ref{fig:LohFusionVMT} shows their proposed BP. Each entry of VMT is a $2^N$ entry mapping table, whose entries are saturating counters. 
Their BP first looks-up constituent predictors and in parallel, selects one mapping table from VMT based on branch history and address bits  (step 1). Then, based on individual predictions, one of the $2^N$ counters of selected mapping table is chosen (step 2). The MSB of this counter provides final prediction. These counters are incremented/decremented based on whether  actual outcome was taken/not-taken, respectively. The constituent predictors of their BP are chosen using a genetic search approach with the constraint of a fixed area budget. Their BP provides high accuracy although it also has a large lookup latency (4 cycles).

\subsection{Multi-hybrid BP designs}\label{sec:multihybrid}
Evers et al. \cite{evers1996using} note that by combining more than two BPs, even higher number of branches can be accurately predicted. For each solo-BP, they use a 2-bit counter which is initialized to the value three. The outcome of BP, whose corresponding counter has a value of three, is finally chosen and ties are broken using a fixed precedence order. As for update, if one of the BPs that had the value three in its counter is found to be correct, the counters for remaining incorrect BPs are decreased by one. Otherwise, the counters for all the correct BPs are increased by one. This ensures that at least one of the counters will have a value of three. Compared to saturating counters, their update strategy can more accurately identify which solo-BPs are presently more accurate for different branches. They use variations of both per-address and global two-level solo-BPs as constituents of their hybrid BP. These solo-BPs have high accuracy by virtue of using a long history. However, due to this, they also have large warm-up time. Since static BPs and small-history dynamic BPs require no and short warm-up periods, respectively, they are also included in the hybrid BP. Their inclusion allows the hybrid BP to show high accuracy during warm-up period. Overall, their hybrid BP provides high accuracy.

\subsection{Prophet-critic hybrid BP}\label{sec:prophet}
\begin{figure} [htbp]
\centering
\includegraphics[scale=0.40]{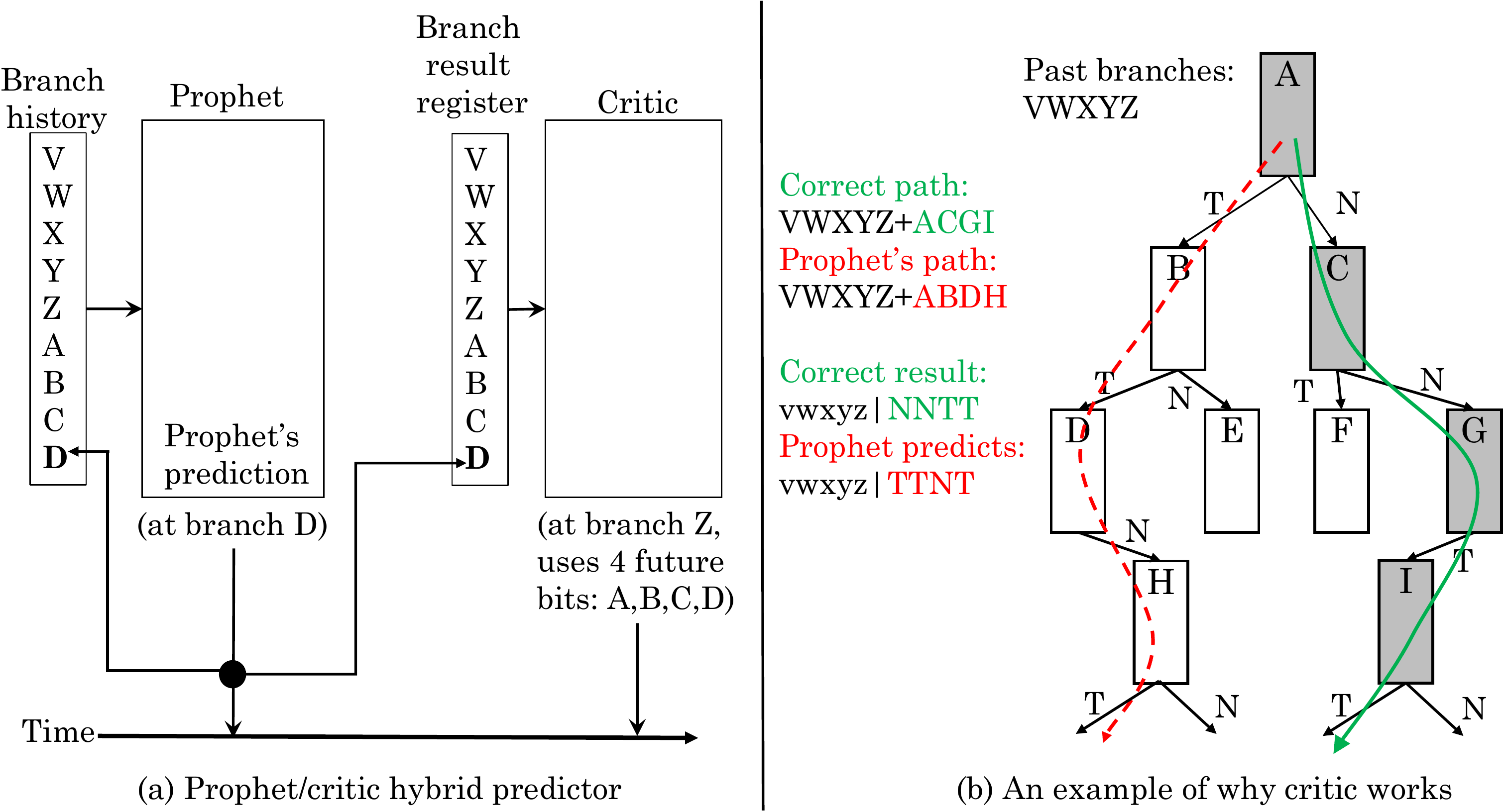}
\caption{ (a) Design of and (b) rationale behind prophet-critic predictor \cite{falcon2004prophet} }\label{fig:prophet}
\end{figure}

Falcon et al. \cite{falcon2004prophet} propose a technique which uses two BPs called prophet and critic, as shown in Figure \ref{fig:prophet}(a).  The prophet makes prediction based on current branch history and goes on the predicted path, making further predictions which form ``branch future'' for the original branch. 
  Based on this information, the critic correlates with both past and future and generates an agree/disagree critique of each prediction of the prophet. The critic's prediction is the ultimate prediction for the branch, since by using future code behavior, it makes more accurate predictions. 
  
  For example, in Figure \ref{fig:prophet}(b), the correct path is shown by the shaded blocks (A, C, G, I). The history of A is formed by the results of its previous branches: V, W, X, Y and Z and this history is used by the prophet to predict A. If all the branches are correctly predicted, branches A, C, G and I (i.e., outcome NNTT where N/T = not-taken/taken) form the future of A. However, if A is mispredicted, the execution proceeds on wrong path shown by red dashed line. Here, the prophet predicts branches A, B, D and H and their predictions are stored in branch result register of the critic. Thus, the critic has both branch history (result of V, W, X, Y, Z) and branch future (result of A, B, D and H).

  If for a branch $B$, the prophet makes prediction after observing $P$ extra branches, the number of future bits used by prophet are said to be $P$, e.g., in Figure \ref{fig:prophet}, $P=4$. With increasing $P$, prophet's accuracy increases since it can detect more progress along the wrong path. When prophet mispredicts a branch for the first time, critic learns this event and when prophet mispredicts the branch again under same situation, critic disagrees with the prediction of the prophet.   Prophet/critic BPs work independently and any existing BP can be used for them. To improve performance, prophet should correctly predict more than 90\% of branches and critic is responsible for predicting only the remaining branches. In both traditional hybrid and overriding BPs, component BPs predict the same branch based on the same information, however, in their BP, two predictions happen at different times. Their technique provides large improvement in  BP accuracy and performance.

\section{Techniques for Improving BP accuracy}\label{sec:improvingaccuracy}

In this section, we discuss several approaches for improving BP accuracy, such as dynamically adapting history or path-length (Section \ref{sec:adaptinglengths}), reducing aliasing (Sections \ref{sec:filteringhistory}-\ref{sec:conflictreducing}), improving accuracy in presence of  multithreading and  short-lived threads (Section \ref{sec:multithreading}) and learning from wrong-path execution (Section \ref{sec:recyclewaste}). \revised{Table \ref{tab:reducingaliasing} summarizes several strategies for reducing aliasing.} 

\begin{table}[htbp]
\centering
\caption{\revised{Strategies for reducing aliasing}}
\label{tab:reducingaliasing}
\begin{tabular}{|l|l|}\hline
Use of tag &  \cite{lai2005improving,eden1998yags,sherwood2000loop,ma2006using,chen2003dynamic,aragon2001selective,seznec2005analysis,jimenez2011optimized,heil1999improving} \\\hline
Intelligent indexing functions &  \cite{ma2006using} \\\hline
Using  multiple indexing functions &  \cite{monchiero2005combined,michaud1997trading} \\\hline
XORing branch address with branch history &  \cite{mcfarling1993combining} \\\hline
Converting harmful interference to beneficial/harmless ones & \cite{sprangle1997agree} \\\hline
Storing only most recent instance of a branch in BP & \cite{gope2014bias} \\\hline
Storing branches with different biases separately & \cite{lee1997bi,eden1998yags} \\\hline
Separate BPs for user and kernel branches &  \cite{li2007aware} \\\hline
\end{tabular}
\end{table}

\subsection{Adapting history-length and path-length}\label{sec:adaptinglengths}
 
Juan et al. \cite{juan1998dynamic} note that using constant history length for all the branches does not perform well since optimum history length depends on input data, code features and context-switch frequency. They propose dynamically varying the number of history bits used for each application and input data in two-level BPs. For gshare BP \cite{mcfarling1993combining}, in different execution intervals, they XOR different number of history bits with PC of branch instruction.  The interval is determined by the fixed number of dynamic branches. The number of mispredictions with each history length is recorded in a table. After each interval, if the number of mispredictions in existing interval is higher than the smallest number in the table, history length is altered (increased or decreased) by one to move towards the optimum history length; otherwise, it is kept the same.   The limitation of their scheme is that after a change in history length, a large portion of PHT state is lost and needs to be generated again. This leads to aliasing and increases mispredictions. To offset its impact, after a history length change, misprediction counters are not updated for one interval. They show that their technique approaches the accuracy obtained using fixed optimum history length in both absence and presence of context-switching. Their technique can be used with any BP that combines global branch history with PC to find the PHT-index, e.g., agree \cite{sprangle1997agree}, bimodal \cite{lee1997bi}, gselect \cite{mcfarling1993combining}, gshare \cite{mcfarling1993combining} and e-gskew \cite{michaud1997trading} BPs.

\begin{figure} [htbp]\centering
\includegraphics[scale=0.40]{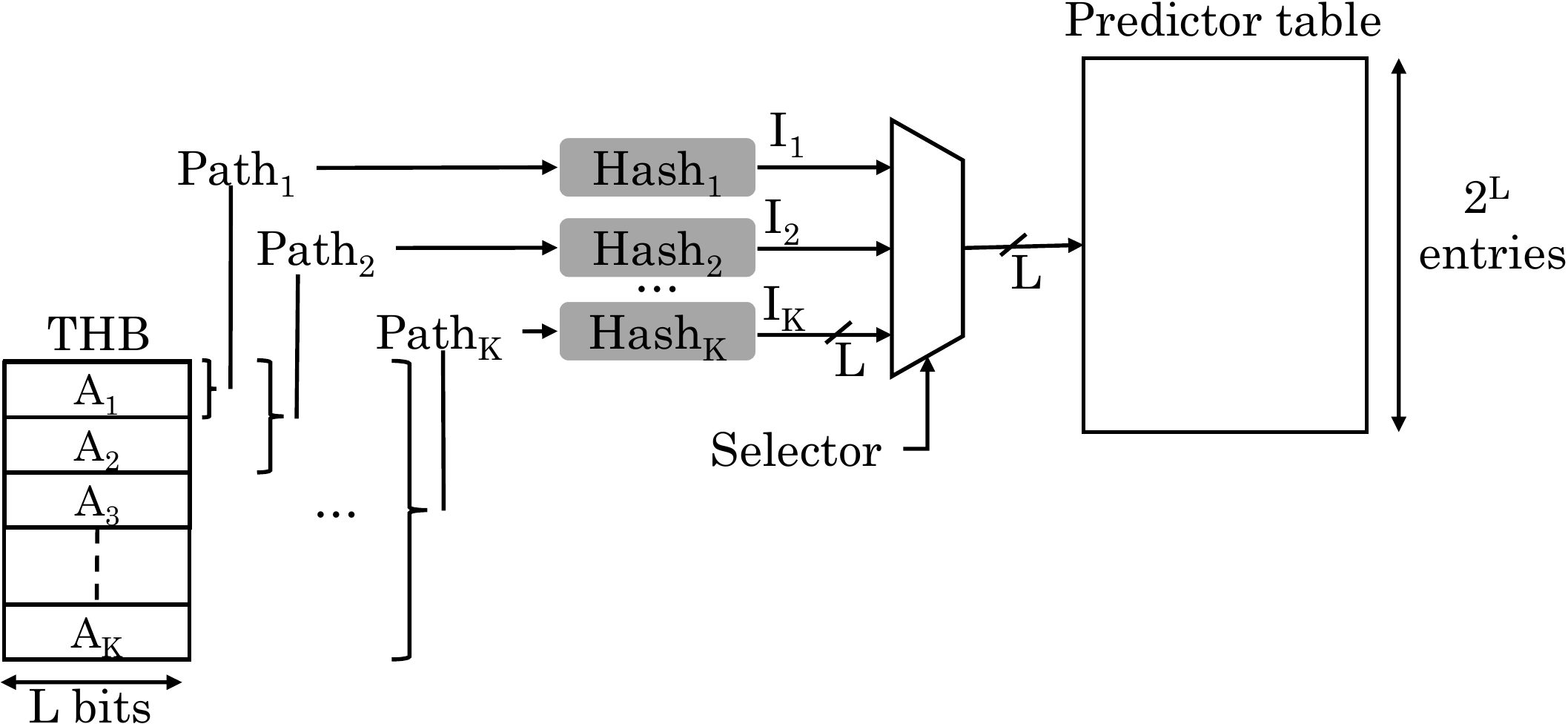}
\caption{Variable path-length based BP  \cite{stark1998variable}}\label{fig:variablepath}
\end{figure}

Stark et al. \cite{stark1998variable} note that in path-based BPs, $K$ most recent target addresses are hashed to obtain the table index for making a prediction. They propose a 2-level BP, which allows using different path length  ($K$) for each branch to achieve high accuracy. The destination addresses of most recent $K$ (e.g., $K=32$) branches form level-1 history, as shown in Figure \ref{fig:variablepath}. The lower $L$ bits of these addresses are stored in ``target history buffer'' (THB). Let A$_j$ refer to $j^{th}$ most-recent address stored in THB. Only those destination addresses that provide useful information about the path leading to the candidate branch are stored in THB. The $L$-bit index obtained from level-1 table is used for accessing level-2 history stored in a ``predictor table''. Each predictor table entry uses a 2-bit saturating counter.  A path of length $j$ (PATH$_j$) has latest $j$ addresses in the THB, thus,  PATH$_j$ has A$_1$ to A$_j$ addresses. $K$ predictor table indices are obtained using $K$ distinct hash functions (HASH$_j$). For example, HASH$_3$ uses A$_1$, A$_2$ and A$_3$ to provide $I_3$. The hash function rotates $j^{th}$ destination address $j-1$ times and then XORs all the addresses in the path. Rotating the address allows remembering the order in which the addresses were seen.  For every branch, based on profiling information, one hash function is chosen which provides highest accuracy for that branch. The hash function can also be chosen dynamically. Their BP reduces training time and aliasing and hence, improves accuracy.

\subsection{Filtering branch history}\label{sec:filteringhistory}  
While increasing the history length generally improves accuracy by allowing exploitation of correlation with the branches in distant past, inclusion of uncorrelated branches can introduce noise in the history. Several techniques seek to reduce this noise by filtering the branch history.

\begin{figure} [htbp]
\centering
\includegraphics[scale=0.40]{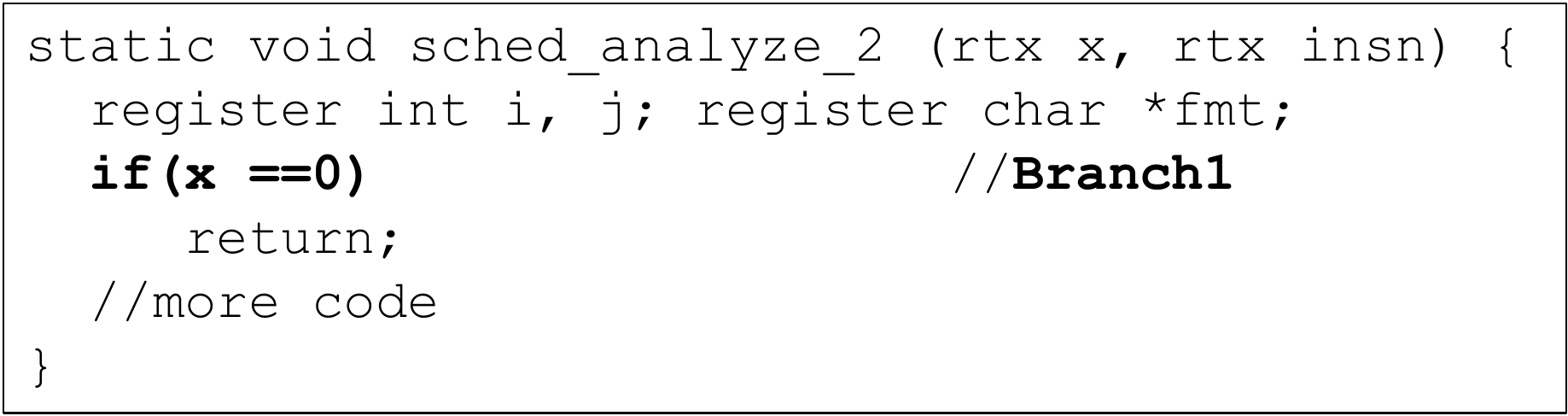}
\caption{\revised{An example of code-zone with limited branch correlation \cite{porter2009creating}} }\label{fig:ZLBC}
\end{figure}

Porter et al. \cite{porter2009creating} note that for code zones with limited branch correlations (ZLBCs), additional information stored in GHR harms accuracy of this and other branches by increasing predictor training time and aliasing. \revised{Figure \ref{fig:ZLBC} shows part of the code from a function in {\tt gcc} benchmark.  Branch1 is the first branch in a function and it checks for null pointer. In the execution of 100M instruction, the branch is executed 2455 times but is never taken. The branch sees 208 different 16-bit histories, and hence, is mipredicted 9\% of times.} To improve accuracy in such scenarios, they propose temporarily removing unbeneficial correlations from GHR to insulate  poorly-correlated branches from the useless correlations, while still allowing  highly-correlated branches to benefit from large histories.

\revised{A ZLBC is a zone where execution moves from one region to another, such that these zones show no correlation}. They identify ZLBCs  using control flow instructions, e.g., backward branches and function calls and returns. For example, the not-taken path of a backward branch generally shows a loop exit and the branches after a loop are expected to have limited correlation with those in the loop. Similarly, branches in a function have limited correlation with those of the preceding and following code zone. On detecting ZLBCs, their technique resets GHR similar to the approach of Choi et al. \cite{choi2008accurate}. For instance, on function calls, the PC of the calling instruction is stored in GHR and on function return, the PC of return instruction is stored in GHR. \revised{ For the example shown in Figure \ref{fig:ZLBC}, using this approach reduces the misprediction rate to just 1\% \cite{porter2009creating}.}  
Their technique can be used with many BPs and  especially benefits simple BPs allowing them to become competitive with  complex BPs.

Xie et al. \cite{xie2013energy} present a technique to reduce the impact of useless branch correlations (i.e., noise) in BPs. They divide the original BP into two sub-predictors: a conventional BP (CBP) and a history-variation BP (VBP). Both BPs work the same, except that their history update mechanism is different: CBP stores branch histories until a branch is committed, whereas VBP uses a history stack to remove branch histories in loops and functions. While entering a loop or function, current history is pushed to the history stack and on exiting the loop, the history is popped back.  RAS and loop predictors present in the processors are augmented with history stack and using them, functions and loops (respectively) are detected. Based on execution histories of earlier branches, either VBP or CBP is chosen using a selector table.

\begin{figure} [htbp]
\centering
\includegraphics[scale=0.40]{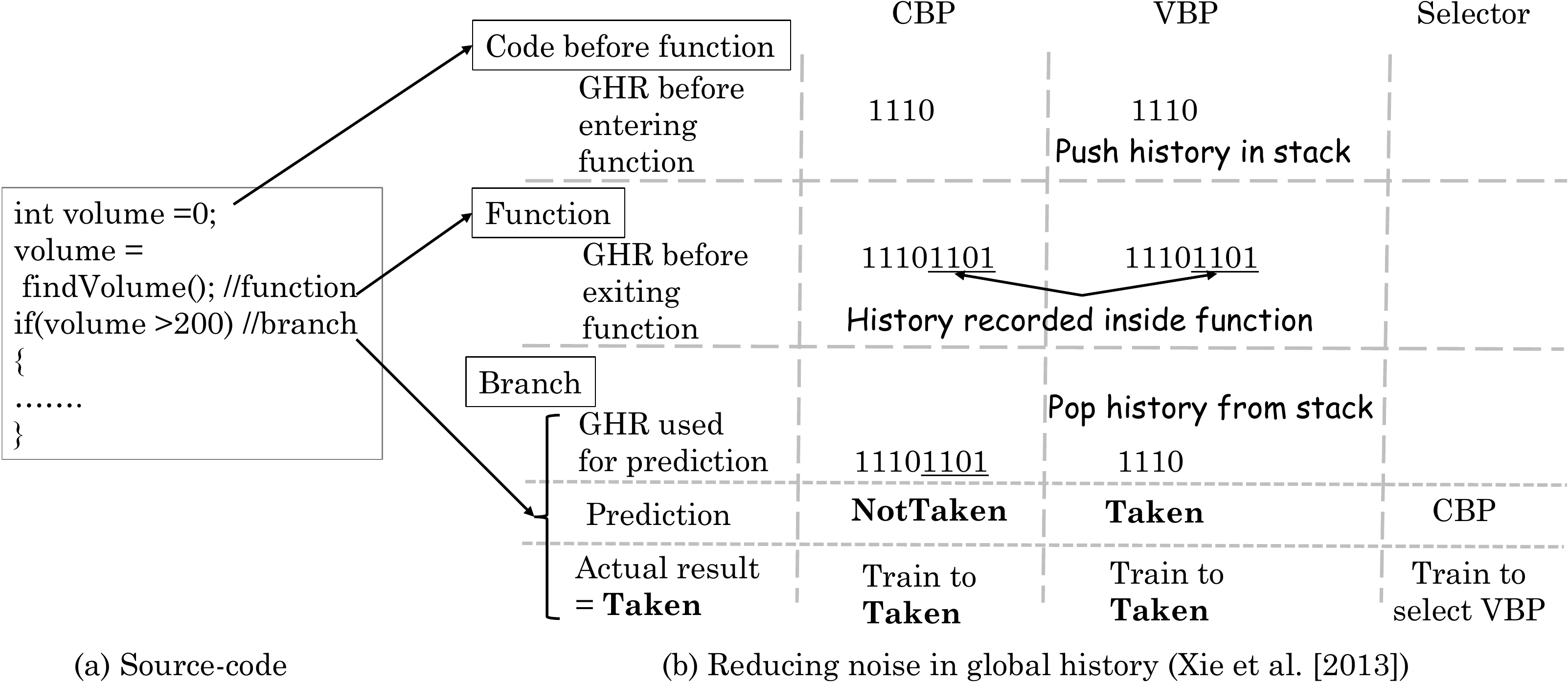}
\caption{Example of working of technique of Xie et al. \cite{xie2013energy}  }\label{fig:NoiseRemovalHAS}
\end{figure}

Figure \ref{fig:NoiseRemovalHAS} summarizes the working of their technique. \revised{Figure \ref{fig:NoiseRemovalHAS}(a) shows the typical source-code. Befor entring the function, the GHR in both CBP and VBP are 1110. The VBP pushes/pops the history in the stack while entering/exiting the function, respectively. During code-execution, CBP works same as a traditional BP. Hence, while predicting the branch, CBP and VBP have different histories and hence, they produce different predictions. The selector decides the final prediction, which, in this case, is `not-taken'. Since actual result happens to be `taken', both CBP and VBP are trained towards `taken' and the selector is trained to choose VBP since it provided correct prediction.}  By using their approach with gshare, perceptron and TAGE predictors, they show that their approach improves BP accuracy.

\begin{figure} [htbp]
\centering
\includegraphics[scale=0.40]{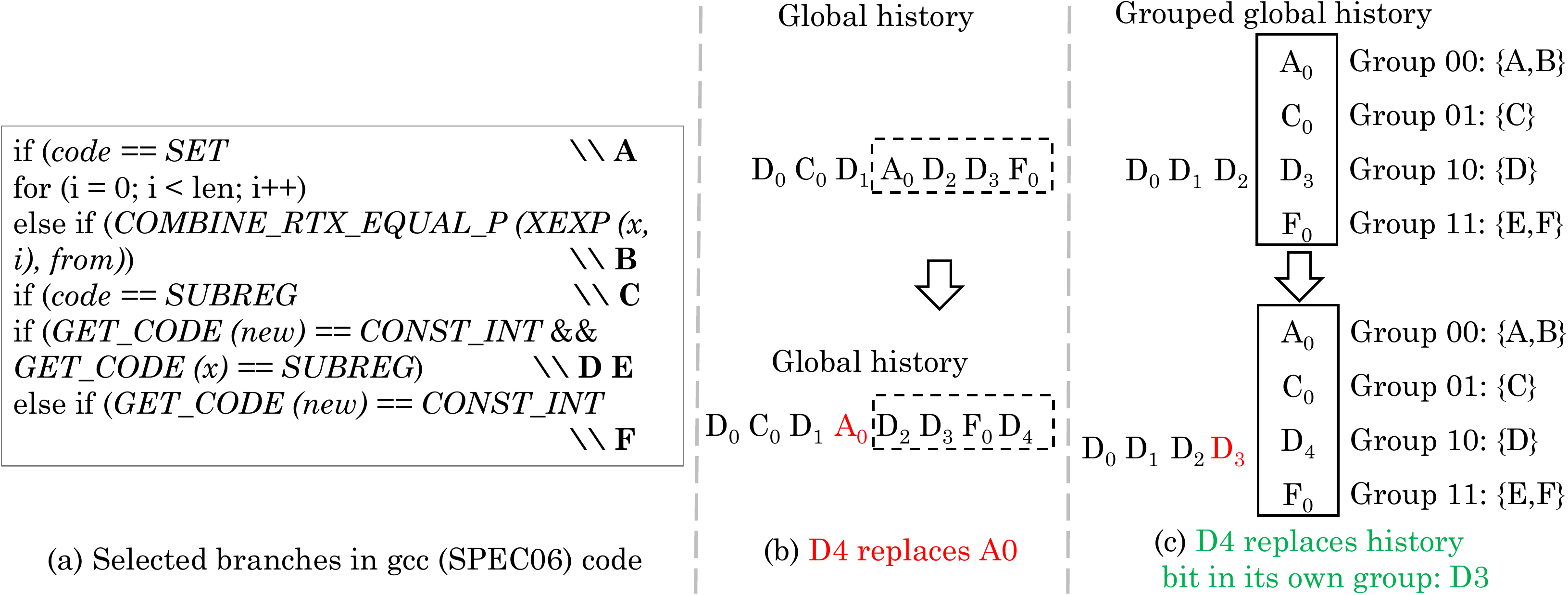}
\caption{Reducing aliasing by dividing global history into groups \cite{huang2015energy}   }\label{fig:groupedglobalhistory}
\end{figure}

Huang et al. \cite{huang2015energy} note that aliasing reduces the effectiveness of BPs significantly, e.g., hot branches in a loop can flush history of outside-loop branches which negates the benefit of maintaining long global history. Different from other BPs that use single global history, they divide global history into multiple groups  based on lower-order bits of branch address to restrict aliasing due to hot branches.   A branch (instance) can replace another branch from its own group only.  By concatenating all history bits in the entries, final history is formed. Their approach allows tracking longer correlation history for the same hardware cost. For example, in Figure \ref{fig:groupedglobalhistory}(b), a 4-bit GHR is used and $Ai$ to $Fi$ refer to the history bits of corresponding branches in the code shown in Figure \ref{fig:groupedglobalhistory}(a). Here, once D4 is committed, it is shifted in GHR and A0 is evicted.  Their technique divides branches in 4 groups based on two low-order bits. Only 1-bit history is stored in each bit, which keeps the total storage cost same. Here, when D4 is committed, it evicts D3 and not A0, as shown in Figure \ref{fig:groupedglobalhistory}(c).    They show that  use of their approach with perceptron predictor reduces the misprediction rate significantly.  The limitation of their technique is that it limits the local information tracked and hence, in some cases, it may perform worse than using local and global history. 

Gope et al. \cite{gope2014bias} propose storing only non-biased branches in branch correlation history, since due to (almost) always having a fixed outcome,  biased branches do not impact the prediction of a subsequent non-biased branch.  Their technique detects biased nature of a branch dynamically using a finite state machine. A conditional branch is initially in {\tt Not-Present} state. When it is committed for the first time, it moves to {\tt Taken} or {\tt Not-Taken} state. If a branch in one of these states executes in opposite direction, it moves to {\tt Not-biased} state. Unless confirmed to be not-biased, a branch is considered as biased and is not included in the    history of upcoming branches.
  
They further note that multiple instances of repeating branches hardly provide additional value. Hence, their technique uses a recency-stack to record only the most-recent instance of a non-biased branch and tries to find correlation with that instance. This reduces the space taken by a single branch in path history.  However, in some cases, different instances of a branch may have different correlations with the latest instance of a branch present in the recency-stack and to take this into account, they also record the position information which shows the distance of the branch from the present branch in the global history.

They also apply their bias-removal approach to perceptron predictor (PP), which is termed as bias-free PP. During training phase, their bias free PP does not work well for strongly-biased branches that do not correlate with remote histories. To resolve this issue, they add a traditional PP component which records correlations for few unfiltered history bits to alleviate mispredictions  during training phase. Further, they use their approach to reduce storage overhead of TAGE predictor. Storing \textit{only one instance of non-biased} branches (instead of all instances of all branches)   allows finding correlated branches from very remote past (e.g., 2000 branches) within a small hardware cost which improves performance.

\textbf{Comparison:} While Gope et al. \cite{gope2014bias} organize the history by the branch instruction address, Huang et al. \cite{huang2015energy} form groups by organizing the history by lower-order bits of the branch PC. Hence, Gope et al.  use associative search for the branch PC, while Huang et al. use table lookup which incurs less overhead. Also, Huang et al. can adapt the number of history bits in each group whereas Gope et al. use a fixed number of history bits for each static branch.

\subsection{Reducing aliasing with kernel-mode instructions}\label{sec:kernelaliasing}
Most research works evaluate BPs for user-mode instructions only.  Chen et al. \cite{chen1996analysis} note that when user-mode instructions account  for more than 95\% of total instructions, user-only and full-system misprediction rates match closely.  However, when user-mode instructions account for less than 90\%, they do not reflect full-system results. Hence, the best BP for  user-only branch traces may not be the same as that for full-system traces. Further, inclusion of kernel branches aggravates aliasing by increasing the number of static branches predicted, which reduces the effective size of BP.   The impact of aliasing is higher in two-level BPs with long histories than in BPs using short depth of local history.  Furthermore,  flushing the branch history  at regular intervals, as suggested by Nair et al. \cite{nair1995dynamic}, does not accurately model the effects of user/kernel interactions since kernel or other processes may not always flush the branch history state. \revised{Also, the flushing is especially harmful for BPs with large table sizes.}

Li et al. \cite{li2007aware} note that branch aliasing between user/kernel codes increases mispredictions in both their executions as their branches show different biases. Also, many branches in kernel mode are weakly biased which are difficult to predict by most BPs. They also note that increasing the predictor size does not alleviate this problem. They propose two techniques to reduce the aliasing. First, they use two branch history registers to separately capture branch correlation information for user and kernel code. However, with this technique, user/kernel aliasing can still happen in PHT.  To address this, their second technique  uses different branch history registers and  PHTs for user and kernel mode. Since the number of active branch sites in kernel code are lower than that in user code, the size of kernel-PHT is kept smaller than that of user-PHT. The limitation of second design is that for the same hardware budget, it can have lower-sized PHTs than that in the baseline or first design. They show that their technique can be used with several BPs and provides large improvement in performance for only small implementation overhead.

\subsection{Using conflict-reducing indexing functions}\label{sec:conflictreducing} 
Ma et al. \cite{ma2006using} evaluate several indexing functions for reducing aliasing in BP tables, which are shown in Table \ref{tab:IndexingFunction}. The BP table has $2^k$ entries. Note that prime-modulo mapping does not utilize all the entries. The a-prime-modulo mapping uses all the table entries although some of the initial entries are used more frequently than the remaining entries. To offset the latency of this mapping, prime-modulo results can be stored in BTB by extending it. 
Further, they evaluate use of different indexing function in different BP banks, similar to skewed-BP design \cite{michaud1997trading}. They evaluate these indexing functions with gshare and perceptron BPs and also specify exact instantiations of these mapping functions for these BPs.

 \begin{figure} [htbp]
\centering
\includegraphics[scale=0.40]{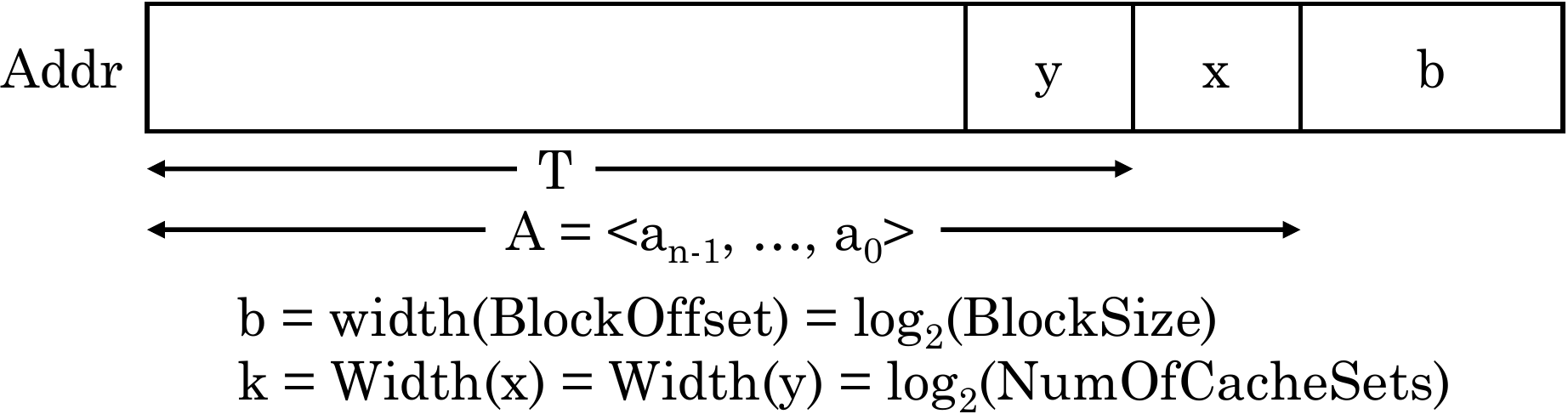}
\caption{ \revised{Address subdivision \cite{ma2006using} and meaning of symbols used in Table \ref{tab:IndexingFunction}} }\label{fig:IndexingFunction}
\end{figure}

\begin{table}[htbp]
  \centering
  \caption{\revised{Mappings and their indexing functions \cite{ma2006using}. The symbols used are defined in Figure \ref{fig:IndexingFunction}.}}
    \begin{tabular}{|p{3.0cm}|p{10.0cm}|}
    \hline
    \multicolumn{1}{|c}{Mapping name} & \multicolumn{1}{|c}{Indexing function} \\\hline
    Modulo & $index = x = A\; mod\; 2^k$ \\
    \hline
    Bitwise-XOR & $index = x \oplus y$ \\
    \hline
    Irreducible polynomial & Binary representation of index is $R(x) = A(x)\; mod\; P(x)$ and is computed as $R(x) = a_{n-1}R_{n-1}(x) + \ldots + a_{1}R_{1}(x) + a_{0}R_{0}(x)$. Here $R_i(x) = x^I\; mod\; P(x)$ can be precomputed after selection of the irreducible polynomial P(x). \\
    \hline
    Prime-modulo  & $index  = A\; mod\; p$. $p$ is a prime closest but smaller than $2^k$ \\
    \hline
    A-prime-modulo & $index = A\; mod\; p\; mod\; 2^k$, where $p$ is a prime number which is nearest to but bigger than the table entry count. \\
    \hline
    Prime-displacement & $index = (T*p+x)\; mod\; 2^k$, where $p$ is a prime number (e.g., $17$) \\
    \hline
    \end{tabular}%
  \label{tab:IndexingFunction}%
\end{table}%

\textit{Results:} The modulo mapping function is the baseline function. For gshare BP, they find that XOR indexing generally improves accuracy, however, irreducible polynomial function increases mispredictions and the prime-modulo function does not consistently provide higher accuracy than simpler mapping functions. The prime-displacement mapping performs better than XOR-mapping for large BP tables but shows lower effectiveness for smaller tables. The workloads having largest branch working set benefit the most from indexing functions.  

For perceptron BP, all indexing functions reduce aliasing significantly, and their improvement is higher for small tables than for large tables due to higher aliasing in small tables. Different mapping functions perform the best at different table sizes. Specifically, using XOR mapping for large tables and prime-displacement mapping for small tables keeps the complexity low while achieving high accuracy. Even when intelligent indexing functions are used, significant amount of aliasing still remains which can be removed by use of tags. Finally, use of multiple indexing functions is also found to be effective in reducing aliasing.

\subsection{Handling multithreading and short threads}\label{sec:multithreading}
With increasing number of threads, the BP size requirements rise further. Hily et al. \cite{hily1996branch} find that in multithreaded workloads,  using a 12-entry RAS with each hardware thread context increases the accuracy of branch prediction significantly and further increasing the number of entries does not provide additional gain. With multiprogrammed workloads (i.e., different applications in a workload have no data-sharing), scaling the size of PHT/BTB tables with number of threads removes all aliasing. For parallel workloads with data-sharing, some BPs (viz., gshare and gselect) benefit slightly due to sharing effect, whereas other BPs (e.g., 2bc) lose accuracy with increasing number of threads. For both types of workloads, on reducing the BTB sizes, conflicts in BTB increase the number of mispredictions.

Choi et al. \cite{choi2008accurate} note that BPs working on large control flow history fail to work well for short threads since GHR does not store thread-specific history information. They propose techniques which predict or re-create the expected GHR, using present or historical information. For example, the child thread inherits the GHR of the parent thread. Also, other techniques proposed by them only provide a consistent starting point for the BP whenever the same thread begins. For example, a unique initial GHR state can be provided to each thread by creating the GHR from the PC of the first instruction of the speculative thread. They show that their technique improves performance by reducing mispredictions for the short threads.

\subsection{Utilizing information from wrong-path execution}\label{sec:recyclewaste}

\begin{figure} [htbp]
\centering
\includegraphics[scale=0.40]{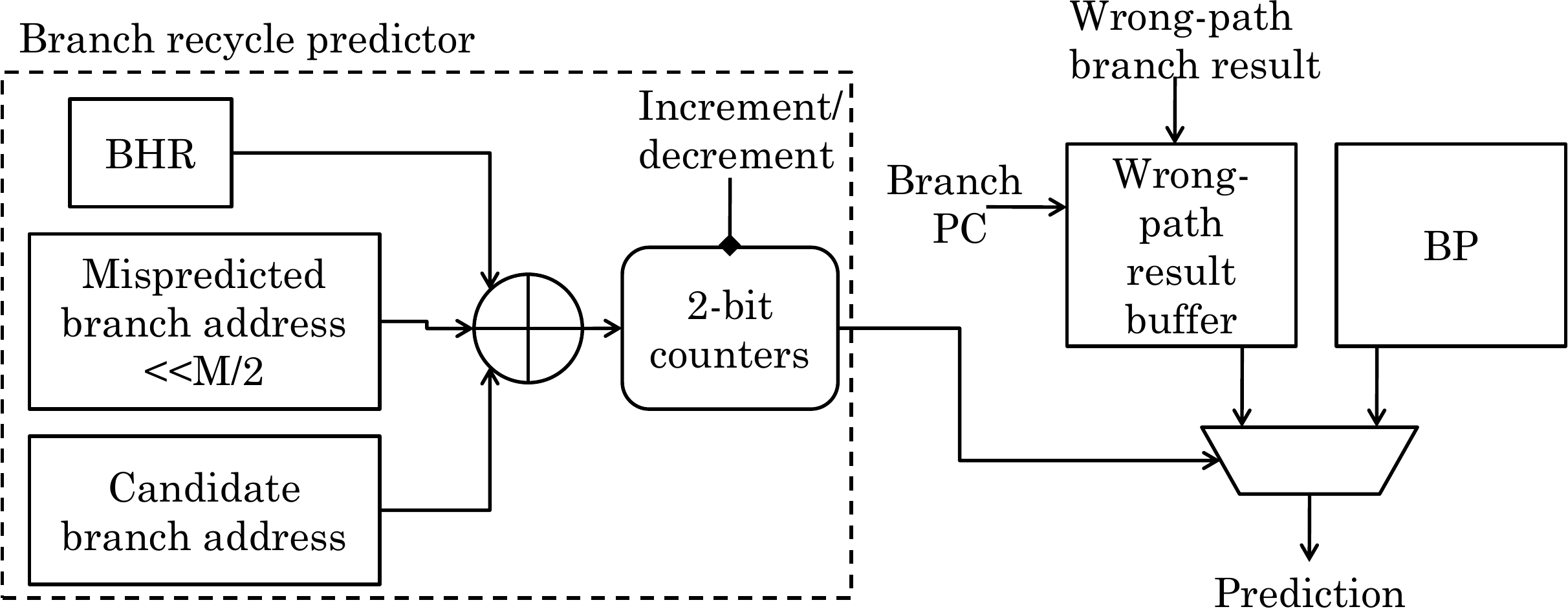}
\caption{The technique for leveraging information on wrong-path \cite{akkary2003recycling}}\label{fig:recyclewaste}
\end{figure}

Akkary et al. \cite{akkary2003recycling} note that in case of control independence, branch results on the wrong path match those on the correct path. However, current BPs do not exploit this information to improve accuracy since they study branch correlation on correct-path only. Their technique finds presence of correlation between the execution of a branch on correct and wrong paths using an algorithm similar to gshare BP \cite{mcfarling1993combining}.   In level-1, outcomes of last $M$ branches seen on correct path are stored  in a BHR. The address of mispredicted branch is shifted left by $M/2$ bits, as shown in Figure \ref{fig:recyclewaste}. Then, BHR, $M$ bits from the branch on correct-path to be predicted and $M$ bits from shifted mispredicted branch address are XORed to obtain index into a table of 2-bit saturating counters. If the counter-value exceeds 1, correlation of a branch with its wrong-path instance is assumed to exist and hence, wrong-path result is finally selected instead of the  prediction of the BP.

As for updating the counter, when results on correct and wrong paths match and BP mispredicts, the counter is increased; but if they mismatch and BP predicts correctly, the counter is decreased; otherwise, no change is made to the counters. Branch results on wrong path are stored in a tagged-buffer.   The branch outcome in this buffer is used at most once and then evicted, thus, it needs to accommodate only the branches in the instruction window. Only outcomes of most-recent wrong-path are maintained since remembering outcomes of multiple wrong-paths provides only marginal improvement. Their technique provides large reduction in mispredictions and improvement in performance. The limitation of their technique is that it cannot improve accuracy for branches that have different outcomes on correct and wrong paths.

\section{Neural BPs and implementation techniques} \label{sec:neuralBPs}
As for neural BPs, researchers have proposed BPs using perceptron \cite{jimenez2001dynamic}, back-propagation \cite{egan2003two,jimenez2002neural,jimenez2001perceptron}, learning vector quantization \cite{egan2003two} and other neural BPs \cite{jimenez2002neural}. We now discuss several neural BP designs.
 
\subsection{Neural BPs}\label{sec:neuralsummary}

\begin{figure} [htbp]
\centering
\includegraphics[scale=0.40]{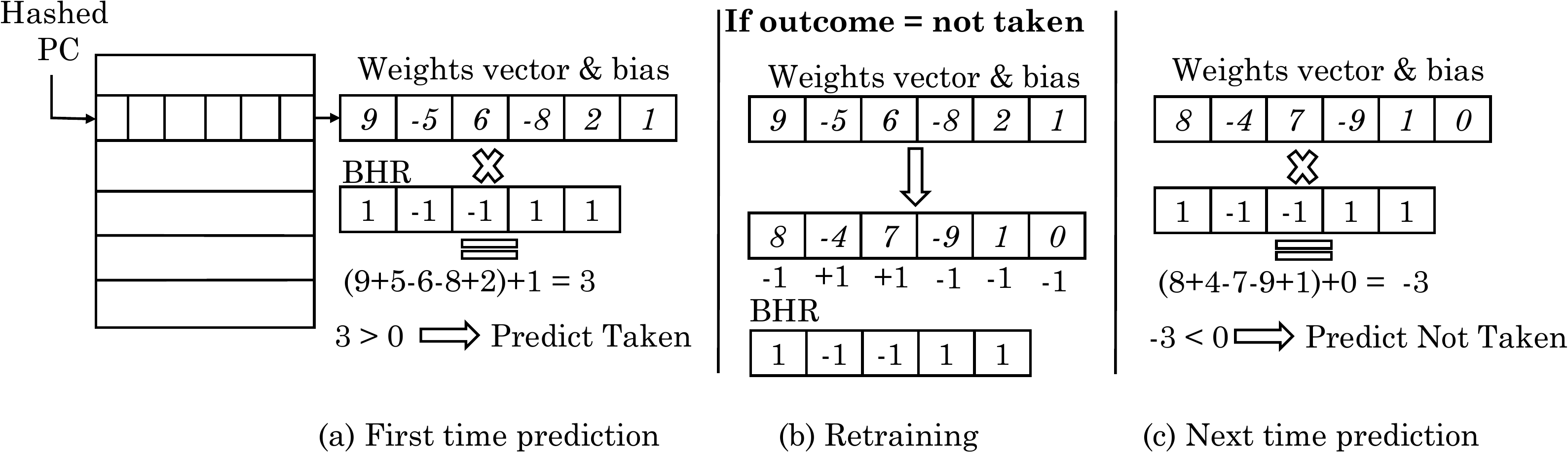}
\caption{Prediction and training process of perceptron predictor \cite{jimenez2001dynamic}  }\label{fig:perceptron}
\end{figure}

Jimenez et al. \cite{jimenez2001dynamic} propose a perceptron-based BP which uses one-layer perceptron, a simple NN, to represent each branch. A perceptron is a vector that stores multiple weights ($w$) showing the degree of correlation between different branches. The outcomes of previous branches (1 = taken, -1 = not-taken) form the input to perceptrons. Input $x_0$, being always 1, provides the bias input. The dot-product of weights and inputs provides the output $Z$, based on the following formula: 
$Z = w_0 + \Sigma w_i x_i $.
A positive or negative  value of $Z$ means taken or non-taken prediction, respectively.  When the actual branch outcome is available, the weights of elements agreeing/disagreeing with the outcome are incremented/decremented, respectively. Figure \ref{fig:perceptron} explains the working of perceptron BP with an example.

The size of the two-level BPs grows exponentially with history length since they index PHT with branch history. By comparison, size of perceptron BP increases only linearly with history lengths \cite{jimenez2002neural}. Hence, their BP can account for much longer histories, which is useful since highly correlated branches may be separated by long distances.  The weights remain close to 1, -1 or 0 in case of positive, negative and no correlation, respectively. Thus, the weights give insight into degree of correlation between branches. Also, the output of their BP shows confidence level of predictions since the difference between the output and zero shows certainty of taking a branch. 

\begin{figure} [htbp]
\centering
\includegraphics[scale=0.40]{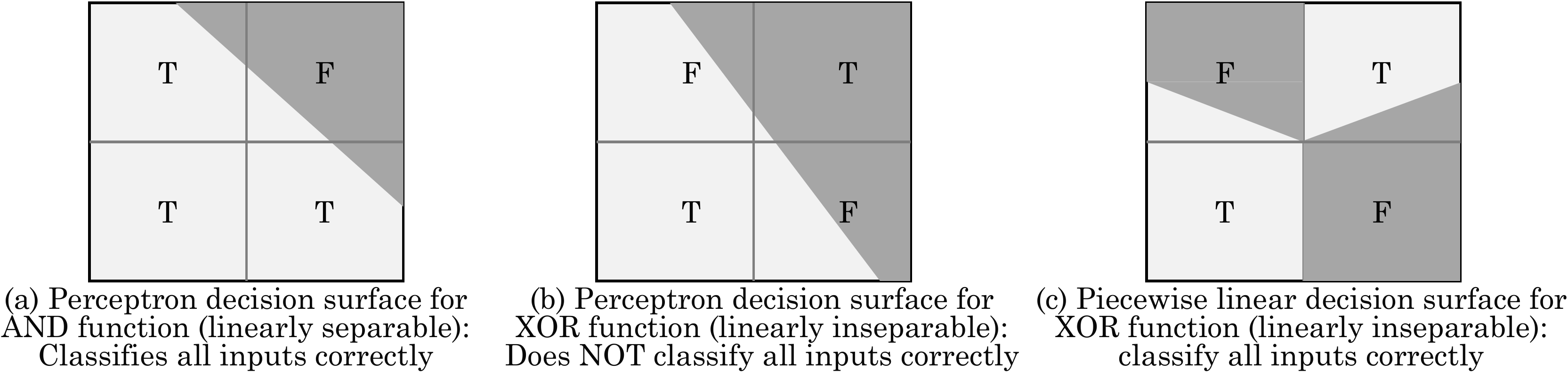}
\caption{\revised{An example of decision surfaces for AND and XOR functions for perceptron and piecewise-linear BPs \cite{jimenez2005piecewise}. X and Y axes represent two input variables. Negative/positive sides show false/true, respectively. The output is shown as T/F (true/false).} }\label{fig:separableinseparable}
\end{figure}

Their BP works well for applications with many linearly separable branches. A Boolean function over variables $x_{1..n}$ is linearly separable if values of $n+1$ weights  can be found that such that all the false instances can be separated from all the true instances using a hyperplane. \revised{For example, AND function is linearly separable whereas XOR function is linearly inseparable. This is illustrated in Figure \ref{fig:separableinseparable}(a)-(b). The perceptron BP works well for AND function, but not for XOR function.} Thus, for linearly inseparable function, perceptron BP has lower accuracy, whereas previous predictors can learn any Boolean function if sufficient time is given to them. Hence, for general applications, their BP is more useful as a component of a hybrid predictor instead of as a solo predictor. Another limitation of their BP is its high latency.

\begin{figure} [htbp]
\centering
\includegraphics[scale=0.40]{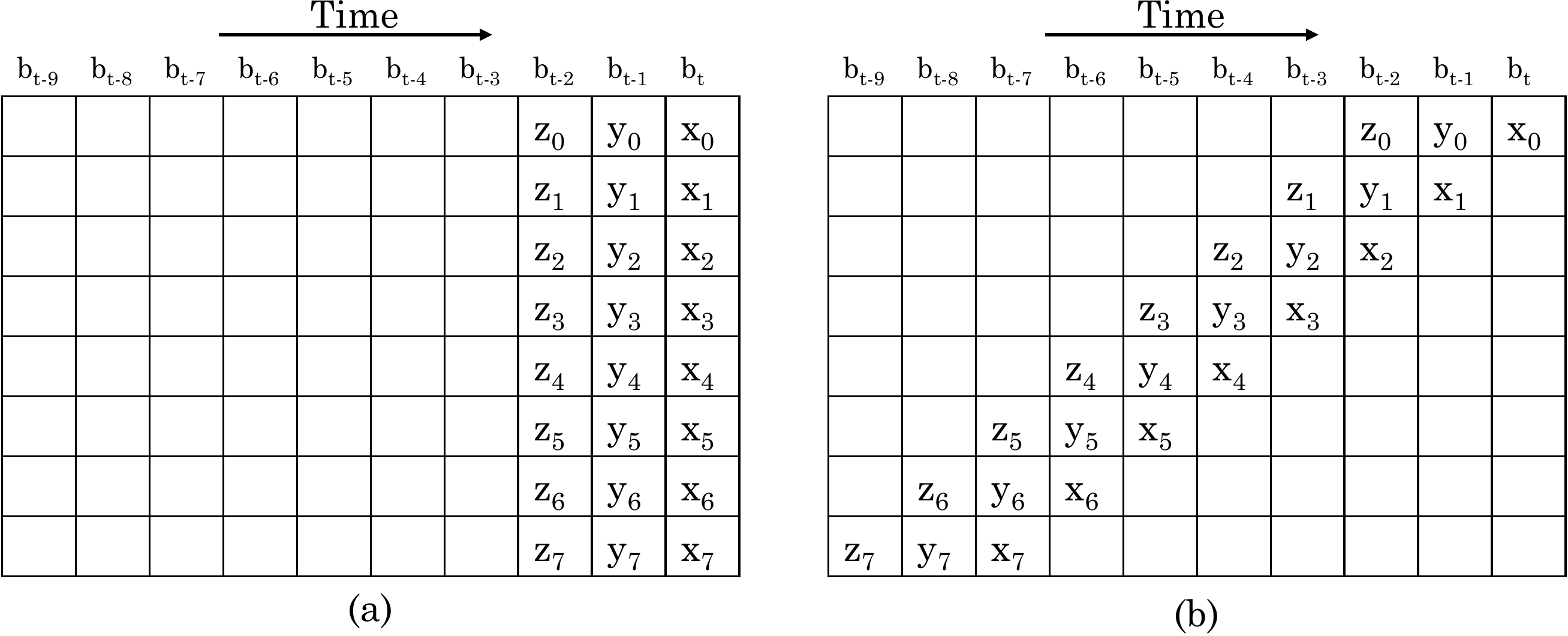}
\caption{Weights used for predicting branch $b_t$ with (a) the perceptron BP \cite{jimenez2001dynamic}  and (b) the path-based neural BP \cite{jimenez2003fast}. The history length is 7 and the vertical columns show the weights vectors. }\label{fig:aheadpipelining}
\end{figure}

Jimenez \cite{jimenez2003fast} presents a path-based neural BP, which uses `ahead pipelining' approach to reduce latency. Their BP chooses its weight vector depending on the path that leads to a branch, instead of only the branch address; and this helps in achieving higher accuracy. Similar to perceptron BP, their path-based BP maintains a matrix of weight vectors. For predicting a branch,  a weight vector is read and only its bias weight needs to be added to a running sum for making a prediction, as shown in Figure \ref{fig:aheadpipelining}. This running sum is updated for multiple previous branches. Thus, their ahead pipelining approach staggers the computation over time and hence, the 
computation can start even before a prediction is made. The prediction can complete in  multiple cycles which removes the requirement of BP overriding. A limitation of their approach is that it does not use PC of a branch to choose its weights which can reduce accuracy. Also, ahead pipelining works by starting the prediction process with stale/partial information and continually mixing latest information in the prediction process, however, only limited new information can be used in the ahead pipelining approach \cite{tarjan2005merging}. Further, branch misprediction requires rolling back a large amount of processor state to a checkpoint.

Jim{\'e}nez \cite{jimenez2005piecewise} presents a piecewise-linear BP which is a generalization of perceptron and path-based BPs   \cite{jimenez2003fast,jimenez2001dynamic}. For every branch B, their BP tracks components of all paths leading to B. It tracks likelihood of a branch at a certain position in the history to agree with the result of B, thus, their BP correlates each element of each  path with the result of B. A prediction is made by aggregating correlations of every component of the current path. Overall, their BP generates multiple linear functions, one for each path to the current branch for separating predicted not-taken from predicted taken branches. \revised{Hence, their BP is effective for linearly inseparable branches, as shown in Figure \ref{fig:separableinseparable}(c).} Assuming no constraints of storage budget or latency, their BP achieves lower misprediction rate than other BPs. They further propose a practical implementation of their BP which uses ahead-pipelining and small values of history length. Further, branch address and address of branch in the path history are stored as modulo $N$ and $M$, respectively. Perceptron-based BP \cite{jimenez2001dynamic} utilizes a single linear function for a branch ($M=1$) and path-based neural BP \cite{jimenez2003fast} utilizes a single global piecewise-linear function for predicting all the branches ($N=1$), and thus, they are special cases of the piecewise-linear BP \cite{jimenez2005piecewise}. They show that the practical version of their BP also outperforms other predictors.

Jimenez et al. \cite{jimenez2002neural} note that unlike (single-layer) perceptron based BP, multi-layer perceptron with back-propagation based BP  can learn linearly inseparable functions and hence, it is expected to provide higher `asymptotic' accuracy than perceptron BP. However, in practice, back-propagation neural BP provides lower accuracy and performance since it  takes much longer time for both training and prediction, which makes back-propagation based BP infeasible and ineffective \cite{jimenez2001perceptron}.
  
Tarjan et al. \cite{tarjan2005merging} propose a hashed perceptron BP which uses the ideas from gshare, path-based and perceptron BPs. They note that perceptron BPs use each weight to measure the branch correlation which leads to linear increase in the number of tables and adders with the amount of history used, and requires the counters for each weight to be large enough so that a weight can override several other weights.
They propose a BP which assigns multiple  branches to the same weight by XORing a segment of the global branch history with the PC. Instead of doing a single match with the global branch/path history, they divide it into multiple segments for performing multiple partial matches. Further, instead of using path history, they use only the branch PC. With hashed indexing, every table works as a small gshare BP and thus, their BP can also predict linearly inseparable branches mapped to the same weight. Also, their BP uses smaller number of  weights for the same history length compared to a path-based BP, and this reduces the chances that multiple weights with no correlation overturn a single weight with strong correlation. Compared to the global or path-based BP, their BP has a shorter pipeline which reduces the amount of state to be checkpointed.  To reduce BP latency, they use ahead pipelining. Compared to path-based and global neural BP, their BP improves accuracy significantly, while reducing the number of adders.

\begin{figure} [htbp]
\centering
\includegraphics[scale=0.40]{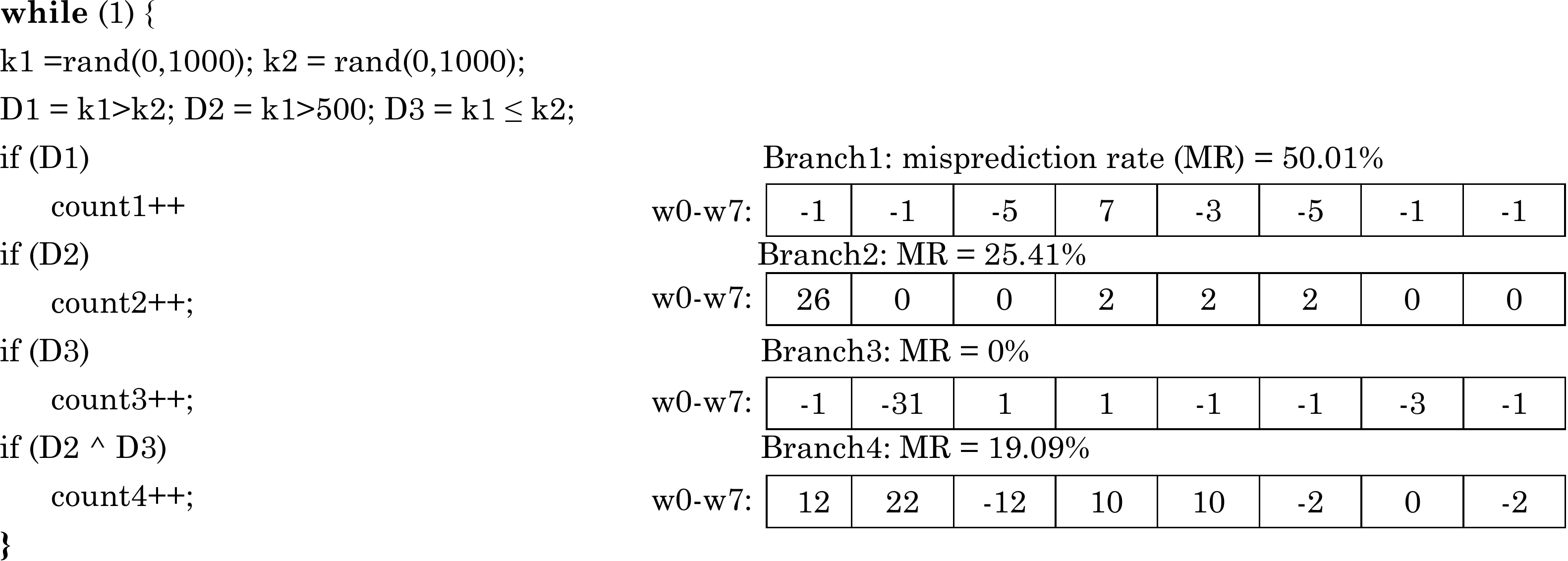}
\caption{ A program to show that perceptron predictor weights provide quantitative measure of branch correlation \cite{gao2005adaptive} }\label{fig:perceptronweights}
\end{figure}

Gao et al. \cite{gao2005adaptive}  note that in perceptron-based BPs, the value of perceptron weights shows the correlation strength. For example, Figure \ref{fig:perceptronweights} shows four branches, where Branch1 is decided by two random variables, Branch2 is correlated with Branch1 since they share the same random variable ({\tt k1}), Branch3 is merely inverse of Branch1 and Branch4 relates to both Branch2 and Branch3 since it uses an XOR function. A perceptron BP is used where  one perceptron with 8-bit GHR is used for each branch. The perceptron weights (w1-w8) and misprediction rates for 100M  instruction simulation  is shown on the right side of Figure \ref{fig:perceptronweights}. Evidently, since Branch1 has no correlation with previous branches, its weights have small random values. Branch2 has large w1 (strong correlation with GHR[0]) due to its correlation with Branch1 and small random w2-w8 values. Branch3 has a single large weight w2 (strong correlation with GHR[1]) since the outcome of Branch3 can be accurately found by knowing that of Branch1. Branch4 has larger weights since it correlates in nonlinear manner with the previous branches.

Further, by reassembling the input to BP, its accuracy can be improved, e.g., if only two recent branches show the most correlation out of $N$, perceptron size can be reduced from $N$ to 2, which improves accuracy due to avoidance of noise. Also, weak correlation can be replaced by stronger ones, e.g., redundant history can be used which helps in handling linearly inseparable branches. They use static profiling for finding a suitable  input vector for each workload type   and this is selected at runtime based on the application type. Further, to account for change in inputs and phase behavior, the correlation strength is periodically tested at runtime and the weakly-correlated inputs are substituted by the strongly-correlated inputs. Their technique improves BP accuracy significantly.  
 
\begin{figure} [htbp]
\centering
\includegraphics[scale=0.40]{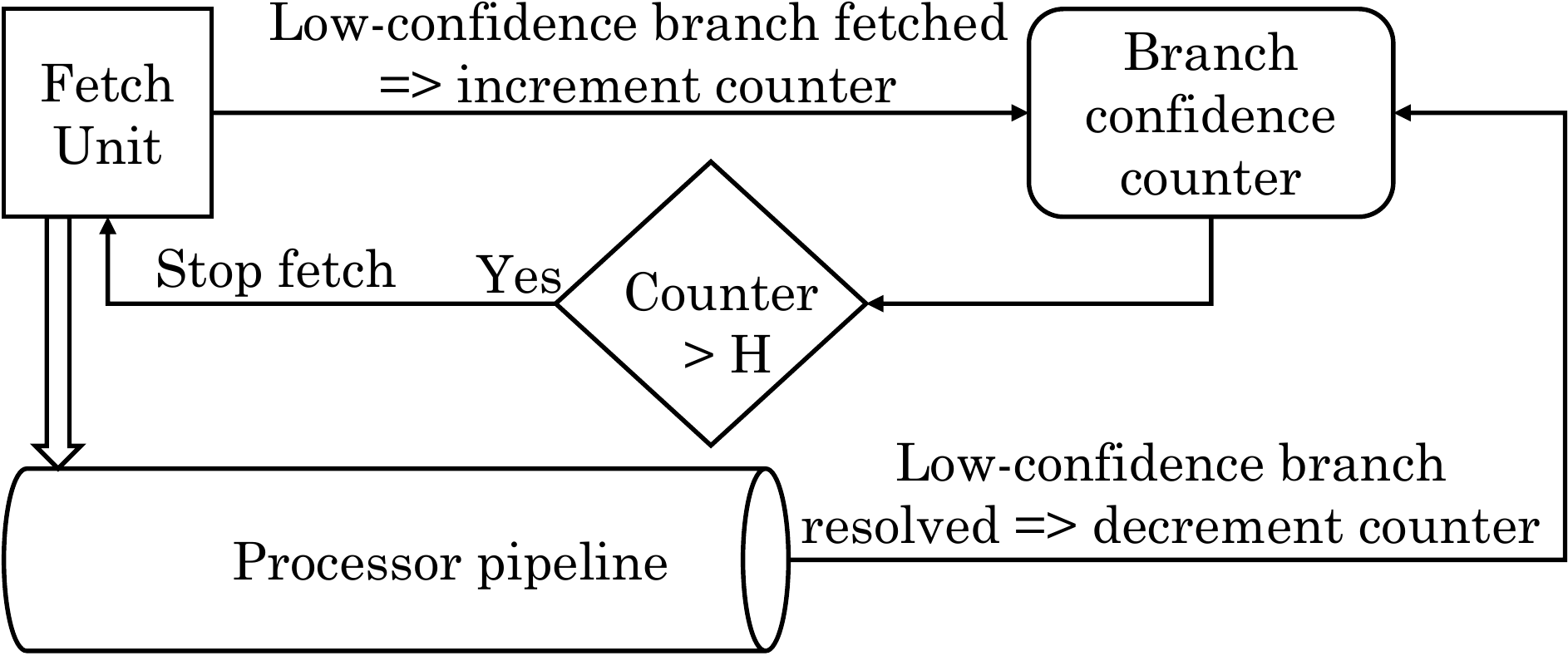}
\caption{ Pipeline gating approach (H shows a threshold)  }\label{fig:pipelinegating}
\end{figure}

Akkary et al. \cite{akkary2004perceptron} propose  a perceptron-based branch confidence estimator (CE). The input to a perceptron is global branch history where T/NT branches are recorded as 1/-1, respectively. Training of confidence estimator happens at retirement (i.e., non-speculatively). If $p$ (=1/-1 for incorrectly/correctly predicted branch) and $c$ (=1/-1 for branches assigned low/high confidence) have different signs, then weights are updated as $w[i]$ $+$$=$ $ p*x[i]$ for all $i$.  Training their BP using right/wrong prediction provides higher accuracy than using taken/not-taken outcome \cite{jimenez2001dynamic}. Also, their CE provides reasonable coverage of mispredicted branches. Further, they divide the non-binary output of their CE in two regions: strongly and weakly low confident and then apply branch reversal and pipeline gating (respectively) to them. This allows improving BP and  pipeline gating using a single hardware. Note that pipeline gating approach, shown in Figure \ref{fig:pipelinegating}, stops fetching instructions when multiple low-confidence branches are fetched. This  prevents misspeculated instructions from entering the pipeline and thus, saves energy.

\subsection{Optimizing Neural BPs}
Despite their higher accuracy, neural BPs also have high overheads. For example, to predict each branch, perceptron BP may require computing a dot product with tens or hundreds of values \cite{amant2008low,jimenez2002neural}, which increases its latency/energy/area and renders it infeasible for implementation in real processors. Path-based neural BP \cite{jimenez2003fast} uses ahead pipelining to offset latency issue, however, it replaces more power/area efficient carry-save adders of original perceptron BP with several carry-completing adders, leading to higher complexity. The piecewise-linear neural BP provides even higher accuracy, but increases the checkpoint/recovery overhead and number of adders significantly. We now discuss works that seek to reduce the implementation overhead of neural BPs.

\begin{figure} [htbp]
\centering
\includegraphics[scale=0.40]{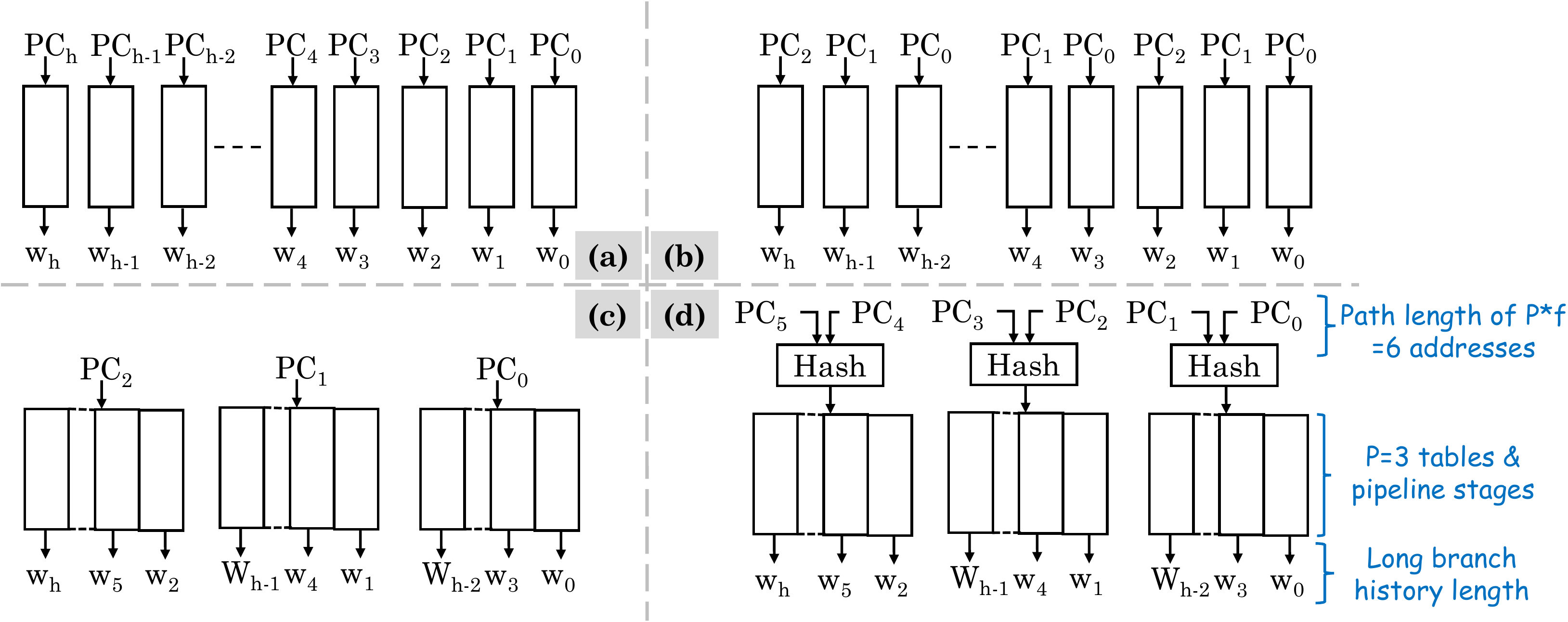}
\caption{(a) Use of $h$ ($=P$) tables in original path-based neural BP (b) logical design and (c) physical design of path-based neural BP with modulo path-history ($P=3$) \cite{jimenez2006controlling} (d) design of path-based neural BP with folded modulo path-history (path-length, history-length and table-counts are all different) \cite{jimenez2006controlling} }\label{fig:ModuloFoldedPath}
\end{figure}

Jimenez et al. \cite{jimenez2006controlling,loh2005reducing} note that branch history length needs to be high for achieving high accuracy, however, keeping the path history length of path-based neural BP same as the history length increases its storage overhead significantly (refer Figure \ref{fig:ModuloFoldedPath}(a)). To resolve this tradeoff, they propose decoupling path and branch history lengths by using ``modulo path-history''. They use only most recent $P$ ($<h$) branch addresses in path history which reduces the number of pipeline stages and tables to $P$ and also reduces the amount of state to be checkpointed. For finding the index for $w_i$, $PC_i\; mod\; P$ is used in place of $PC_i$. Figure \ref{fig:ModuloFoldedPath}(b) shows this design assuming $P=3$. Every $P^{th}$ weight is indexed using the same branch address and this allows interleaving the weights to reduce the number of tables required to $P$, as shown in  Figure \ref{fig:ModuloFoldedPath}(c).

Since reduction in path-history length prohibits exploiting correlation with branches more than $P$ address away, they propose a ``folded modulo path'' design for decoupling path-history length from the predictor pipeline depth. This design uses path history length of $P\times f$ while still using $P$ tables ($f$ = folding factor). While traditional PBNP indexes each table with one branch address, their design hashes $f$ addresses to obtain the index. Figure \ref{fig:ModuloFoldedPath}(d) illustrates this design with $P=3$ and $f=2$ for keeping a path-length of 6 with only 3 tables. Their history-folding approach can also be used with other BPs. 
 
Loh et al. \cite{loh2005reducing} note that for  biased branches, tracking $h$ weights and doing their dot-product is unnecessary. Their technique considers a branch whose bias weight ($w_0$) has reached the minimum or maximum value as a biased branch. For such branches, only bias weight is used for making prediction and no dot-product is performed. In case of correct prediction, no update is performed which saves energy and reduces aliasing for remaining branches. Their proposed BP reduces energy consumption and area with negligible reduction in accuracy.

\subsection{Analog implementations of neural BPs}
To lower the implementation overhead of neural BPs, some works use the properties of analog domain  \cite{amant2008low,wang2013practical,saadeldeen2013memristors} or memristors   \cite{saadeldeen2013memristors,wang2013practical}. While these implementations  reduce energy consumption, they require conversion between digital and analog domains. We now discuss these techniques.

Amant et al. \cite{amant2008low} present a neural BP which employs analog circuitry for implementing power-hungry parts of BP. Digital to analog converters change digital weights into analog currents and current summation is used to combine them. By steering positive/negative weights to different wires and finding the wire with higher current provides a dot-product result as a single-bit, naturally working as an analog-to-digital converter. Their BP uses two features for improving accuracy. First, given the fast computation speed of analog circuitry, their BP obviates the need of using ahead pipelining and thus, can use up-to-date PCs to generate the weights. Second, since recent weights generally correlate more strongly with the branch result than older weights, their BP scales the weights based on this correlation to improve accuracy. In analog domain, this only requires changing transistor sizes whereas in digital domain, this would have required performing several energy-hungry multiply operations. Their BP is fast, highly energy efficient and nearly as accurate as L-TAGE \cite{seznec2007256} predictor.  Also, its accuracy is only slightly lower compared to an infeasible digital design.

Jimenez \cite{jimenez2011optimized} proposes several optimizations to analog neural predictor \cite{amant2008low} to exercise tradeoff between accuracy and implementation overhead.  First, instead of using only global history, they use both global and per-branch history for prediction and training, which improves accuracy. Second, since recent branches show higher correlation than older branches, their weights are also larger and hence, they represent weight matrix such that row-size changes with the column-index. Third, depending on correctness of a partial prediction, the corresponding coefficient is tuned. Fourth, the minimum value of perceptron outputs below which perceptron learning is issued, is changed dynamically. Next, to reduce aliasing between bias weights, a cache is maintained which stores partially tagged branch addresses and bias weight for conditional branches. Finally, they note that the accuracy of their BP becomes low when its output  is close to zero. For such cases, they use two other predictors, viz., gshare and PAg (if confidence of gshare is low). The proposed optimizations improve the accuracy significantly.

Wang et al. \cite{wang2013practical} present an analog neural BP designed with memristor for saving BP energy. They store perceptron weights in a 2D table of ``multi-level cell'' (MLC) of memristor device. Each MLC has two analog outputs and their relative difference shows the weight stored in the MLC. To negate the weight, only the roles of two weights needs to be exchanged using a history bit. This allows multiplying with 1/-1 in much less time than in digital domain. For making a prediction, a row of weights is indexed using PC. Also, the corresponding global history bit is read and if it is 1, one analog current signal of the cell is put on positive line and another on negative line and vice versa if the history bit is 0. All the current signals on the negative (positive) line  are automatically added using Kirchoff's current law which is much faster than addition in digital domain. If total current on  positive line is greater than that on negative line, branch prediction is `taken' and vice versa. Compared to a digital implementation, their  implementation improves accuracy due to both characteristics of analog computation and memristor which helps in resolving some irregular dependent branches. A limitation of using memristor is relative lack of commercial maturity of memristor devices   which may reflect in high noise, latency, process variation, etc.


\section{Techniques for reducing BP latency and energy}\label{sec:latencyenergy}
We now discuss techniques for reducing BP latency (Sections \ref{sec:latencyreduction}-\ref{sec:multiplebranches}) and storage/energy (Sections \ref{sec:bhrpartitioning}-\ref{sec:noredundant}). Since a decrease in BP accuracy only affects performance and not application correctness, BP energy can be aggressively optimized.  Table \ref{tab:classification2} summarizes the techniques for reducing latency and energy/storage overhead of BPs. Some works perform partial update of BP to improve accuracy and/or save energy. Some works update BP tables speculatively (and not at retirement stage) for  improving accuracy.



\begin{table}[htbp]
\centering
\caption{Techniques for reducing latency and storage overhead}
\label{tab:classification2}
\begin{tabular}{|p{6cm}|p{7.5cm}|}\hline
\multicolumn{2}{|c|}{Techniques for reducing latency}   \\\hline
Ahead pipelining &  \cite{seznec2006case,loh2006revisiting,jimenez2005piecewise,loh2005simple,jimenez2003fast,seznec2005analysis,tarjan2005merging,gao2007pmpm} \\\hline
Overriding BP &  \cite{jimenez2000impact,loh2006revisiting,manne1999branch} \\\hline
Caching &  \cite{jimenez2000impact} \\\hline
Cascading lookahead & \cite{jimenez2000impact} \\\hline
Multiple predictions in single cycle & \cite{yeh1993increasing,jimenez2003reconsidering,seznec1996multiple} \\\hline
Dividing large table and concurrently accessing sub-tables & \cite{skadron2000taxonomy} \\\hline
Accessing slow BPs only for hard-to-predict branches & \cite{kampe2002fab} \\\hline
\multicolumn{2}{|c|}{Techniques for reducing storage/energy overhead}   \\\hline
BP virtualization &  \cite{alotoom2010exact,sadooghi2012toward} \\\hline
Storing history in modulo-$N$ manner &  \cite{jimenez2005piecewise,jimenez2006controlling,loh2005reducing}  \\\hline

Power-gating based on temporal/spatial locality & \cite{baniasadi2002branch,hu2002applying,yang2006power,chaver2003branch} \\\hline
Avoiding access to BP & \revised{for biased branches  \cite{parikh2004power} and cache blocks with no control-flow instructions \cite{parikh2004power}} \\\hline

Partial update policies. Updating only & 
 
\revised{Selected sub-banks or predictor components \cite{lai2005improving,michaud1997trading,seznec1999dealiased,loh2005simple,lee1997bi,seznec2002design,eden1998yags},  if BP does not have sufficient information for accurate prediction  \cite{baniasadi2004sepas,seznec2011new}, on a misprediction \cite{heil1999improving}, on a misprediction or when confidence in prediction is low  \cite{seznec2005analysis} } \\\hline

Speculative update of BP & \cite{jimenez2003reconsidering,yeh1992alternative} \\\hline
Reducing BP size by intelligently managing global history &  \cite{ayoub2009filtering,xie2013energy,huang2015energy,schlais2016badgr} \\\hline
Banked-design of BP &   \cite{parikh2004power,hu2002applying} \\\hline
Reducing BP complexity  & \cite{loh2005reducing} \\\hline
Analog  designs & \cite{amant2008low,wang2013practical,saadeldeen2013memristors} \\\hline
Use of memristor & \cite{saadeldeen2013memristors,wang2013practical} \\\hline 
\end{tabular}
\end{table}

\subsection{BP Pipelining}\label{sec:latencyreduction}
\begin{figure} [htbp]
\centering
\includegraphics[scale=0.40]{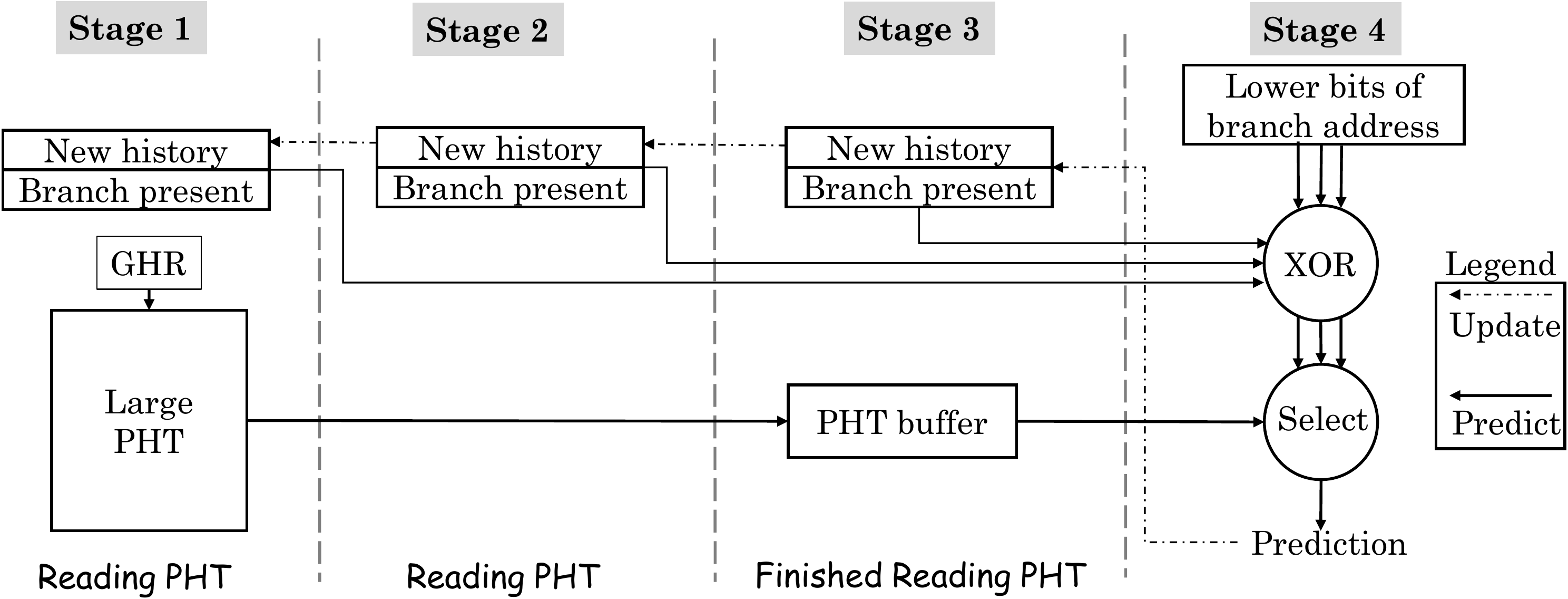}
\caption{Pipelined gshare implementation \cite{jimenez2003reconsidering}}\label{fig:pipelinedgshare}
\end{figure}

Jimenez  \cite{jimenez2003reconsidering} proposes pipelining the BP to bring its effective latency to one cycle and demonstrates his approach by pipelining  gshare  BP. The proposed design uses a 4-stage pipelined BP, 3 stages for reading PHT and 1 stage for making a prediction. To record recent speculative global history from the beginning of  PHT access, the first three stages use a `Branch present' latch  and a `New History Bit'. The `branch present' latch shows whether fetching and prediction of the branch was done in that cycle and the  `new history bit' shows the corresponding speculative global history bit shifted in from the subsequent stage of pipeline, as shown in Figure \ref{fig:pipelinedgshare}.

A branch fetched in cycle $T$ is predicted as follows: (a) In cycle T-3 at stage 1, GHR is used to start fetching eight two-bit counters into eight-entry PHT buffer. If a branch was fetched in stage 2, its `new history' bit is shifted from stage 2 and `branch present' bit is set, else `branch present' bit is reset. (b) In cycle T-2 at stage 2 and in cycle T-3 at stage 3, history and branch present bits are sent to previous stage in first half-cycle and these bits of next stage are shifted in the second half-cycle. (c) Additionally, data read from PHT is placed in PHT buffer at the end of cycle T-1. (d) For a branch fetched in cycle T, lower bits of its address are XORed with lower bits of GHR, shifted left and merged with at most 3 `new history' bits from earlier stages. From this, the PHT buffer is indexed and the entry obtained gives the final prediction. By using  a larger buffer size, their BP can make multiple predictions in a cycle. They show that despite having lower accuracy  compared to perceptron, 2bc+gskew and hybrid BPs, their BP provides slightly better performance due to its 1-cycle latency.

\subsection{BP caching and overriding}

\begin{figure} [htbp]
\centering
\includegraphics[scale=0.40]{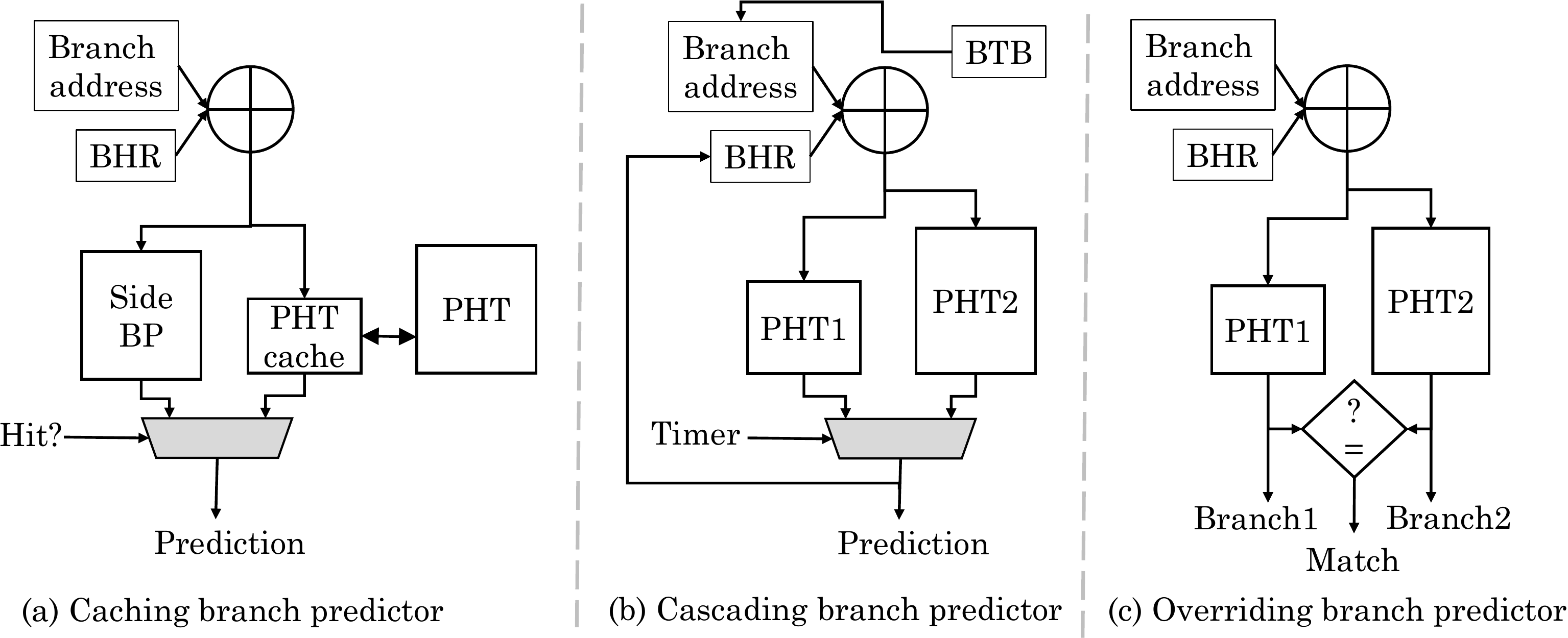}
\caption{Strategies for mitigating BP latency  \cite{jimenez2000impact} }\label{fig:LatencyMitigation3Strategies}
\end{figure}

Jimenez et al. \cite{jimenez2000impact} present three strategies for improving BP accuracy without increasing its latency. The common key idea of these strategies is to use a small predictor for making fast prediction and a large predictor for boosting accuracy. They illustrate their strategies for gshare BP.  In first strategy shown in Figure \ref{fig:LatencyMitigation3Strategies}(a), BP table entries are cached in a small table that can be accessed in one cycle. The output of XOR gate is sent to both PHT-cache and a side predictor. The number of entries in PHT is the number of possible combinations of addresses produced by XOR function. PHT-cache stores a subset of these entries. A hit in PHT-cache provides its prediction, whereas on a miss, the prediction from the side predictor is used.  The actual branch outcome updates both PHT and PHT-cache. The effectiveness of this strategy depends on side BP and ability of PHT-cache to capture branch locality.

 The second ``cascading lookahead prediction'' strategy works on the idea that if the distance between two branches is more than a cycle, then the multiple-cycle latency of BP can be hidden by predicting for the expected future branch.   The subsequent prediction is determined by predicted history   and last predicted branch target. The last prediction is appended to BHR and the predicted branch target is taken from BTB. This strategy uses tables with increasing size and latency, as shown in Figure \ref{fig:LatencyMitigation3Strategies}(b). Both tables (PHT1 and PHT2) are simultaneously accessed. If PHT2 provides prediction before arrival of next-branch, its prediction is used, otherwise, that of PHT1 is used. This design can be extended  to more than two levels of tables. The effectiveness of this design depends on inter-branch distance,  latency of PHT2 and BTB accuracy. 
 
 In the third ``overriding BP'' strategy, two predictions are provided by a fast table (PHT1) and a slow and more-accurate table (PHT2), respectively (refer Figure \ref{fig:LatencyMitigation3Strategies}(c)). Execution proceeds based on the prediction from PHT1, however, if the prediction from PHT2 obtained later differs from that of PHT1, then the prediction of PHT2 overrides that of PHT1 and execution on wrong path is discarded. Pipelining of BP ensures that a branch need not wait till PHT2 access is completed for the earlier branch.  \textit{Results:} overriding BP performs the best, whereas cascading BP performs reasonably well. The cached BP provides no improvement over the original BP since the tags required for the caching strategy incur more storage cost than the cache itself and hence, limit effectiveness and capacity of the cache.

\textbf{Challenges of BP overriding: } A limitation of BP overriding is that the speedup from it may not be commensurate with the implementation overheads since in case of disagreement between fast and slow BPs, high latency penalty  is incurred \cite{jimenez2002neural}. Further, BP overriding increases complexity of fetch and recovery engine significantly. The prediction of fast BP needs to be stored and later compared with that of slow BP and in case of disagreement between them, either all or selected (i.e., only those that depend on the prediction) instructions on wrong path need to be discarded. The former approach degrades performance and wastes energy due to squashing of large number of instructions, whereas latter approach requires tracking instructions dependent on the prediction \cite{santana2003latency}. 

One strategy to avoid the need of overriding is to increase the length of prediction unit, e.g., predicting instruction stream and instruction traces. This helps in keeping execution engine busy for multiple cycles with initial prediction, and during this time, the second prediction can be generated which allows hiding the latency of second prediction \cite{santana2003latency}.

\subsection{BP virtualization}\label{sec:bpvirtualization}
 
 Predictor virtualization approach seeks to   increase the effective capacity of predictors without using a single large and slow monolithic design. It records the full BP metadata in a virtual table stored in memory hierarchy (e.g., L2 cache) and brings only active entries in a smaller on-chip table. When the working set exceeds the on-chip table capacity, the data is spilled to and brought from the virtual table.

\begin{figure} [htbp]
\centering
\includegraphics[scale=0.40]{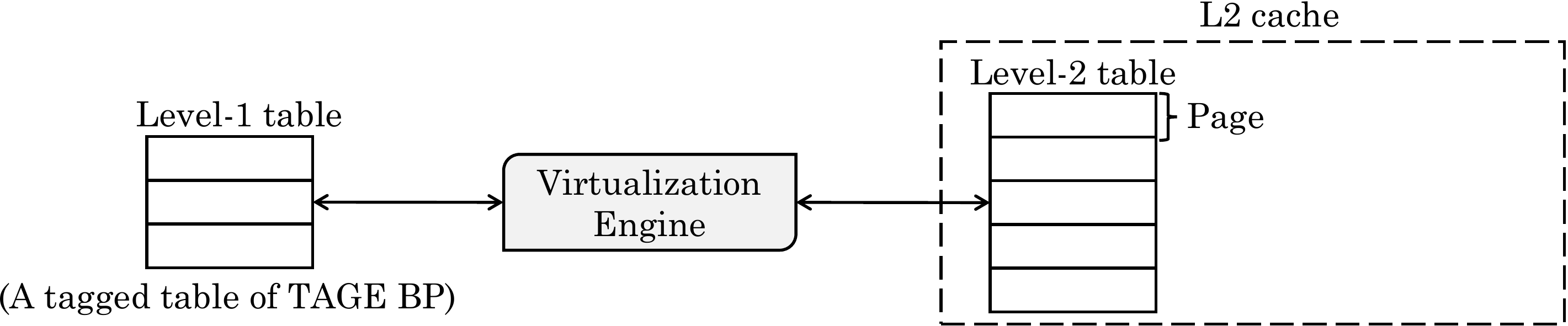}
\caption{\revised{BP virtualization approach   \cite{sadooghi2012toward} (figure not drawn to scale)}}\label{fig:BPvirtualization}
\end{figure}

Sadooghi et al. \cite{sadooghi2012toward} propose a technique for virtualizing BP. They apply their technique to TAGE multi-level BP. They augment one of the tagged predictor table with a second-level table which is stored in L2 cache. \revised{This is illustrated in Figure \ref{fig:BPvirtualization}.} To hide the latency of accessing second-level table, multiple correlated entries need to be fetched on each miss. Since higher effective capacity of 2-level table design already reduces aliasing, they lower the randomness of predictor access stream to increase its locality. For this, branch PC bits are directly used as higher portion of the index and only the lower portion of the index is randomized. Thus, predictor table is partitioned into multiple sub-tables (called `pages') such that each branch accesses a unique page, as shown in Figure \ref{fig:BPvirtualization}. Thus, the stream of pages, which is BP access stream, show temporal and spatial locality which is already present in instructions. The level-1 table caches recently-referenced pages of level-2 table. Based on their PCs, branches are assigned to unique pages, which has the disadvantage of increasing the contention. Their technique requires changing only the index hash function. The prediction mechanism needs to access only the first-level table. Swapping of pages between two levels is done by a separate module. Compared to the non-virtualized design, their virtualized BP design reduces storage requirement while maintaining accuracy.  \revised{The limitation of virtualization is that for applications with poor locality, a high number of swapping operations are introduced.}

\subsection{Predicting multiple branches in each cycle}\label{sec:multiplebranches}
\begin{figure} [htbp]
\centering
\includegraphics[scale=0.40]{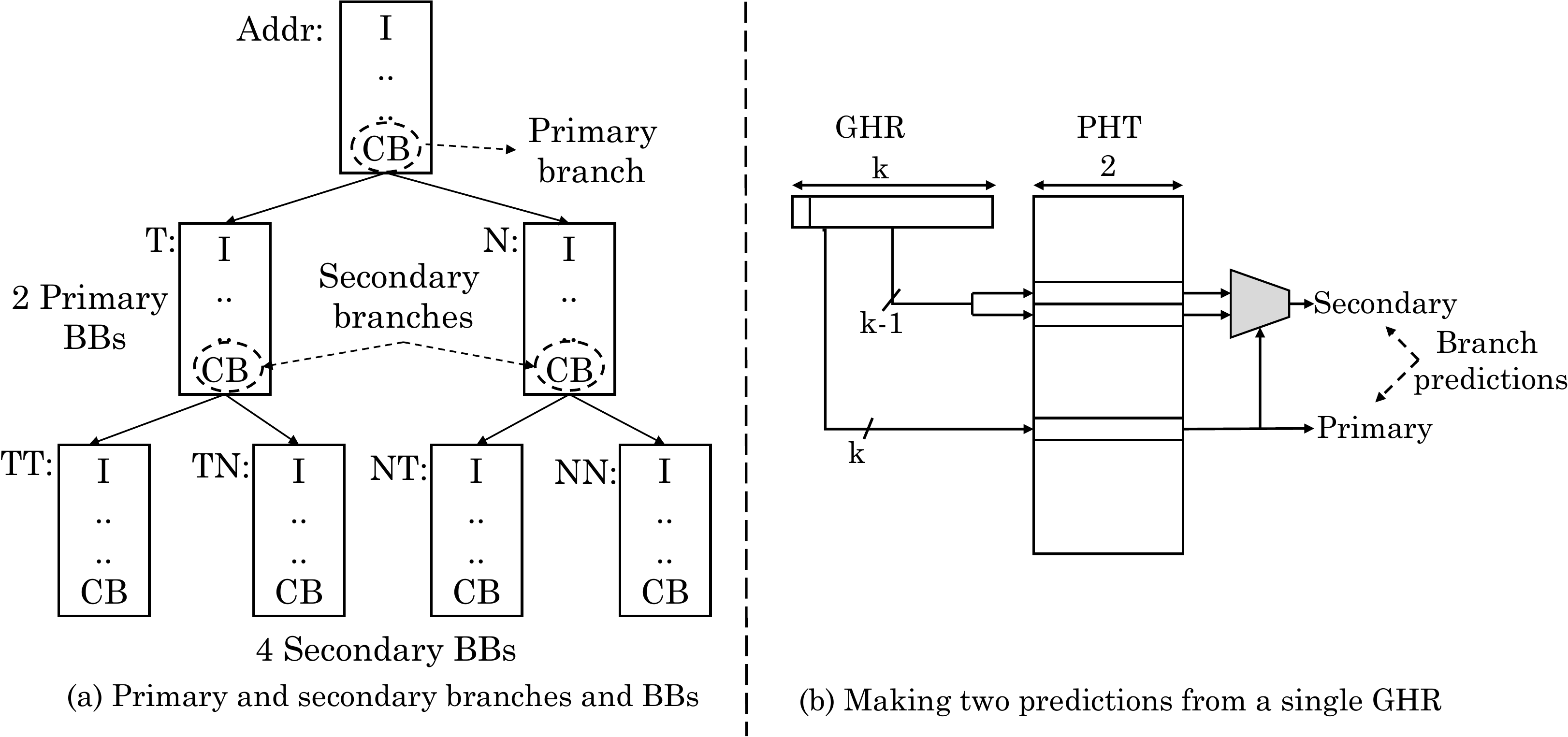}
\caption{(a) Illustration of primary and secondary branch (b) making multiple predictions each cycle (CB = conditional branch) \cite{yeh1993increasing} }\label{fig:MultiplePredictionEachCycle}
\end{figure}

Yeh et al. \cite{yeh1993increasing} note that being able to predict only one branch and fetch only one set of successive instructions from I-cache in each cycle limits fetch capability to one basic block per cycle. To address this challenge, they propose a BP which can predict  multiple branches and fetch multiple non-consecutive BBs in every cycle. For example, if every BB has 5 instructions and two branch paths can be correctly predicted in each cycle, then 10 instructions can be fetched in each cycle. For two branch predictions, they use the terms primary/secondary branches/BBs as shown in Figure \ref{fig:MultiplePredictionEachCycle}(a). Their BP is extended from two-level GAg BP \cite{yeh1993comparison} such that the prediction is extrapolated to subsequent branches also.  

For predicting the primary branch, the index for PHT is obtained from all $k$ bits in the BHR table. For the second branch, right-most $k-1$ bits are used to index two consecutive entries in PHT from which one is selected based on the primary branch prediction (refer Figure \ref{fig:MultiplePredictionEachCycle}(b)). Similarly, for third prediction, $k-2$ bits are used and four consecutive entries are obtained, of which one is finally selected using primary and secondary predictions. Thus, $k$-bit branch history information is used by all predictions.

 To record the branch addresses, they use a branch address cache (BAC) which is accessed using fetch address in parallel with I-cache. Number of fetch addresses used to access BAC is same as the number of BBs fetched. A hit in BAC implies that recently fetched instructions have a branch.  BAC provides starting addresses of BBs following the multiple predicted branches. For primary, secondary and third BB, there are 2, 4 and 8 fetch addresses, respectively. A miss in BAC implies that there is a large BB and hence, I-cache bandwidth is used entirely for fetching sequential instructions. To allow fetching multiple BBs, I-cache needs to provide low miss-rate and high bandwidth. Hence, they use an I-cache with multiple interleaved banks which has lower overhead than using multiple ports. Their technique provides large performance improvement, although with increasing number of BBs fetched, the hardware cost of their technique increases    exponentially.

\subsection{Reducing PHT size by BHR partitioning}\label{sec:bhrpartitioning}

\begin{figure} [htbp]
\centering
\includegraphics[scale=0.40]{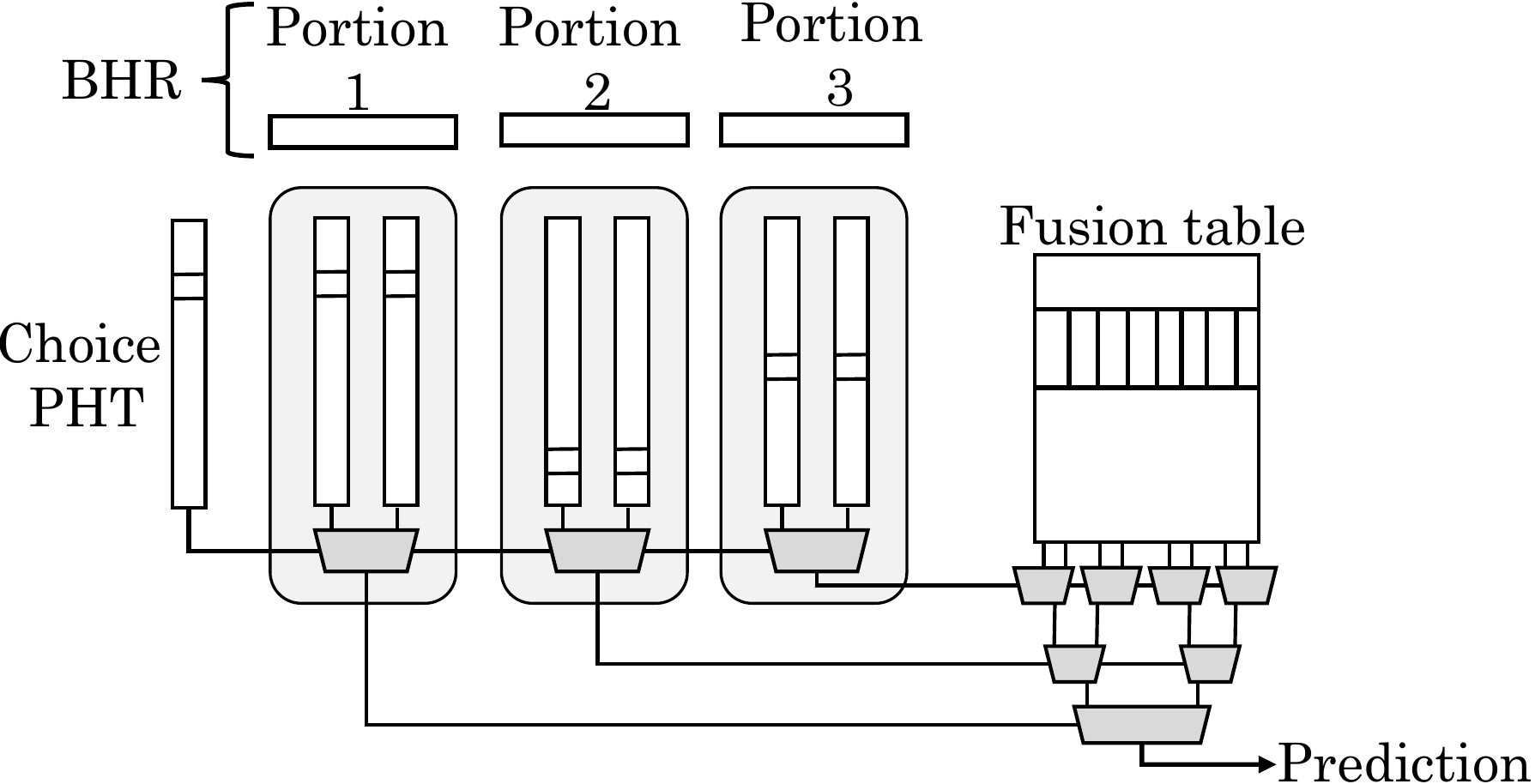}
\caption{Fusion-based hybrid BP \cite{loh2005simple} }\label{fig:fusionhybrid}
\end{figure}

Loh et al. \cite{loh2005simple} note that the number of entries required by PHT based predictor increases exponentially with history length, compared to only linear increase in the neural predictor \cite{jimenez2001dynamic}. They propose dividing the branch history register (BHR) into smaller portions and using a separate PHT to handle each portion as shown in the left side of Figure \ref{fig:fusionhybrid}. Each PHT uses bi-modal design to reduce harmful aliasing. Another table-based predictor fuses \cite{loh2002predicting} per-portion predictions to make the final prediction as shown in the right side of Figure \ref{fig:fusionhybrid}. To achieve low latency, they use ahead-pipelining \cite{jimenez2003fast} for bimodal predictors and prediction fusion scheme which allows achieving an effective latency of one-cycle. Although dividing the branch history loses correlation across different segments, its impact on accuracy remains small since strong correlations tend to remain clustered and many correlations are redundant due to which including only few correlations is sufficient. Their BP  provides better performance than GH perceptron BP \cite{jimenez2001dynamic} for all sizes and path-based neural BP \cite{jimenez2003fast} at 16KB or higher sizes.

\subsection{Leveraging branch locality}  \label{sec:locality}

Hu et al. \cite{hu2002applying} note that the spatial and temporal locality present in applications provides opportunities for saving BP leakage energy by deactivating (decaying) unused entries. They note that a BP with $K$ 2-bit entries is implemented as a square structure of $\sqrt{K}$ rows, each of which connects to $2\times \sqrt{K}$ bits. Hence, they perform  deactivation at the granularity of each row in the array instead of individual entries since deactivating at granularity of 2-bit entries would incur prohibitively high metadata overhead. Periodically, the predictor rows that were not accessed for the duration of one `decay interval'  are deactivated since they are unlikely to be reused soon. On access to a deactivated row, the row is activated and the default state of weakly `not-taken' is used.

As for the reason for presence of spatial locality, they note that since only few neighboring regions of a program code are expected to be active for a short time-duration, the PC of branch access is also expected to vary in a small range. Further, several taken branches of `if-else' statements perform short-jumps only. Code locality leads to spatial locality in BP rows, especially for the BPs indexed only by PC, such as bimodal BP \cite{lee1997bi}. Hence, consecutive conditional branches are highly likely to fall in the same row. Also, for applications with few static branches, accesses to bimodal BP show high temporal locality since each static branch in the bimodal BP accesses a single BP entry only. In BPs where PC is XORed with global branch history (e.g., gshare), one branch can access multiple BP table entries. Still, accesses to these BPs shows sufficient temporal locality, albeit lower than that in bimodal BP.

They further study a global+local hybrid BP similar to that in Alpha 21264 processor and note that its organization provides additional opportunities for energy saving. Here, the indices of selector and global-history predictor are always the same due to their similar designs. Hence, a BP row has the same state (active or inactive) in both these components. When only one of local or global component row is active, its prediction is used. Further, if this active row is in the global component, then the selector row will also be active and hence, if the selector's choice is the local component, then the local component is activated. But if the global (and hence, selector) row is inactive, neither of global and selector rows are activated. In essence, by utilizing the decision of the selector, unnecessary activations are avoided which increases leakage energy saving compared to a naive policy which always activates an inactive row. Decay intervals larger than 64K cycles are found to provide large leakage energy saving without substantially increasing mispredictions or performance loss. Their technique brings large reduction in BP leakage energy.

Banisadi et al. \cite{baniasadi2002branch} propose a technique to save energy in hybrid (e.g., gshare+bimodal)  BPs. They note that branch instructions show temporal locality,  such that more than 50\% branches appear within 8 branches and more than  80\% branches appear within 64 branches. Also, a branch generally uses the same sub-predictor for prediction, still hybrid BPs access both sub-predictors and the meta-predictor for each branch which wastes energy. Based on these observations, they use a FIFO (first-in first-out) buffer, where each buffer entry uses PC of a recently used branch as the tag and sub-predictors used by the two subsequent  branches in the dynamic execution stream. Assigning sub-predictor hints to a preceding branch helps in avoiding increase in prediction latency, as the hints are available at least one  cycle before the actual prediction.

The processor fetches at most two branches in each cycle for a total of 8 instructions. Then, it compares the last fetched branch with any buffer entry. On a match, sub-predictor hints
of the next two buffer entries are retrieved and assumed to be same for the next two branches. Based on this, the unused sub-predictor  and meta-predictor are both power-gated and the used predictor is directly accessed. No power-gating is performed (and hence, all sub-predictors are accessed) if (1) no match was not found in the buffer or (2) the buffer indicates that the branch was previously mispredicted and hence, confidence in the previously used sub-predictor is low. Their technique saves power for both small and large hybrid BPs. Also, using their technique leads to lower power consumption than using a single sub-predictor. However, in absence of temporal locality of branches, their technique can increase power consumption by choosing wrong sub-predictors.

Parikh et al. \cite{parikh2004power} present three techniques to save BP energy. First, they use banking to reduce access latency and dynamic energy since only a portion of BP needs to be kept active at a time. Second, they use a structure called ``prediction probe detector'' (PPD) which has same number of entries as the I-cache. PPD  stores pre-decode bits to detect whether the cache block has conditional branches which avoids access to BP and if cache block has no control-flow instructions, then, access to BTB itself is avoided. Thus, instead of BTB, only PPD needs to be accessed in each cycle which is much smaller in size. However, since PPD entry corresponds to I-cache block, their design increases complexity particularly with set-associative I-cache. Third, they use profiling to identify highly-biased branches which do not require static prediction. They also note that even on using an adaptive control strategy,  pipeline gating does not provide energy saving, which is due to inaccuracy of pipeline gating itself and wrong-path instructions waste only small energy before being squashed on a misprediction.

\subsection{Avoiding redundant reads and updates}\label{sec:noredundant} 
 
Baniasadi et al. \cite{baniasadi2004sepas}  note that for several branches, after completion of the learning phase, the collected information remains same for a long time (e.g., a loop with large number of iterations). Thus, in their steady-state phase, these branches are easy to predict. Their technique identifies branches which were taken for a threshold number of times and were predicted accurately and does not access BP for them. Also, if BP already has sufficient information for accurate prediction, BP is not updated. They further note that several other branches can be accurately predicted with only one of the component predictors in a hybrid predictor and these branches generally use the same predictor as they used the last time. For such branches, only one component-predictor is accessed. Their technique reduces BP accesses and energy consumption with minor impact on  performance. 

Seznec et al. \cite{seznec2011new} propose strategies to reduce hardware overhead of TAGE and improve its accuracy by enhancing it with side predictors. They note that updating predictor tables at retire time avoids pollution by wrong path, however, this late update leads to extra mispredictions over ideal policy of fetch-time update. Any branch on correct path requires three accesses to predictor: reads at  prediction and retire stages and write at retire stage. This necessitates complex bank-interleaved or multiported designs. They show that unlike in other predictors (e.g., gshare \cite{mcfarling1993combining}, GEHL \cite{seznec2005analysis}), in TAGE predictor, avoiding retire-time read leads to only minor loss in accuracy. Combining this with redundant-update avoidance scheme \cite{baniasadi2004sepas} leads to significant reduction in BP accesses, e.g. only 1.13 accesses for each retired branch on average. This allows designing TAGE predictor as a 4-way bank-interleaved design with single-ported memory banks, which provides large saving in area and energy. The bank-interleaving approach can be used with most global history BPs with only small accuracy loss, although using it local history BPs leads to large losses.

    To  further reduce mispredictions due to late  
updates, they propose integrating side predictors with TAGE. The ``immediate update mimicker'' (IUM)  reduces such mispredictions by predicting branches which have in-flight non-retired instances. If two branches B1 and B2 are predicted by the same table and same table entry, and B1 has already executed but has not retired, then outcome of branch B1 is used as the prediction for branch B2 instead of the output of the predictor. Further, TAGE cannot predict loop exits for loops with irregular control flow. To predict such loop exits, they use a loop predictor \cite{gao2005adaptive}. Since TAGE fails to accurately predict biased branches which are uncorrelated with branch path or history, they also use a statistical-corrector predictor \cite{seznec2011isltage}. Statistical-corrector predictor can also utilize local history to improve effectiveness of TAGE. Overall, their enhanced TAGE predictor achieves high accuracy.

\section{Conclusion and Future Outlook}\label{sec:conclusion}
In this paper, we presented a survey of dynamic branch prediction techniques. We classified the research works to present a bird's-eye view of the field. We reviewed several research proposals to bring forth their key insights and relative merits. We hope that real contribution of this survey will be to stimulate further research in this area and make definite impact on the BPs used in the next-generation processors. We close the paper with a brief mention of future challenges in this area.

Since BP is a speculative technique which wastes energy, its use in highly energy-constrained systems such as battery-operated mobile devices is challenging \cite{mittal2014SurveyEmbeddedPowerManagement}.  To reduce BP energy consumption, it can be designed with non-volatile memory  or aggressive energy management techniques such as power-gating can be used \cite{mittal2014surveycache}. Also, developing techniques for improving BP accuracy and reducing energy wastage on wrong-path execution  will be a major research challenge for computer architects. 
 
\revised{BPs are useful not only for predicting program control flow to improve performance, but also to predict occurrence of interesting event to guide power-management of processors. Recently, Bhattacharjee  \cite{bhattacharjee2017using} has proposed using BP for predicting brain activity in brain-machine implants. Since these implants need to achieve extremely high energy efficiency, they propose running the embedded processor at normal frequency in case of an interesting activity and transitioning it to low-power mode otherwise. They use multiple BPs, e.g., gshare \cite{mcfarling1993combining}, two-level BPs \cite{yeh1991two}, Smith BP \cite{smith1981study} and perceptron BP \cite{jimenez2002neural} for   predicting neuronal activity in the cerebellum. They observe that perceptron BP can predict cerebellar activity with high accuracy (up to 85\%). Based on this, the processor can be intelligently transitioned to low-power modes which saves power.   The reason for high accuracy of perceptron BP is its ability to track correlations with much longer history than other BPs. These and similar techniques are also effective in mitigating energy-overhead of BP. Evidently, BPs are versatile and powerful tools and can find use in processors used for health/medical, space (e.g., saving energy in robots operated in remote environments) and military applications.}

The lack of detailed information about BPs in real processors and subsequent differences in the simulation frameworks used in academia and industry \cite{loh2005simulation} may threaten the validity of BP studies. To address this, a closer collaboration between academia and industry is required. Further, BPs evaluated with old workloads with small footprints (e.g., SPEC92) may not be optimized for recent workloads \cite{sechrest1996correlation} since in SPEC92 workloads, a few branches are executed very frequently and handling them improves overall accuracy whereas this strategy may not reduce overall accuracy in recent workloads with large number of static branches. Clearly, evaluation of BPs with up-to-date workloads is important to obtain meaningful conclusions. 

Due to  aggressive performance optimization features present in modern processors, different branches have different misprediction overheads \cite{loh2002predicting}.  Hence, an improvement in BP accuracy may not translate into corresponding performance improvement. Yet, several works only report BP accuracy  \cite{seznec2006case,albericio2014wormhole,seznec2015inner,jimenez2001dynamic,gao2005adaptive,loh2005reducing,jimenez2011optimized,seznec2002design,lee1997bi,sadooghi2012toward,schlais2016badgr,porter2009creating,pan1992improving,seznec2005analysis,yeh1991two,yeh1993comparison,seznec2011new,sechrest1996correlation,egan2003two,chang1996improving,gao2007pmpm,ma2006using,lai2005improving,seznec1999dealiased,juan1998dynamic,kampe2002fab,stark1998variable,monchiero2005combined} without reporting system-performance and hence, they may provide misleading impression of the potential of a BP management technique. To address this issue, researchers need to adopt comprehensive evaluation methodology where both accuracy and performance impact of BPs are evaluated.


\ifCLASSOPTIONcaptionsoff
  \newpage
\fi



%
{\footnotesize
\bibliographystyle{IEEEtran1}
\bibliography{References}

\begin{thebibliography}{100}
\providecommand{\url}[1]{#1}
\csname url@samestyle\endcsname
\providecommand{\newblock}{\relax}
\providecommand{\bibinfo}[2]{#2}
\providecommand{\BIBentrySTDinterwordspacing}{\spaceskip=0pt\relax}
\providecommand{\BIBentryALTinterwordstretchfactor}{4}
\providecommand{\BIBentryALTinterwordspacing}{\spaceskip=\fontdimen2\font plus
\BIBentryALTinterwordstretchfactor\fontdimen3\font minus
  \fontdimen4\font\relax}
\providecommand{\BIBforeignlanguage}[2]{{%
\expandafter\ifx\csname l@#1\endcsname\relax
\typeout{** WARNING: IEEEtran.bst: No hyphenation pattern has been}%
\typeout{** loaded for the language `#1'. Using the pattern for}%
\typeout{** the default language instead.}%
\else
\language=\csname l@#1\endcsname
\fi
#2}}
\providecommand{\BIBdecl}{\relax}
\BIBdecl

\bibitem{seznec2002design}
A.~Seznec, S.~Felix, V.~Krishnan, and Y.~Sazeides, ``{Design tradeoffs for the
  Alpha EV8 conditional branch predictor},'' in \emph{ISCA}, 2002, pp.
  295--306.

\bibitem{jimenez2011optimized}
D.~A. Jim{\'e}nez, ``An optimized scaled neural branch predictor,'' in
  \emph{ICCD}, 2011, pp. 113--118.

\bibitem{jimenez2003reconsidering}
D.~A. Jim{\'e}nez, ``Reconsidering complex branch predictors,'' in \emph{HPCA},
  2003, pp. 43--52.

\bibitem{jimenez2000impact}
D.~A. Jim{\'e}nez, S.~W. Keckler, and C.~Lin, ``The impact of delay on the
  design of branch predictors,'' in \emph{MICRO}, 2000, pp. 67--76.

\bibitem{smith1981study}
J.~E. Smith, ``A study of branch prediction strategies,'' in \emph{ISCA}, 1981,
  pp. 135--148.

\bibitem{skadron2000taxonomy}
K.~Skadron, M.~Martonosi, and D.~W. Clark, ``A taxonomy of branch
  mispredictions, and alloyed prediction as a robust solution to wrong-history
  mispredictions,'' \emph{PACT}, 2000.

\bibitem{mittal2017ValueLocality}
S.~Mittal, ``A survey of value prediction techniques for leveraging value
  locality,'' \emph{Concurrency and Computation: Practice and Experience},
  2017.

\bibitem{young1995comparative}
C.~Young, N.~Gloy, and M.~D. Smith, ``A comparative analysis of schemes for
  correlated branch prediction,'' \emph{ISCA}, pp. 276--286, 1995.

\bibitem{loh2005simple}
G.~H. Loh, ``A simple divide-and-conquer approach for neural-class branch
  prediction,'' in \emph{PACT}, 2005, pp. 243--254.

\bibitem{evers1996using}
M.~Evers, P.-Y. Chang, and Y.~N. Patt, ``Using hybrid branch predictors to
  improve branch prediction accuracy in the presence of context switches,'' in
  \emph{ISCA}, 1996, pp. 3--11.

\bibitem{bonanno2013two}
J.~Bonanno, A.~Collura, D.~Lipetz, U.~Mayer, B.~Prasky, and A.~Saporito, ``Two
  level bulk preload branch prediction,'' in \emph{HPCA}, 2013, pp. 71--82.

\bibitem{fog2011microarchitecture}
A.~Fog, ``{The microarchitecture of Intel, AMD and VIA CPUs},'' Technical
  University of Denmark, Tech. Rep., 2011.

\bibitem{juan1998dynamic}
T.~Juan, S.~Sanjeevan, and J.~J. Navarro, ``Dynamic history-length fitting: A
  third level of adaptivity for branch prediction,'' in \emph{ISCA}, 1998, pp.
  155--166.

\bibitem{milenkovic2002demystifying}
M.~Milenkovic, A.~Milenkovic, and J.~Kulick, ``Demystifying intel branch
  predictors,'' in \emph{Workshop on Duplicating, Deconstructing and
  Debunking}, 2002.

\bibitem{gwennap1996digital}
L.~Gwennap \emph{et~al.}, ``Digital 21264 sets new standard,''
  \emph{Microprocessor report}, vol.~10, no.~14, pp. 11--16, 1996.

\bibitem{yeh1993comparison}
T.-Y. Yeh and Y.~N. Patt, ``A comparison of dynamic branch predictors that use
  two levels of branch history,'' in \emph{ISCA}, 1993, pp. 257--266.

\bibitem{gope2014bias}
D.~Gope and M.~H. Lipasti, ``Bias-free branch predictor,'' in
  \emph{International Symposium on Microarchitecture (MICRO)}.\hskip 1em plus
  0.5em minus 0.4em\relax IEEE, 2014, pp. 521--532.

\bibitem{loh2005reducing}
G.~H. Loh and D.~A. Jimenez, ``Reducing the power and complexity of path-based
  neural branch prediction,'' in \emph{Workshop on Complexity Effective Design
  (WCED)}, 2005, pp. 1--8.

\bibitem{sprangle1997agree}
E.~Sprangle, R.~S. Chappell, M.~Alsup, and Y.~N. Patt, ``The agree predictor: A
  mechanism for reducing negative branch history interference,'' in
  \emph{ISCA}, 1997, pp. 284--291.

\bibitem{seznec2006case}
A.~Seznec and P.~Michaud, ``A case for (partially)-tagged geometric history
  length predictors,'' \emph{Journal of Instruction Level Parallelism}, 2006.

\bibitem{evers1998analysis}
M.~Evers, S.~J. Patel, R.~S. Chappell, and Y.~N. Patt, ``An analysis of
  correlation and predictability: What makes two-level branch predictors
  work,'' in \emph{ACM SIGARCH Computer Architecture News}, vol.~26, no.~3,
  1998, pp. 52--61.

\bibitem{gao2005adaptive}
H.~Gao and H.~Zhou, ``Adaptive information processing: An effective way to
  improve perceptron branch predictors,'' in \emph{JILP}, 2005.

\bibitem{tarjan2005merging}
D.~Tarjan and K.~Skadron, ``Merging path and gshare indexing in perceptron
  branch prediction,'' \emph{ACM transactions on architecture and code
  optimization (TACO)}, vol.~2, no.~3, pp. 280--300, 2005.

\bibitem{akkary2004perceptron}
H.~Akkary, S.~T. Srinivasan, R.~Koltur, Y.~Patil, and W.~Refaai,
  ``Perceptron-based branch confidence estimation,'' in \emph{HPCA}, 2004, pp.
  265--265.

\bibitem{cooper2011engineering}
K.~Cooper and L.~Torczon, \emph{Engineering a compiler}.\hskip 1em plus 0.5em
  minus 0.4em\relax Elsevier, 2011.

\bibitem{nair1995dynamic}
R.~Nair, ``Dynamic path-based branch correlation,'' in \emph{MICRO}, 1995, pp.
  15--23.

\bibitem{seznec2005analysis}
A.~Seznec, ``{Analysis of the O-GEometric History Length branch predictor},''
  in \emph{ISCA}, 2005, pp. 394--405.

\bibitem{jimenez2002neural}
D.~A. Jim{\'e}nez and C.~Lin, ``Neural methods for dynamic branch prediction,''
  \emph{ACM Transactions on Computer Systems (TOCS)}, vol.~20, no.~4, pp.
  369--397, 2002.

\bibitem{loh2006revisiting}
G.~H. Loh, ``Revisiting the performance impact of branch predictor latencies,''
  in \emph{ISPASS}, 2006, pp. 59--69.

\bibitem{seznec201164}
A.~Seznec, ``{A 64 Kbytes ISL-TAGE Branch Predictor},'' in \emph{JWAC-2:
  Championship Branch Prediction}, 2011.

\bibitem{seznec1999dealiased}
A.~Seznec and P.~Michaud, ``De-aliased hybrid branch predictors,'' Ph.D.
  dissertation, INRIA, 1999.

\bibitem{mittal2014SurveyDataCenter}
S.~Mittal, ``Power management techniques for data centers: A survey,'' Oak
  Ridge National Laboratory, USA, Tech. Rep. ORNL/TM-2014/381, 2014.

\bibitem{eden1998yags}
A.~N. Eden and T.~Mudge, ``{The YAGS branch prediction scheme},'' in
  \emph{MICRO}, 1998, pp. 69--77.

\bibitem{aragon2001selective}
J.~L. Arag{\'o}n, J.~Gonz{\'a}lez, J.~M. Garc{\'\i}a, and A.~Gonz{\'a}lez,
  ``Selective branch prediction reversal by correlating with data values and
  control flow,'' in \emph{ICCD}, 2001, pp. 228--233.

\bibitem{bhattacharjee2017using}
A.~Bhattacharjee, ``Using branch predictors to predict brain activity in
  brain-machine implants,'' in \emph{International Symposium on
  Microarchitecture}, 2017, pp. 409--422.

\bibitem{sethuram2007neural}
R.~Sethuram, O.~I. Khan, H.~V. Venkatanarayanan, and M.~L. Bushnell, ``A neural
  net branch predictor to reduce power,'' in \emph{VLSID}, 2007, pp. 679--684.

\bibitem{ayoub2009filtering}
R.~Ayoub and A.~Orailoglu, ``{Filtering global history: Power and Performance
  Efficient Branch Predictor},'' in \emph{International Conference on
  Application-specific Systems, Architectures and Processors}, 2009.

\bibitem{xie2013energy}
Z.~Xie, D.~Tong, and X.~Cheng, ``An energy-efficient branch prediction
  technique via global-history noise reduction,'' \emph{ISLPED}, 2013.

\bibitem{huang2015energy}
M.~Huang, D.~He, X.~Liu, M.~Tan, and X.~Cheng, ``An energy-efficient branch
  prediction with grouped global history,'' in \emph{ICPP}, 2015, pp. 140--149.

\bibitem{schlais2016badgr}
D.~J. Schlais and M.~H. Lipasti, ``{BADGR: A practical GHR implementation for
  TAGE branch predictors},'' in \emph{ICCD}, 2016, pp. 536--543.

\bibitem{gao2008address}
H.~Gao, Y.~Ma, M.~Dimitrov, and H.~Zhou, ``Address-branch correlation: A novel
  locality for long-latency hard-to-predict branches,'' in \emph{HPCA}, 2008,
  pp. 74--85.

\bibitem{wang2013practical}
J.~Wang, Y.~Tim, W.-F. Wong, and H.~H. Li, ``A practical low-power
  memristor-based analog neural branch predictor,'' in \emph{ISLPED}, 2013, pp.
  175--180.

\bibitem{amant2008low}
R.~S. Amant, D.~A. Jim{\'e}nez, and D.~Burger, ``Low-power, high-performance
  analog neural branch prediction,'' in \emph{MICRO}, 2008, pp. 447--458.

\bibitem{baniasadi2004sepas}
A.~Baniasadi and A.~Moshovos, ``{SEPAS: A highly accurate energy-efficient
  branch predictor},'' in \emph{ISLPED}, 2004, pp. 38--43.

\bibitem{yang2006power}
C.~Yang and A.~Orailoglu, ``Power efficient branch prediction through early
  identification of branch addresses,'' in \emph{CASES}, 2006, pp. 169--178.

\bibitem{sendag2008low}
R.~Sendag, J.~Y. Joshua, P.-f. Chuang, and D.~J. Lilja, ``Low power/area branch
  prediction using complementary branch predictors,'' in \emph{IPDPS}, 2008,
  pp. 1--12.

\bibitem{monchiero2005combined}
M.~Monchiero and G.~Palermo, ``The combined perceptron branch predictor,''
  \emph{Euro-Par 2005 Parallel Processing}, p. 487–496, 2005.

\bibitem{jimenez2001dynamic}
D.~A. Jim{\'e}nez and C.~Lin, ``Dynamic branch prediction with perceptrons,''
  in \emph{HPCA}, 2001, pp. 197--206.

\bibitem{jimenez2003fast}
D.~A. Jim{\'e}nez, ``Fast path-based neural branch prediction,'' in
  \emph{MICRO}, 2003, p. 243.

\bibitem{saadeldeen2013memristors}
H.~Saadeldeen, D.~Franklin, G.~Long, C.~Hill, A.~Browne, D.~Strukov,
  T.~Sherwood, and F.~T. Chong, ``Memristors for neural branch prediction: a
  case study in strict latency and write endurance challenges,'' in
  \emph{International Conference on Computing Frontiers}, 2013, p.~26.

\bibitem{jimenez2005piecewise}
D.~A. Jim{\'e}nez, ``Piecewise linear branch prediction,'' in \emph{ISCA},
  2005, pp. 382--393.

\bibitem{egan2003two}
C.~Egan, G.~Steven, P.~Quick, R.~Anguera, F.~Steven, and L.~Vintan, ``Two-level
  branch prediction using neural networks,'' \emph{Journal of Systems
  Architecture}, vol.~49, no.~12, pp. 557--570, 2003.

\bibitem{jimenez2001perceptron}
D.~A. Jim{\'e}nez and C.~Lin, ``Perceptron learning for predicting the behavior
  of conditional branches,'' in \emph{Neural Networks, 2001. Proceedings.
  IJCNN'01. International Joint Conference on}, vol.~3.\hskip 1em plus 0.5em
  minus 0.4em\relax IEEE, 2001, pp. 2122--2127.

\bibitem{chaver2003branch}
D.~Chaver, L.~Pi{\~n}uel, M.~Prieto, F.~Tirado, and M.~C. Huang, ``Branch
  prediction on demand: an energy-efficient solution,'' in \emph{ISLPED}, 2003,
  pp. 390--395.

\bibitem{falcon2004prophet}
A.~Falcon, J.~Stark, A.~Ramirez, K.~Lai, and M.~Valero, ``Prophet/critic hybrid
  branch prediction,'' in \emph{ISCA}, 2004, p. 250.

\bibitem{kampe2002fab}
M.~Kampe, P.~Stenstrom, and M.~Dubois, ``{The FAB predictor: Using fourier
  analysis to predict the outcome of conditional branches},'' in \emph{HPCA},
  2002, pp. 223--232.

\bibitem{baniasadi2002branch}
A.~Baniasadi and A.~Moshovos, ``Branch predictor prediction: A power-aware
  branch predictor for high-performance processors,'' in \emph{ICCD}, 2002, pp.
  458--461.

\bibitem{evers2000improving}
M.~Evers, ``Improving branch prediction by understanding branch behavior,''
  Ph.D. dissertation, The University of Michigan, 2000.

\bibitem{seznec2011new}
A.~Seznec, ``A new case for the tage branch predictor,'' in \emph{MICRO}, 2011.

\bibitem{seznec2015inner}
A.~Seznec, J.~S. Miguel, and J.~Albericio, ``The inner most loop iteration
  counter: a new dimension in branch history,'' in \emph{MICRO}, 2015, pp.
  347--357.

\bibitem{albericio2014wormhole}
J.~Albericio, J.~S. Miguel, N.~E. Jerger, and A.~Moshovos, ``Wormhole: Wisely
  predicting multidimensional branches,'' in \emph{MICRO}, 2014, pp. 509--520.

\bibitem{heil1999improving}
T.~H. Heil, Z.~Smith, and J.~E. Smith, ``Improving branch predictors by
  correlating on data values,'' in \emph{MICRO}, 1999, pp. 28--37.

\bibitem{yeh1993increasing}
T.-Y. Yeh, D.~T. Marr, and Y.~N. Patt, ``Increasing the instruction fetch rate
  via multiple branch prediction and a branch address cache,'' in \emph{ICS},
  1993, pp. 67--76.

\bibitem{chang1994branch}
P.-Y. Chang, E.~Hao, T.-Y. Yeh, and Y.~Patt, ``Branch classification: a new
  mechanism for improving branch predictor performance,'' in \emph{MICRO},
  1994, pp. 22--31.

\bibitem{parikh2004power}
D.~Parikh, K.~Skadron, Y.~Zhang, and M.~Stan, ``Power-aware branch prediction:
  Characterization and design,'' \emph{IEEE Transactions on Computers},
  vol.~53, no.~2, pp. 168--186, 2004.

\bibitem{parikh2002power}
D.~Parikh, K.~Skadron, Y.~Zhang, M.~Barcella, and M.~R. Stan, ``Power issues
  related to branch prediction,'' in \emph{HPCA}, 2002, pp. 233--244.

\bibitem{stark1998variable}
J.~Stark, M.~Evers, and Y.~N. Patt, ``Variable length path branch prediction,''
  in \emph{ASPLOS}.\hskip 1em plus 0.5em minus 0.4em\relax ACM, 1998, pp.
  170--179.

\bibitem{loh2002predicting}
G.~H. Loh and D.~S. Henry, ``Predicting conditional branches with fusion-based
  hybrid predictors,'' in \emph{PACT}, 2002, pp. 165--176.

\bibitem{sechrest1996correlation}
S.~Sechrest, C.-C. Lee, and T.~Mudge, ``Correlation and aliasing in dynamic
  branch predictors,'' in \emph{ISCA}, 1996, pp. 22--32.

\bibitem{chen1996analysis}
N.~Gloy, C.~Young, J.~B. Chen, and M.~D. Smith, ``An analysis of dynamic branch
  prediction schemes on system workloads,'' in \emph{ISCA}, 1996.

\bibitem{li2007aware}
T.~Li, L.~K. John, A.~Sivasubramaniam, N.~Vijaykrishnan, and J.~Rubio,
  ``Os-aware branch prediction: Improving microprocessor control flow
  prediction for operating systems,'' \emph{IEEE Transactions on Computers},
  vol.~56, no.~1, 2007.

\bibitem{chang1996improving}
P.-Y. Chang, M.~Evers, , and Y.~Patt, ``Improving branch prediction accuracy by
  reducing pattern history table interference,'' \emph{PACT}, 1996.

\bibitem{sherwood2000loop}
T.~Sherwood and B.~Calder, ``Loop termination prediction,'' in
  \emph{International Symposium on High Performance Computing}.\hskip 1em plus
  0.5em minus 0.4em\relax Springer, 2000, pp. 73--87.

\bibitem{alotoom2010exact}
M.~Al-Otoom, E.~Forbes, and E.~Rotenberg, ``{EXACT: Explicit Dynamic-Branch
  Prediction with Active Updates},'' in \emph{Computing Frontiers}, 2010.

\bibitem{chen2003dynamic}
L.~Chen, S.~Dropsho, and D.~H. Albonesi, ``Dynamic data dependence tracking and
  its application to branch prediction,'' in \emph{High-Performance Computer
  Architecture, 2003. HPCA-9 2003. Proceedings. The Ninth International
  Symposium on}.\hskip 1em plus 0.5em minus 0.4em\relax IEEE, 2003, pp. 65--76.

\bibitem{porter2009creating}
L.~Porter and D.~M. Tullsen, ``Creating artificial global history to improve
  branch prediction accuracy,'' in \emph{ICS}, 2009, pp. 266--275.

\bibitem{yeh1991two}
T.-Y. Yeh and Y.~N. Patt, ``Two-level adaptive training branch prediction,'' in
  \emph{MICRO}, 1991, pp. 51--61.

\bibitem{yeh1992alternative}
T.-Y. Yeh and Y.~N. Patt, ``Alternative implementations of two-level adaptive
  branch prediction,'' in \emph{ISCA}, 1992, pp. 124--134.

\bibitem{pan1992improving}
S.-T. Pan, K.~So, and J.~T. Rahmeh, ``Improving the accuracy of dynamic branch
  prediction using branch correlation,'' in \emph{ASPLOS}, 1992, pp. 76--84.

\bibitem{lee1997bi}
C.-C. Lee, I.-C.~K. Chen, and T.~N. Mudge, ``The bi-mode branch predictor,'' in
  \emph{MICRO}, 1997, pp. 4--13.

\bibitem{mcfarling1993combining}
S.~McFarling, ``Combining branch predictors,'' Digital Western Research
  Laboratory, Tech. Rep., 1993.

\bibitem{chen1996analysisdata}
I.-C.~K. Chen, J.~T. Coffey, and T.~N. Mudge, ``Analysis of branch prediction
  via data compression,'' \emph{ACM SIGPLAN Notices}, vol.~31, no.~9, pp.
  128--137, 1996.

\bibitem{cleary1984data}
J.~Cleary and I.~Witten, ``Data compression using adaptive coding and partial
  string matching,'' \emph{IEEE transactions on Communications}, vol.~32,
  no.~4, pp. 396--402, 1984.

\bibitem{gao2007pmpm}
H.~Gao and H.~Zhou, ``Pmpm: Prediction by combining multiple partial matches,''
  \emph{Journal of Instruction-Level Parallelism}, vol.~9, pp. 1--18, 2007.

\bibitem{michaud2005ppm}
P.~Michaud, ``{A PPM-like, tag-based branch predictor},'' \emph{Journal of
  Instruction Level Parallelism}, vol.~7, no.~1, pp. 1--10, 2005.

\bibitem{michaud1997trading}
P.~Michaud, A.~Seznec, and R.~Uhlig, ``Trading conflict and capacity aliasing
  in conditional branch predictors,'' in \emph{ISCA}, 1997, pp. 292--303.

\bibitem{seznec1993case}
A.~Seznec, ``A case for two-way skewed-associative caches,'' in \emph{ACM
  SIGARCH Computer Architecture News}, vol.~21, no.~2.\hskip 1em plus 0.5em
  minus 0.4em\relax ACM, 1993, pp. 169--178.

\bibitem{ma2006using}
Y.~Ma, H.~Gao, and H.~Zhou, ``Using indexing functions to reduce conflict
  aliasing in branch prediction tables,'' \emph{IEEE Transactions on
  Computers}, vol.~55, no.~8, pp. 1057--1061, 2006.

\bibitem{seznec2011storage}
A.~Seznec, ``{Storage free confidence estimation for the TAGE branch
  predictor},'' in \emph{HPCA}, 2011, pp. 443--454.

\bibitem{seznec2007256}
A.~Seznec, ``{A 256 kbits L-TAGE branch predictor},'' \emph{Journal of
  Instruction-Level Parallelism (JILP)}, vol.~9, 2007.

\bibitem{seznec2011isltage}
A.~Seznec, ``{A 64-Kbytes ISL-TAGE branch predictor},'' in \emph{JWAC-2:
  Championship Branch Prediction}, 2011.

\bibitem{lai2005improving}
C.~Lai, S.-L. Lu, Y.~Chen, and T.~Chen, ``Improving branch prediction accuracy
  with parallel conservative correctors,'' in \emph{Computing frontiers}.\hskip
  1em plus 0.5em minus 0.4em\relax ACM, 2005, pp. 334--341.

\bibitem{mittal2016SurveyCPURF}
S.~Mittal, ``{A Survey of Techniques for Designing and Managing CPU Register
  File},'' \emph{Concurrency and Computation: Practice and Experience}, 2016.

\bibitem{thomas2003improving}
R.~Thomas, M.~Franklin, C.~Wilkerson, and J.~Stark, ``Improving branch
  prediction by dynamic dataflow-based identification of correlated branches
  from a large global history,'' in \emph{ISCA}, 2003, pp. 314--323.

\bibitem{choi2008accurate}
B.~Choi, L.~Porter, and D.~M. Tullsen, ``Accurate branch prediction for short
  threads,'' \emph{ASPLOS}, vol.~36, no.~1, 2008.

\bibitem{hily1996branch}
S.~Hily and A.~Seznec, ``Branch prediction and simultaneous multithreading,''
  in \emph{Parallel Architectures and Compilation Techniques, 1996.,
  Proceedings of the 1996 Conference on}.\hskip 1em plus 0.5em minus
  0.4em\relax IEEE, 1996, pp. 169--173.

\bibitem{akkary2003recycling}
H.~Akkary, S.~T. Srinivasan, and K.~Lai, ``Recycling waste: Exploiting
  wrong-path execution to improve branch prediction,'' in \emph{ICS}, 2003.

\bibitem{jimenez2006controlling}
D.~A. Jimenez and G.~H. Loh, ``Controlling the power and area of neural branch
  predictors for practical implementation in high-performance processors,'' in
  \emph{SBAC-PAD}, 2006, pp. 55--62.

\bibitem{manne1999branch}
S.~Manne, A.~Klauser, and D.~Grunwald, ``Branch prediction using selective
  branch inversion,'' in \emph{Parallel Architectures and Compilation
  Techniques, 1999. Proceedings. 1999 International Conference on}.\hskip 1em
  plus 0.5em minus 0.4em\relax IEEE, 1999, pp. 48--56.

\bibitem{seznec1996multiple}
A.~Seznec, S.~Jourdan, P.~Sainrat, and P.~Michaud, ``Multiple-block ahead
  branch predictors,'' in \emph{ASPLOS}, 1996, p. 116–127.

\bibitem{sadooghi2012toward}
M.~Sadooghi-Alvandi, K.~Aasaraai, and A.~Moshovos, ``Toward virtualizing branch
  direction prediction,'' in \emph{DATE}, 2012, pp. 455--460.

\bibitem{hu2002applying}
Z.~Hu, P.~Juang, K.~Skadron, D.~Clark, and M.~Martonosi, ``Applying decay
  strategies to branch predictors for leakage energy savings,'' in \emph{ICCD},
  2002, pp. 442--445.

\bibitem{santana2003latency}
O.~J. Santana, A.~Ramirez, and M.~Valero, ``Latency tolerant branch
  predictors,'' in \emph{Innovative Architecture for Future Generation
  High-Performance Processors and Systems, 2003}.\hskip 1em plus 0.5em minus
  0.4em\relax IEEE, 2003, pp. 30--39.

\bibitem{mittal2014SurveyEmbeddedPowerManagement}
S.~Mittal, ``{A Survey of Techniques For Improving Energy Efficiency in
  Embedded Computing Systems},'' \emph{International Journal of Computer Aided
  Engineering and Technology (IJCAET)}, vol.~6, no.~4, pp. 440--459, 2014.

\bibitem{mittal2014surveycache}
S.~Mittal, ``A survey of architectural techniques for improving cache power
  efficiency,'' \emph{Elsevier Sustainable Computing: Informatics and Systems},
  vol.~4, no.~1, pp. 33--43, 2014.

\bibitem{loh2005simulation}
G.~H. Loh, ``Simulation differences between academia and industry: A branch
  prediction case study,'' in \emph{Performance Analysis of Systems and
  Software, 2005. ISPASS 2005. IEEE International Symposium on}.\hskip 1em plus
  0.5em minus 0.4em\relax IEEE, 2005, pp. 21--31.

\end{thebibliography}
}

%



\end{document}